\documentclass[11pt,a4paper]{article}

\usepackage[english]{babel}
\usepackage[utf8x]{inputenc}
\usepackage[T1]{fontenc}
\usepackage{times}
\usepackage[a4paper,top=3cm,bottom=2cm,left=3cm,right=3cm,marginparwidth=1.75cm]{geometry}
\usepackage{amsmath, amssymb}
\usepackage{graphicx} 
\usepackage{bm}
\usepackage[colorinlistoftodos]{todonotes}
\usepackage[colorlinks=true, allcolors=blue]{hyperref}
\usepackage{paralist}       
\usepackage{enumerate,xstring}
\usepackage{xfrac}
\usepackage{array}
\usepackage{arydshln}
\usepackage{threeparttable}
\usepackage{pifont}
\usepackage{wrapfig}
\usepackage{aas_macros}
\usepackage[authoryear]{natbib}

\usepackage{subfigure}

\def\comment#1{}

\usepackage{stackengine}
\usepackage{amsmath}
\usepackage{hyperref}

\count\footins = 1000 

\DeclareMathOperator{\arcsinh}{arcsinh}

\newcommand{\dd}{\mathrm{d}}
\newcommand{\gsim}{\;\mbox{\raisebox{-0.5ex}{$\stackrel{>}{\scriptstyle{\sim}}$}
}\;}
\newcommand{\lsim}{\;\mbox{\raisebox{-0.5ex}{$\stackrel{<}{\scriptstyle{\sim}}$}
}\;}
\newcommand{\mpl}{M_{\rm pl}}
\newcommand{\rs}{{r_{\rm s}}}
\newcommand{\ms}{M_{\rm s}}
\newcommand{\rv}{r_{\rm V}}

\def\>{\right\rangle}
\def\<{\left\langle}
\def\be{\begin{equation}}
\def\ee{\end{equation}}
\def\bea{\begin{eqnarray}}
\def\eea{\end{eqnarray}}
\def\d{\delta}
\def\vn{\vec{\nabla}} 
\def\v#1{\vec{#1}} 
\def\refeq#1{Eq.~(\ref{eq:#1})}

\usepackage{titlesec}
\titleclass{\subsubsubsection}{straight}[\subsection]

\titleformat{\paragraph}
{\normalfont\normalsize\bfseries}{\theparagraph}{1em}{}
\titlespacing*{\paragraph}{0pt}{3.25ex plus 1ex minus .2ex}{1.5ex plus .2ex}

\newcounter{subsubsubsection}[subsubsection]
\renewcommand\thesubsubsubsection{\thesubsubsection.\arabic{subsubsubsection}}
\renewcommand\theparagraph{\thesubsubsubsection.\arabic{paragraph}} 

\titleformat{\subsubsubsection}
  {\normalfont\normalsize\bfseries}{\thesubsubsubsection}{1em}{}
\titlespacing*{\subsubsubsection}
{0pt}{3.25ex plus 1ex minus .2ex}{1.5ex plus .2ex}

\makeatletter
\renewcommand\paragraph{\@startsection{paragraph}{5}{\z@}%
  {3.25ex \@plus1ex \@minus.2ex}%
  {-1em}%
  {\normalfont\normalsize\bfseries}}
\renewcommand\subparagraph{\@startsection{subparagraph}{6}{\parindent}%
  {3.25ex \@plus1ex \@minus .2ex}%
  {-1em}%
  {\normalfont\normalsize\bfseries}}
\def\toclevel@subsubsubsection{4}
\def\toclevel@paragraph{5}
\def\toclevel@paragraph{6}
\def\l@subsubsubsection{\@dottedtocline{4}{7em}{4em}}
\def\l@paragraph{\@dottedtocline{5}{10em}{5em}}
\def\l@subparagraph{\@dottedtocline{6}{14em}{6em}}
\makeatother

\newcommand{\bq}{\begin{eqnarray}}
\newcommand{\eq}{\end{eqnarray}}

\title{The Novel Probes Project --- Tests of Gravity on Astrophysical Scales}
\author{Tessa Baker$^{1}$,
Alexandre Barreira$^2$,
Harry Desmond$^3$\thanks{\href{mailto:harry.desmond@physics.ox.ac.uk}{harry.desmond@physics.ox.ac.uk}},
Pedro Ferreira$^3$,\\
Bhuvnesh Jain$^4$,
Kazuya Koyama$^5$,
Baojiu Li$^6$ ,
Lucas Lombriser$^{7,8}$,\\
Andrina Nicola$^9$,
Jeremy Sakstein$^{4}$\thanks{\href{mailto:sakstein@physics.upenn.edu}{sakstein@hawaii.edu}},
Fabian Schmidt$^2$
}

\date{ 
\small
$^1$ School of Physics and Astronomy, Queen Mary University of London, Mile End Road, London, E1 4NS, UK\\
$^2$ Max-Planck-Institut f{\"u}r Astrophysik, Karl-Schwarzschild-Str.~1, 85741 Garching, Germany\\
$^3$ Astrophysics, University of Oxford, Denys Wilkinson Building, Keble Road, Oxford OX1 3RH, UK\\
$^4$ Center for Particle Cosmology,
Department of Physics and Astronomy,
University of Pennsylvania,
209 S. 33rd St., Philadelphia, PA 19104, USA\\
$^5$ Institute of Cosmology $\&$ Gravitation, University of Portsmouth, Dennis Sciama Building, Burnaby Road, Portsmouth, PO1 3FX, UK\\
$^6$ Institute for Computational Cosmology, Department of Physics, Durham University, South Road, Durham DH1 3LE, UK\\
$^7$ D\'epartement de Physique Th\'eorique, Universit\'e de Gen\`eve, 24 quai Ernest Ansermet, 1211 Gen\`eve 4, Switzerland\\
$^8$ Institute for Astronomy, University of Edinburgh,
Royal Observatory, Blackford Hill, Edinburgh, EH9 3HJ, UK\\
$^9$ Department of Astrophysical Sciences, Princeton University, Princeton, NJ 08544, USA
}

\begin{document}
\maketitle

\begin{abstract}
We introduce The Novel Probes Project, an initiative to advance the field of astrophysical tests of the dark sector by creating a forum that connects observers and theorists. This review focuses on tests of gravity and is intended to be of use primarily to observers, as well as theorists with interest in the development of experimental tests. It is twinned with a separate upcoming review on dark matter self-interactions.

Our focus is on astrophysical tests of gravity in the weak-field regime, ranging from stars to quasilinear cosmological scales. This regime is complementary to both strong-field tests of gravity and background and linear probes in cosmology. In particular, the nonlinear screening mechanisms which are an integral part of viable modified gravity models lead to characteristic signatures specifically on astrophysical scales. The potential of these probes is not limited by cosmic variance, but comes with the challenge of building robust theoretical models of the nonlinear dynamics of stars, galaxies, and large scale structure.

In this review we lay the groundwork for a thorough exploration of the weak-field, nonlinear regime, with an eye to using the current and next generation of observations for tests of gravity. We begin by setting the scene for how gravitational theories beyond GR are expected to behave, focusing primarily on screening mechanisms. We describe the analytic and numerical techniques for exploring the relevant astrophysical regime, as well as the pertinent observational signals. With these in hand we present a range of astrophysical tests of gravity, and discuss prospects for future measurements and theoretical developments.
\end{abstract}
\clearpage
\tableofcontents

\section{The Novel Probes Project}
\label{sec:npp}

The Novel Probes Project aims to bring together theorists and experimentalists to address questions about the dark sector of the universe by means of astrophysical observables. It has two arms, one on tests of gravity as described here and another on dark matter self-interactions (Adhikari et al 2020, in prep). At the heart of the Project are Slack\texttrademark-hosted discussion forums intended to foster collaboration between groups with differing expertise, and provide platforms on which to ask questions to experts. The gravity forum may be accessed through the project website -- \url{https://www.novelprobes.org} -- and we welcome the participation of interested researchers, regardless of expertise in the field of astrophysical tests of gravity. The present document outlines the topics we intend the forum to cover, and will be ``living'' in the sense that it will be regularly updated as theories develop and observational constraints improve.

The forum will be partitioned into the following eight channels, each monitored by 4-5 experts in the field:
\begin{enumerate}[(i)]
\item Theory, including cosmic acceleration and screening mechanisms (Secs.~\ref{sec:GWs} and~\ref{sec:theory} of this review)
\item The use of galaxy surveys (Sec.~\ref{sec:surveys})
\item Analytic methods for the nonlinear regime, including degeneracies with other types of physics (e.g. baryons; Sec.~\ref{sec:nonlinear})
\item The use of simulations (Sec.~\ref{sec:simulations})
\item Cosmological probes (Sec.~\ref{sec:cosmo_constraints})
\item General astrophysical tests of modified gravity and screening mechanisms (Secs.~\ref{sec:RSD}--\ref{sec:screening_maps})
\item Specific tests of thin-shell screening (e.g. chameleons \& symmetrons; Sec.~\ref{sec:chameleon_tests})
\item Specific tests of Vainshtein screening (Sec.~\ref{sec:vainshtein_tests} and Sec.~\ref{sec:vainshtein_breaking})
\end{enumerate}

\section{Introduction}



\noindent Modern instruments and methods for probing gravity have inaugurated a second golden age of General Relativity (GR). The first results from the Laser Interferometer Gravitational Wave Observatory (LIGO) have not only allowed us to confirm some of the most fundamental predictions of GR -- gravitational waves and black holes -- but have also led to the first constraints on gravity in the strong-field regime. The exquisite timing of millisecond pulsars is allowing us to dramatically increase the precision with which we can measure a number of fundamental parameters that characterize deviations from GR in the transition from the weak- to strong-field regime. And, for the first time, we have obtained images of the event horizon of the black hole at the centre of M87 using radio interferometric measurements from the consortium of telescopes known as the Event Horizon Telescope. Sagittarius A*, the black hole at the centre of our own galaxy, will be similarly imaged in the future, and these images will be backed up by detailed measurements of stellar orbits using the Gravity programme run by the European Southern Observatory (ESO). With these new observations we have every reason to expect our understanding of gravity to drastically improve.
    
Observations of the large-scale structure of the universe -- from maps of the Cosmic Microwave Background (CMB) to surveys of galaxies and measurements of weak gravitational lensing -- have revolutionized cosmology. It is now possible to find accurate constraints on fundamental parameters such as the curvature of space, the fractional density of dark matter and dark energy, and the mass of the neutrino. Such has been the success of these observational programmes that a slew of new observatories and satellites have been planned, and are under construction, that will substantially improve the amount and quality of data available to map out the structure of space-time on the largest scales. 
 
A key goal of this research is to test the fundamental assumptions that underpin our current model of cosmology. Given that GR plays such a crucial role, there have been a number of proposals for testing deviations from its predictions on these previously unexplored scales. The resultant constraints would be complementary to those obtained on smaller scales (for example, in the Solar System, or with millisecond pulsars) and in strong-field regimes. These test different aspects of GR in very different gravitational environments, and involve almost orthogonal measurement techniques and systematics.
%

The success of GR -- and the cosmological $\Lambda$ cold dark matter ($\Lambda$CDM) model that it supports -- begs the question: why ought we devote significant effort to testing it? There are many reasons. GR is notoriously resistant to incorporation into a quantum theory of gravity, so that small and large scale phenomena are currently described by qualitatively different frameworks. There is some hope that a modified theory of gravity could be more commensurable with quantum mechanics. Phenomenologically, GR requires the addition of dark matter and dark energy to explain astrophysical and cosmological observations, neither of which have non-gravitational support. This raises the possibility that these phenomena may be artifacts arising from the application of an incorrect theory of gravity, and many of the theories that we discuss in this review were motivated explicitly by the hope of cosmic self-acceleration. Finally, from an effective theory point of view there is no reason to expect a rank-2 tensor to be the only gravitational degree of freedom operative at large scales, and UV-complete theories such as string theory naturally produce additional scalars, vectors and tensors. The program of testing GR may therefore be viewed as the search for more general low-energy degrees of freedom in the universe, which may in turn provide clues to quantum gravity at the Planck scale.

While much of the success of modern cosmology in testing GR and $\Lambda$CDM has relied on the accuracy and ease with which one can calculate predictions in linear theory, there are limitations to this approach. In particular, relying on large-scale modes of both the gravitational and matter fields introduces a cosmic variance limit: there is a finite amount of information one can access, limited by our cosmic horizon, which translates into clear limits on the potential strength of constraints. It makes sense, therefore, to start exploring smaller scales where nonlinear gravitational collapse plays a major role.  By looking at ``astrophysical'' structures -- from clusters of galaxies all the way down to stars -- it should be possible to probe the weak-field regime of gravity in a range of environments far more varied than those accessed on either very large or very small scales.

\begin{figure}[ht]
  \centering
     \includegraphics[scale=0.45]{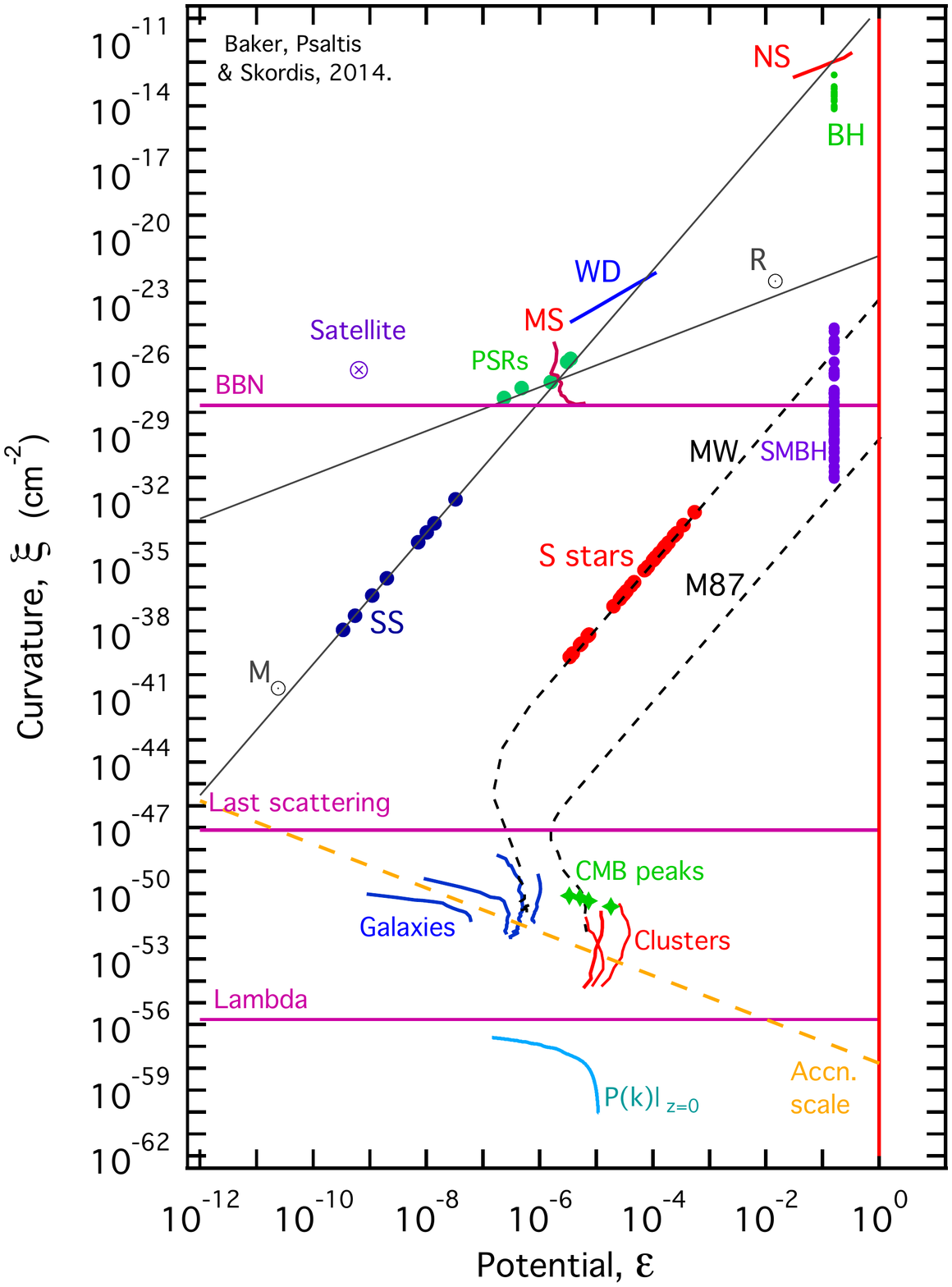}
     \caption{The landscape of gravitational environments probed by known systems. The x-axis indicates the characteristic gravitational potential of the system, and the y-axis approximately quantifies the typical spacetime curvature probed by that system (note that the y-axis quantity is \textit{not} exactly the Ricci curvature, as this vanishes for vacuum systems). For full details of this figure and its implications, see ~\citet{2015ApJ...802...63B}.
     }
  \label{fig:landscape1}
  \end{figure}

\begin{figure}[ht]
   \centering
  \includegraphics[scale=0.45]{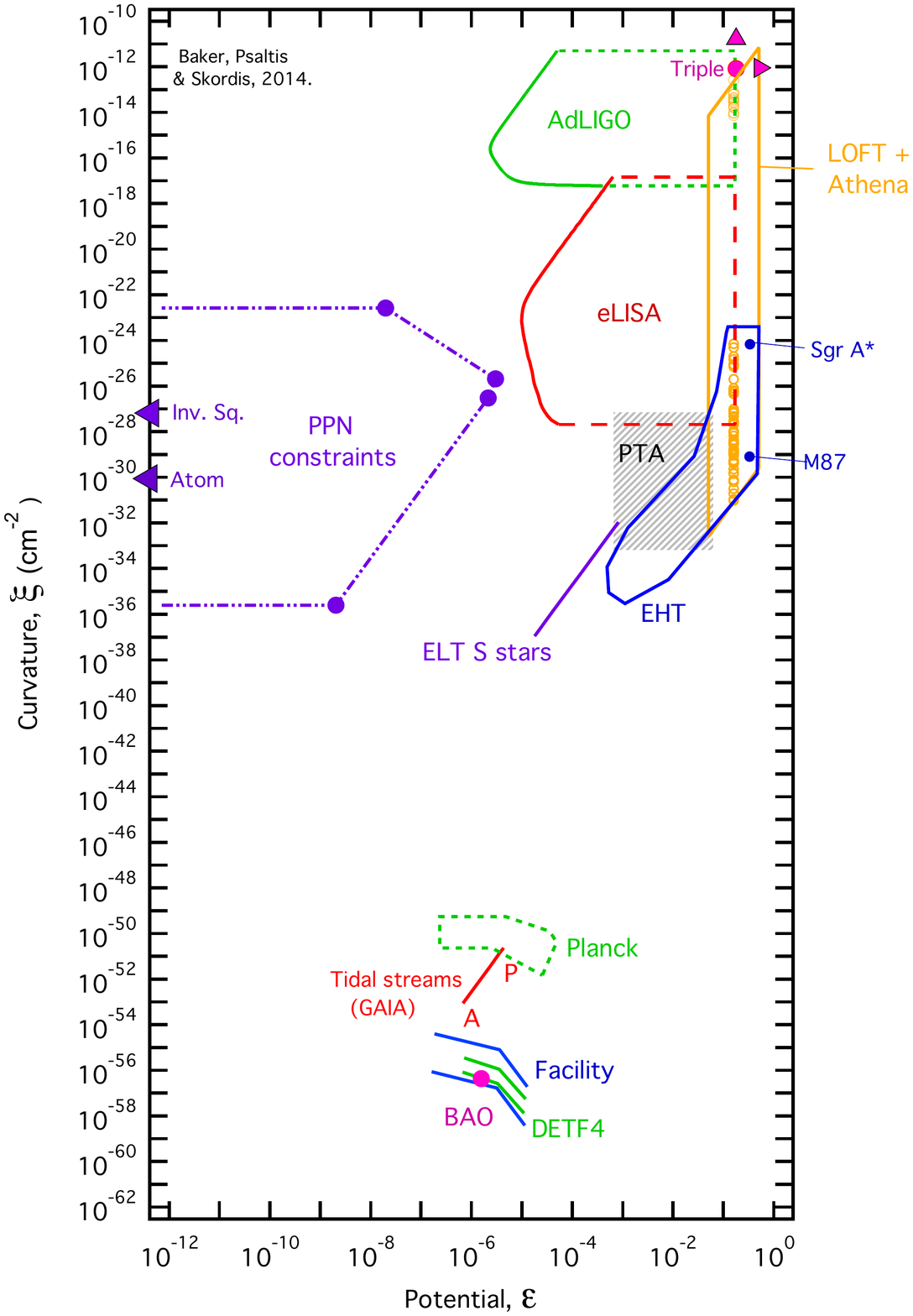}
  \caption{As per Fig.\ref{fig:landscape1}, but with experimental or observational tests corresponding to each region.}
  \label{fig:landscape2}
\end{figure}

Figs.~\ref{fig:landscape1} and ~\ref{fig:landscape2} show a two-dimensional representation of this range of environments, quantified according to two `yardsticks': the typical Newtonian gravitational potential of a system (in units where $c \equiv 1$), and a measure of its spacetime curvature (see ~\citealt{2015ApJ...802...63B} for full details). Fig. ~\ref{fig:landscape1} indicates that cosmological observables such as the matter power spectrum (denoted $P(k)$) and angular power spectrum of the CMB anisotropies probe low-curvature regimes, whilst the Solar System probes a curvature regime intermediate between that of cosmology and compact objects. Note that there is a discernible `desert' in our observations shown in Fig.~\ref{fig:landscape2}, spanning curvatures $\sim 10^{-52} - 10^{-38} \rm{cm}^{-2}$. Fig.~\ref{fig:landscape1} shows that this region is inhabited only by galaxies, for which we have few reliable probes of gravitational physics. This raises the possibility that some kind of transition scale resides in this desert, and the question of whether we can identify astrophysical tests that could reveal its presence.

Programs for testing gravity using astrophysical objects in the cosmological, weak field regime are poorly developed. The large sample sizes they afford can only be exploited if the great complexity of relevant physical processes can be modelled or controlled for. The fact that such objects necessarily involve mildly to strongly nonlinear gravitational collapse (albeit in the weak-field regime) is problematic, as the numerical and semi-analytic methods that have been developed to study this collapse tend to have poorer control over systematics than models for both large (linear) or small (Solar System and laboratory) scales. Only recently have these methods begun to be applied to tests of gravity and fundamental physics, besides ``galaxy formation'' physics in $\Lambda$CDM. Furthermore, non-gravitational physics can play a significant role in the formation and resulting morphology of astrophysical structures. The interaction of gas, plasma and radiation leads to a slew of baryonic effects capable of either suppressing or enhancing the gravitational collapse that one would expect from N-body dynamics alone. If one adds to this the effect of feedback from energetic astrophysical phenomena such as supernovae or active galactic nuclei (AGN), it becomes very difficult to extract purely gravitational information from observations of collapsed objects.

Gravitational theory in the nonlinear regime is typically intricately connected with the phenomenon of ``screening''. In a number of extensions of GR with (typically scalar) extra degrees of freedom, nonlinear corrections will suppress modifications close to sources or in high density environments. Thus, deviations from GR are masked (and may become negligible) in regions where gravity is stronger. There are a few different types of screening mechanisms, and while there are attempts to create a unified framework, no approach has proved satisfactory in all respects. This complicates the program of extracting information from this regime in a controlled manner, but at the same time offers the prospect of testing a great range of modified gravity theories by means of just a few screening mechanisms.

Given the complications that arise when testing gravity in the nonlinear regime, it is not surprising that it has yet to be targeted systematically. This makes this regime ripe for exploration, both theoretically and observationally. Given the abundance of present and upcoming measurements, it behooves us to create and fully develop frameworks for extracting the gravitational information. On the one hand, this might involve adapting some of the tried and tested statistical methods that are used to study the large-scale structure of the universe: correlation functions of various orders in both configuration and Fourier space contain some of the primary information. But one ought also to consider statistical methods which could tease out information beyond the moments of the correlation function, or deviations from GR in particular environments such as clusters, dwarf galaxies or voids. The morphology and dynamics of these objects may be acutely sensitive to gravitational features below the resolution limit of -- or washed out in -- conventional statistical methods.

In this review we will lay the groundwork for a thorough exploration of the weak-field, nonlinear regime of gravitational collapse (with a clear emphasis on cosmological scales) with an eye to exploiting the current and next generation of cosmological surveys. Our review should be seen as complementary to current reviews on constraining gravity with gravitational waves, compact objects and black holes, more focused on the strong field regime. We begin with a summary of the impact of the recent LIGO detection of a neutron star merger and gamma-ray counterpart (GW170817 and GRB 170817A), which restricts -- but does not eliminate -- the viable parameter space of modified gravity, and comment on the prospect of a dynamical origin to the acceleration of the universe's expansion in light of this (Sec.~\ref{sec:GWs}). We then introduce and explain the screening mechanisms used to hide novel effects within the Solar System (Sec.~\ref{sec:theory}). We describe the current and planned surveys providing useful information on gravity in the nonlinear regime (Sec.~\ref{sec:surveys}), before discussing the features of this regime more generally (Sec.~\ref{sec:nonlinear}), including the degeneracy between gravitational and ``galaxy formation'' physics and the estimators that have been devised to overcome it. In Sec.~\ref{sec:simulations} we present simulations with modified gravity on both cosmological and smaller scales, and in Sec.~\ref{sec:cosmo_constraints} we discuss observational tests on cosmological scales. The bulk of the review is contained in Sec.~\ref{sec:astro_tests}, where we present a suite of astrophysical tests that target screening. We split these by the type of screening mechanism covered, and include probes from the scale of stars to galaxy clusters. Finally, in Sec.~\ref{sec:prospects} we discuss future theoretical and observational prospects.

This review primarily targets observers seeking novel applications of their data to tests of fundamental physics (although it is also intended to be useful to theorists). We therefore keep theoretical complexity to a minimum, and exclude aspects of the models that are of purely formal or theoretical interest. We will update the review regularly as new ideas are developed and proposed tests carried out, and couple it with a discussion forum hosted by Slack (see \url{https://www.novelprobes.org}) where the authors will answer questions from, and collaborate with, interested parties.

\vspace{5mm}

\noindent \textbf{A note on scope:} We consider ``astrophysical'' probes to be those pertaining to scales in the universe that are not currently linear, i.e. $r \lesssim \mathcal{O}(10^2) \: \text{Mpc}$. We include galaxy surveys within this class, as well as galaxy clusters and voids. We restrict ourselves to the weak-field regime in order to focus on physics that is relevant cosmologically; we therefore discuss the physics of compact objects only insofar as they are relevant for cosmological modified gravity theories. We further exclude tests of gravity within the Solar System and on Earth, which involve very different techniques and observables to those on larger scales as well as probing objects which most viable modified gravity objects would consider screened. We consider a probe ``novel'' if it is not well established as a principal science target of current surveys or experiments. Finally, we exclude modified gravity theories such as Modified Newtonian Dynamics (MOND) which do not fit comfortably into modern field-theoretic frameworks and have limited domains of validity; MOND itself is extensively reviewed in \citet{Famaey:2011kh}.

\section{Modified gravity after GW170817}
\label{sec:GWs}
\subsection{Bounds on the speed of gravitational waves}

The first direct detection of gravitational waves (GWs) from a binary neutron star (BNS) merger was made by the LIGO-Virgo Collaboration on 17th August 2017, and announced publicly on 16th October 2017 \citep{2017PhRvL.119p1101A}. A multi-wavelength electromagnetic follow-up campaign captured the electromagnetic counterpart of the merger, as the kilonova emission rapidly brightened and faded over the next tens of days \citep{LIGO_multi}.

This event, GW170817, presented a unique opportunity to place bounds on the speed of gravitational waves. In General Relativity gravitational waves travel at the speed of light but, as we will detail below, many modified gravity theories alter this prediction. A common model-independent parameterisation used to describe the speed of GWs is \citep{Bellini:2014fua}:
\begin{align}
c_T^2&=c^2\left[1+\alpha_T(t)\right]
\label{alphadef}
\end{align}
where $c_T$ is the propagation speed of tensor modes, and $t$ is physical time. The function $\alpha_T(t)$ can adopt both positive and negative values in principle, although negative values were bounded even prior to GW170817 by the lack of observed gravi-Cerenkov radiation from ultra high-energy cosmic rays ($\sim 10^{10}$ GeV; \citealt{1980AnPhy.125...35C,2001JHEP...09..023M}). Note that the energy scale of these bounds is much higher than that relevant to our discussion here. 

For our purposes, the crucial feature of GW170817 was that the first electromagnetic signals of the merger, gamma-rays, arrived at Earth 1.74 seconds after the arrival time of GWs corresponding to the merger. Because the source was comparatively local in cosmological terms, located at a physical distance $d= 40$ Mpc ($z\sim 0.01$), a simple Euclidean treatment of distances is sufficient.  Then we can write the difference in the arrival times of the photons and gravitational waves as \citep{2014PhRvD..90d4048N,Lombriser:2015sxa,2017PhRvL.119y1301B,Sakstein:2017xjx,2017PhRvL.119y1302C,PhysRevLett.119.251304}:
\begin{align}
\Delta t &= t_{\rm int} + \delta t = t_{\rm int}+t_\gamma - t_{\rm GW} \nonumber\\
&= t_{\rm int}+\frac{d}{c}-\frac{d}{c_T}\nonumber\\
&= t_{\rm int}+\frac{d}{c}\left(1-\frac{1}{\sqrt{1+\alpha_T}}\right)\nonumber\\
&\simeq t_{\rm int}+\frac{d}{c}\frac{\alpha_T}{2}
\end{align}
Here $t_\gamma$ and $t_{GW}$ are the times taken for the photons and GWs to travel distance $d$, respectively. $t_{\rm int}$ is an intrinsic time delay between the emission of the two signals, i.e. a delay occurring \textit{at the source}. According to current modelling of BNS mergers, this intrinsic delay could be up to $\sim 10 s$ in duration for short GRB (sGRB) associations, and potentially up to hour timescales for long GRB associations ~\citep{PhysRevD.60.121101,2012ApJ...760...12A,2019FrPhy..1464402Z}. In principle it could be of either sign, i.e. it is possible that the photons could be emitted before the GWs. Of the time intervals above, the only one we can measure is $\Delta t = 1.74$s, indicating that these events are consistent with a BNS-sGRB association.

Rearranging the above expression, we finally obtain:
\begin{align}
\alpha_T &\simeq \frac{2c}{d}\left(\Delta t - t_{\rm int}\right)
\end{align}
If we assume the GW and photon emission to be exactly simultaneous, setting $t_{\rm int}=0$, this results in a bound of $\alpha_T \lesssim 10^{-15}$. If we allow for a hundred seconds of intrinsic delay in either direction, this weakens the bound to $|\alpha_T|\lesssim 10^{-13}$. Note that really we are only bounding the value of the $\alpha_T(t)$ function at $t\simeq 0$, which in principle does not rule out non-trivial evolution in the past. Prior to GW170817, the most stringent LIGO/Virgo bounds on the propagation speed of GWs was $0.55<c_T<1.42$, which was found by \citet{2017PhRvL.119p1102C} using Bayesian methods applied to the first three binary black hole merger detections. Clearly, GW170817 improved these by several orders of magnitude.

Even when adopting the most conservative bound above, this result remains an impressively stringent constraint on a key deviation from GR. The straightforward interpretation of this result is that any viable theory of gravity must possess tensor modes that propagate at the speed of light \textit{exactly} (and that $t_{\rm int}\simeq 1.74$s for this event); this has general implications for the structure of gravity theories, as we will detail in the next subsection. However, a loophole remains: it is theoretically possible for a gravity theory to be structured such that $0<\alpha_T<10^{-13}$. Without some physical principle or symmetry to enforce such a near cancellation, this would seem to require a significant degree of fine-tuning of the theory. 

It was pointed out in \citet{deRham:2018red} that, by coincidence, the frequency of the binary neutron star merger lies close to a generic strong coupling frequency of low energy effective field theories (EFTs) of modified gravity \citep{2013JCAP...02..032G, 2013JCAP...08..025G,  Gleyzes:2014dya, Gleyzes:2015rua}. Recall that an EFT utilises an expansion in low energy scales to keep only the most relevant Lagrangian operators of a theory. However, the coefficients of these operators can run with energy scale; when they reach of order unity, the theory is said to be strongly coupled. It is then no longer safe to assume that operators originally neglected in the low-energy theory are irrelevant. They must be accounted for, which is frequently not possible (since the UV completion of a theory is often unknown). As a result, the low-energy EFT has broken down and can no longer be used.

\citet{deRham:2018red} argued that, when a dark energy EFT has typical parameter values needed to give it interesting dynamics on cosmological scales, the energy scale of GW170817 lies potentially within its strongly coupled regime. This calls into question the validity of straightforwardly applying the time delay results above to Horndeski gravity (introduced below) or the associated low-energy effective theories. Furthermore, the authors argue that if a low-energy effective theory is to admit a Lorentz-invariant completion, one would actually \textit{expect} the action of operators above the strong coupling scale to return the speed of tensor modes to $c$. 
This argument ultimately depends on the specifics of a particular gravity model. Hence in what follows, we will present the results of a straightforward application of the binary neutron star constraints to modified gravity theories, assuming no strong coupling scales come into play. However, one should carefully check whether future theoretical models are subject to this caveat.

\subsection{Consequences for existing theories}

\subsubsubsection*{Scalar--tensor theories}

To understand the implications of the above bounds, it is helpful to study the deviations $c_T\neq c$ in some example models. One of the most useful to pursue is Horndeski gravity, which is the most general theory of a gravitational metric and a scalar field $\phi$ that propagates three (one scalar and two tensor) degrees of freedom \citep{Horndeski1974,2011PhRvD..84f4039D,2011PThPh.126..511K}. The Horndeski Lagrangian is  equivalent to that obtained when applying the EFT approach referenced above to a theory of a metric and a scalar field\footnote{Though the EFT approach also permits extension to include higher-derivative `Beyond Horndeski' operators, whose higher derivatives cancel in the field equations. These operators are not present in the original formulation of the Honrndeski Lagrangian shown in eqs.(\ref{HDLagrangian}-\ref{HDL5}).}. The original formulation of Horndeski gravity is constructed as a sum of four Lagrangian terms
\begin{align}
S=\int \!\mathrm{d}^4x \sqrt{-g}\left\{\sum_{i=2}^5{\cal L}_i[\phi,g_{\mu\nu}]\right\},
\label{HDLagrangian}
\end{align}
where 
\begin{align}
{\cal L}_2&= K(\phi, X) \\
{\cal L}_3&=  -G_3(\phi, X) \Box\phi \\
{\cal L}_4&=   G_4(\phi, X)R+G_{4,X}\left\{(\Box \phi)^2-\nabla_\mu\nabla_\nu\phi \nabla^\mu\nabla^\nu\phi\right\} \label{HDL4},  \\
{\cal L}_5&= G_5(\phi, X)G_{\mu\nu}\nabla^\mu\nabla^\nu\phi
-\frac{1}{6}G_{5,X}\big\{ (\nabla\phi)^3
-3\nabla^\mu\nabla^\nu\phi\nabla_\mu\nabla_\nu\phi\Box\phi 
 \nonumber \\ 
 & +2\nabla^\nu\nabla_\mu\phi \nabla^\alpha\nabla_\nu\phi\nabla^\mu\nabla_\alpha \phi
\big\}  \label{HDL5} \,,
\end{align}
 and $X\equiv-\nabla^\nu\phi\nabla_\nu\phi/2$. Here $K$ and $G_i$ are four functions that control the contribution from each sub-Lagrangian. Note that the derivatives of $G_4$ and $G_5$ are also relevant; we have suppressed the arguments of these differentiated functions ($G_{4,X}$ and $G_{5,X}$) for clarity. A linearised calculation shows that the function $\alpha_T$ introduced in eq.(\ref{alphadef}) is related to $G_4$ and $G_5$ by \citep{2011PThPh.126..511K,Bellini:2014fua,2017PhRvL.119y1301B,Sakstein:2017xjx,2017PhRvL.119y1302C,PhysRevLett.119.251304}:
\begin{align}
\alpha_T&\equiv\frac{2X}{M^2_*}\left[2G_{4,X}-2G_{5,\phi}-\left(\ddot{\phi}-\dot{\phi}H\right)G_{5,X}\right] \label{eq:alphaTH}
\end{align}
where $
M^2_*\equiv2\left(G_4-2XG_{4,X}+XG_{5,\phi}-{\dot \phi}HXG_{5,X}\right)
$. When combined with the bound \\ \mbox{$\alpha_T\big|_{z=0}\lesssim 10^{-13}$} derived in the previous subsection, eq.(\ref{eq:alphaTH}) has strong implications for theories within the broad Horndeski family. 
As mentioned above, there are two broad routes to interpret eq.(\ref{eq:alphaTH}). In the next subsection we will enumerate methods to render $\alpha_T\big|_{z=0}$ small but non-zero. In this subsection we discuss the implications of requiring $\alpha_T = 0$ identically.

The simplest way to ensure that eq.(\ref{eq:alphaTH}) vanishes is to set each of the function derivatives $G_{4,X}$, $G_{5,X}$ and $G_{5,\phi}$ to zero individually. Feeding this information back into eq.(\ref{HDLagrangian}) reduces ${\cal L}_4$ to a conformal coupling (i.e. a simple function $G_4(\phi))$ to the Ricci scalar~\citep{2012JCAP...07..050K,2016JCAP...11..006M}. The remaining piece of the ${\cal L}_5$ Lagrangian can be eliminated entirely by integrating by parts, and using the Bianchi identity ($\nabla^\mu G_{\mu\nu}=0$).

After this exercise we are left with the following template for scalar-tensor theories:
\begin{align}
\label{template}
S=\int \!\mathrm{d}^4x \sqrt{-g}\left\{G_4(\phi) R +K(\phi,X)-G_3(\phi,X)\Box\phi\right\}+S_M,
\end{align}
where $S_M$ is the matter Lagrangian. Some examples of theories which fit onto this template -- and hence produce $c_T = c$   -- are $f(R)$ gravity, the cubic galileon, Kinetic Gravity Braiding (`KGB') and the scalar-tensor limit of the non-self-accelerating branch of the Dvali-Gabadadze-Porrati braneworld model (nDGP) \citep{2010RvMP...82..451S,Nicolis:2008in, KGB2010,2004JHEP...06..059N}. For example, $f(R)$ gravity corresponds to the choices $G_4=\phi\equiv df(R)/dR = f_R$, $K = f(R)-Rf_R$ and $G_3=0$; $f(R)$ models are strongly constrained (but not eliminated) by electromagnetic data sets \citep{2016PhRvD..93h4016D}. The cubic galileon is recovered from the Horndeski action by setting $G_4 = 1$, $K=-c_2X$ and $G_3 = c_3 X/M^3$, where $c_3$ and $M$ are free parameters of the galileon model\footnote{All parameters in the galileon field equations appear as ratios with $c_2$. It is common practice to fix $c_2=-1$ and constrain the remaining free parameters under this choice.}; the cubic galileon was powerfully constrained by~\citet{Renk:2017rzu} by galaxy-ISW cross-correlation. KGB and nDGP remain viable \citep{Kimura:2011td, Lombriser:2009xg, 2016PhRvD..94h4022B}, although the latter requires some form of dark energy to produce accelerated expansion. In contrast, other models such as the quartic and quintic galileons, which invoke the full complexity of the ${\cal L}_4$ and ${\cal L}_5$, are now ruled out -- at least, in this straightforward interpretation of the results. 

Of course, there are other ways to make eq.(\ref{eq:alphaTH}) vanish identically. One would be to posit that all its three component terms are finely balanced such as to cancel one another out. However, it would seem difficult to enforce such a cancellation for all redshifts, given the appearance of the background-dependent quantities $H$ and $\dot\phi$. Hence observations of a second, similar binary neutron star merger at higher redshift should be able to easily confirm or refute this hypothesis. It is also highly likely that inhomogeneities will spoil these tunings.
%

Several other ways of evading the bound on $|\alpha_T|$ within Horndeski theory have been explored. Through careful analysis, \citet{2019PhRvL.122f1301C} found a subtle loophole in the derivation of the GW propagation speed that allows for special choices of the quartic and quintic Horndeski Lagrangians to persist. With these special forms, the deviation of the GW speed from unity vanishes \textit{dynamically} on cosmological backgrounds, when the scalar equation of motion is used. Unfortunately this trick of rescuing a particular class of theories only works on a homogeneous cosmological background; the anomalous GW speed contribution reappears at an unviable when cosmological perturbations are taken into account.

Another possibility would be to require $G_{5,X}=0$, but $G_{4,X}=G_{5,\phi}$, i.e. insist upon a specific connection between the functional forms appearing in ${\cal L}_4$ and ${\cal L}_5$. Further relations of this kind are possible in extensions of Horndeski gravity, namely Beyond Horndeski gravity and Degenerate Higher-Order Scalar-Tensor (DHOST); see \citet{
Zumalacarregui:2013pma, Gleyzes:2014dya, Gleyzes:2014qga, LangloisNoui2016, Langlois2017,Langlois:2017dyl,2018arXiv180202728K,2018arXiv180310510K,2018arXiv180307476C,Koyama:2015oma,Crisostomi:2017lbg,CrisostomiKoyama2018} for further details. However, the Beyond Horndeski extension itself is now effectively ruled out by analyses revealing that it permits GWs to decay into fluctuations of the dark energy scalar field \citep{Creminelli:2018xsv,2019arXiv190607015C}. Such a process would likely occur rapidly and prevent any detectable GWs from reaching Earth, rendering the theory non-viable. Weaker bounds on the remaining parameter space of Horndeski gravity from related considerations were recently presented in \citet{2019arXiv191014035C}.

\subsubsubsection*{Vector--tensor and tensor--tensor theories}
A calculation analogous to the one above can be repeated for general vector-tensor theories. The most general Lorentz-invariant, second-order vector-tensor theory currently known is the (Beyond) Generalised Proca theory \citep{2014JHEP...04..067T,2014JCAP...05..015H}, which has a structure of derivative interactions similar to Horndeski gravity (note that the vector field must be massive for this to be possible). However, unlike Horndeski theory, it has not yet been shown that Beyond Generalised Proca contains \textit{all} possible terms resulting in second-order equations of motion. The final result of the Generalised Proca calculation has the same structure as eq.(\ref{template}), except that the equivalent of $G_4$ is fixed to be a constant, and the equivalent $K$ and $G_3$ are functions of $X=-1/2 A_\mu A^\mu$ only. Further details can be found in \citet{2018JCAP...03..021L}.

Analogous to the DHOST extensions of Horndeski described above, there exists a DHOST-like extension of Generalised Proca \citep{KimuraNaruko2017}. This family of models contains members which are consistent with the bounds from GW170817 \citep{KaseKimura2018}.

A related, but distinct, branch of work pursues Lorentz-violating vector-tensor theories, in which the vector defines a preferred direction in the cosmological background. One example is Einstein-Aether gravity, in which the preferred direction (which must be time-like to preserve spatial isotropy) is enforced via a Lagrange multiplier. Of the four $c_i$ parameters defining the Einstein-Aether action, the GW speed is controlled by two of them as $c_T^2=c^2/(1-c_1-c_3)$ (note that these are \emph{not} the same $c_i$ defining galileon theories -- the repeated use of notation is unfortunate). The results from event GW170817 imply the bound $|c_1+c_3|<10^{-15}$, see \citet{2018arXiv180204303O} for further details. 

Another example is Ho\u{r}ava-Lifschitz gravity (a small family containing a few subcases; \citealt{Horava2009,Blas2010,Cliftonreview2010}): here, local Lorentz invariance is recovered as an approximate symmetry at low energies, but broken at (extremely) high energies. The impact of GW170817 on Ho\u{r}ava-Liftschitz theories was detailed in \citet{Gumrukcuoglu2018}; in short, the constraint on $|\alpha_T|$ maps directly into a tight constraint on the parameter $\beta$ that appears in the low-energy
limit of the Ho\u{r}ava-Lifschitz action.
 
The logical continuation of these results is to ask what the implications are for tensor-tensor theories, i.e. bigravity.  Here, the GW timings can be more physically interpreted as bounding the massive graviton mode. The constraint from GW170817 results in a bound $m_g \lesssim  10^{-22}$eV, where $m_g$ is the graviton mass. Although this may initially seem restrictive, the bound is already surpassed by existing tests of gravity in the Solar System, which imply $m_g \lesssim  10^{-30}$eV \citep{2017RvMP...89b5004D}. In essence, the status of massive graviton theories has been unaffected by the results of GW170817. 

Finally, we note that the multifield tensor-vector-scalar theory, TeVeS, was closely scrutinized in the light of the measurements of GW speed. Whilst the original formulation of TeVeS is ruled out by the GW170817 results \citep{2018PhRvD..97d1501B,PhysRevD.97.084040,2018Univ....4...84H}, \citet{Skordis2019} found a class of extended TeVeS-like theories that yield $c_T = c$, and hence are consistent with the GW bounds.

\subsection{Other survivors}
As we have detailed above, a straightforward interpretation of the GW170817 observations puts firm restrictions on the form of viable, cosmologically-relevant scalar-tensor and vector-tensor theories of gravity. However, there remain some more subtle ways to maintain consistency with the GW results. We will enumerate here the ones we are aware of at present:
\begin{itemize}
\item {\bf Tuned cancellations.} As explained in the previous section, if the Lagrangian functions appearing in either the (Beyond) Horndeski or Generalised Proca models are related in specific ways, a suppression of the value of $\alpha_T\big|_{z=0}$ can be arranged. The most finely-tuned models of this kind can be ruled out by a second set of BNS merger observations at a higher redshift than GW170817. From a theoretical perspective, such models are highly fine-tuned and are radiatively-unstable unless there is some symmetry enforcing an infinite set of tunings. We are aware of no such symmetries.
\item {\bf DHOST-like extensions of both scalar-tensor and vector-tensor theories.} These are higher-derivative extensions of Horndeski and Generalised Proca, which include the Beyond Horndeski and Beyond Generalised Proca models as sub-cases. In the original DHOST family (extensions of Horndeski), all three terms in eq.(\ref{HDL4}) have independent amplitudes, and new terms containing more than three copies of the field are present. This enhanced flexibility allows the constraint $\alpha_T=0$ to be satisfied by `using up' fewer functions; hence more of the full Lagrangian survives. The disadvantage is that the theory is defined by more functions, and hence more challenging to constrain observationally. This extended class of theories has received much attention since the GW results \citep{Langlois:2017dyl,CrisostomiKoyama2018, 2018arXiv180310510K}. The equivalent results for DHOST-like extensions of Generalised Proca can be found in \citet{KaseKimura2018}.

Recently, \citet{Creminelli:2018xsv,2019arXiv190607015C} studied the decay of GWs into fluctuations of a dark energy field. Decay of GWs is usually forbidden in GR by Lorentz invariance, but the presence of a dark energy field effectively acts as a `medium' through which the GWs propagate, spontaneously breaking Lorentz variance. Decays of the form $\gamma\rightarrow \pi\pi$ and $\gamma\rightarrow \gamma\pi$ are then possible in \textit{some} theories, where $\gamma$ is a graviton and $\pi$ is a fluctuation of the scalar dark energy field. Such decays were found to occur in Beyond Horndeski and DHOST theories. They can also occur in the original Horndeski theory, but are controlled by the $G_{4,X}$ operator and derivatives of the $G_5$ operator that are already constrained by GW170817; hence the surviving sector of Horndeski shown in eq.(\ref{template}) does not suffer from graviton decay effects.

The fact that GWs have been observed to reach Earth, clearly without having decayed entirely into the dark energy field, imposes further constraints on DHOST and Beyond Horndeski theories.

%
\item {\bf Mass Scales.} Note that the numerator of eq.(\ref{eq:alphaTH}) contains only derivatives of functions, whilst the denominator further contains an undifferentiated instance of $G_4$. Consider a toy model where $G_4$ has the form\footnote{Note that this example is for demonstratory purposes only; it is not intended to represent a viable gravity theory.}:
\begin{align}
G_4(\phi,X)&=M^2 +\frac{X}{m^2}
\end{align}
Here $m$ is a mass scale. Note that $G_4$ and $XG_{4,X} = X/m^2$ now differ by a constant factor of $M^2$. Depending on the hierarchy of $M$, $m$ and the mass of the scalar field, it is possible that $G_4\gg XG_{4,X}$ (for example, if $M$ were to be the Planck mass and $m\gg H$). In this case, the denominator of eq.(\ref{eq:alphaTH}) dominates the numerator, leading to a suppression of $\alpha_T$. Depending on the precise mass scales involved, this kind of gravity theory may lead to fifth force signatures on sub-Hubble scales, of the kind constrained in \citet{Sakstein:2014isa,Sakstein:2014aca,Ip:2015qsa,Sakstein:2015jca,Burrage:2017qrf,Burrage:2016bwy}.
%
\item {\bf Scale suppression.} 
In a similar vein to the point above, \citet{2018arXiv180209447B} highlight that the modes associated with GW170817 and the modes typically associated with cosmic acceleration are separated by some nineteen orders of magnitude. In the `Equation of State' dark sector parameterization~\citep{EoS2013,EoS2014} (a title not to be confused with general usage of the words `equation of state') this factor of $10^{19}$ can act to suppress corrections to GR in the observables probed by GW170817 (such as the gravitational wave group velocity), even though the model possesses significant effects on cosmological scales.
\item {\bf Frozen background fields.} Note that the numerator of eq.(\ref{eq:alphaTH}) is multiplied by the kinetic term of the scalar field, $X=\dot{\phi}^2/2$ (on an FRW background). Therefore, if the scalar field sits at the minimum of its potential (say), $X\ll 1$ and the entire RHS of eq.(\ref{eq:alphaTH}) is suppressed. A quintessence or k-essence model with an appropriately tuned potential could display this behavior~\citep{Copeland2006}.
\item {\bf Non-universal couplings.} The near-simultaneous arrival of GWs and photons indicates that the standard model matter sector is coupled to the same metric that constitutes the Einstein-Hilbert part of the gravitational action. However, it is still possible that dark matter could be non-minimally coupled to this particular metric (effectively, it couples to a different metric conformally or disformally related to one featuring in the Ricci scalar $R$). Hence, dark energy models with non-minimal coupling to dark matter could still have significant effects on large-scale structure whilst maintaining consistency with the GW results, provided that $c_T=c$ in the Einstein frame. Examples of these kinds of models -- which violate the Weak Equivalence Principle -- can be found in \citet{2018arXiv180306368A,2018PhRvD..97b3506V,2019JCAP...10..013D}. 
\end{itemize}


\subsection{Producing cosmic acceleration through gravity}
\label{sec:cosmicacceleration}

\noindent The constraints on the gravitational wave propagation speed
narrow down the landscape of modified gravity models, but leave untouched a subset that can be targeted with future cosmological experiments. The majority of the models we discuss in this review were originally motivated as possible explanations of cosmic acceleration. The hope was that either new fundamental fields or infrared corrections to GR on cosmological scales could naturally drive the expansion of the  universe at late times,
thereby removing the need for a finely-tuned small cosmological constant.

This hope has not been realized for most models in the current literature. Often this is due to the competing demands of fitting observations probing both the `background' and `perturbative' universe. The region of parameter space that allows a model to yield viable acceleration will often not overlap with the parameter region allowed by measures of large-scale structure or the CMB. 
For instance, the minimal modification of gravity required in Horndeski scalar-tensor theories to provide a cosmic acceleration that is genuinely different from that of a potential or kinetic dark energy contribution and satisfies $c_{\rm T}=1$ has been shown to provide a $3\sigma$ worse fit to cosmological data than a cosmological constant~\citep{Lombriser:2016yzn}.

Because of this incompatibility, some modified gravity models still require a cosmological constant identical to that of $\Lambda$CDM to fit observations. In other families of theories, the cosmological constant may find an alternative, more subtle presentation, e.g. consider an $f(R)$ model where $f(R)\rightarrow $ const. for small $R$, e.g. \citet{Hu:2007nk}. Ultimately, the theory still contains a constant that must be fine-tuned to the observed value. Another example is that of bigravity theories \citep{2016JPhA...49r3001S}, where the structure of the interaction potential for the two tensor fields contains two constants\footnote{This arises because the potential is constructed from a set of symmetric polynomials, the lowest-order of which is a constant.}. One of these acts like a cosmological constant for the regular spacetime metric (to which matter couples), whilst the second is effectively a cosmological constant for the second dynamical metric metric (which does not couple explicitly to matter).

At times there has been an even grander hope, that corrections to GR might -- as well as explaining cosmic acceleration -- additionally alleviate the need for dark matter to explain observations. Such an idea originally found footing with the success of MOND in explaining galactic rotation curves without dark matter.
A significant step forward was the construction of a fully covariant gravity theory, TeVeS, that contains a MOND limit \citep{Bekenstein} (see \citealt{Skordis2009} for a review). 
However, MOND was found to struggle to reproduce this success for larger systems such as galaxy clusters \citep{2003MNRAS.342..901S}, whilst TeVeS is constrained by Planck measurements of the CMB \citep{PhysRevD.92.083505} and the $E_G$ statistic\footnote{The $E_G$ statistic is a ratio constructed from galaxy clustering observables, galaxy-galaxy lensing, and galaxy velocities extracted from redshift space distortions. The combination of observables is designed such that $E_G$ is, in principle, insensitive to galaxy bias (though see \citealt{Leonard} for a reality-check of this). In $\Lambda$CDM $E_G\simeq 0.4$ on all scales; in most modified gravity models this prediction is altered by non-equality of the two metric potentials in eq.(\ref{eq:Phipsi}) below, and scale-dependent growth of structure.} \citep{reyes/etal:2010}. Both MOND\footnote{Whilst MOND is not a relativistic framework, \citet{2018PhRvD..97d1501B} tested it within a class of `dark matter emulator' models, which predict that GWs and photons/neutrinos move on different geodesics.} and \textit{some variants} of TeVeS are in significant tension with the recent gravitational wave detections \citep{2018PhRvD..97d1501B,PhysRevD.97.084040,2018Univ....4...84H}, though surviving models remain, e.g. the special TeVeS theory of \citep{Skordis2019}, and the bimetric and non-local formulations of MOND \citep{2009PhRvD..80l3536M,2011PhRvD..84l4054D}. A related vector-tensor model, Einstein-Aether gravity, likewise produces unacceptably large modifications to the matter power spectrum when required to act as a dark matter candidate; it has more success (though still constrained) when acting purely as a dark energy candidate \citep{2010PhRvD..81j4015Z, 2018arXiv180204303O}. Other ideas for a unified dark sector, such as entropic gravity and Chaplygin gases \citep{2006tmgm.meet..840G,2016arXiv161102269V,2017PhRvD..95l4018H,2008arXiv0802.1798P}, likewise seem to generally fare worse, not better, than models requiring a standard cold dark matter component (though see \citealt{2018arXiv181009474F} for an interesting new development). 

Although the original goal of cosmological modified gravity has not so far been met, this certainly does not remove the need to test alternatives to GR. Cosmology operates on distance scales 16 orders of magnitude larger than those on which GR has been stringently tested using Solar System experiments and binary pulsars. In order to test this extreme extrapolation of GR to large scales, we need sensible and consistent mathematical alternatives to compare against. The theories we discuss here provide a description of perturbation dynamics that modify the standard GR relations between our four main cosmological probes (background expansion rate, the growth of structure, the deflection of light, propagation speed of gravitational waves) in a testable way. 

The construction of such models has also greatly deepened our understanding of the theoretical underpinnings of gravity. For example, we have learned how to build a full, nonlinear theory of two coupled tensors -- something previously thought impossible \citep{deRham2014}. Likewise, we have found theories whose effects are strongly enhanced or suppressed by their environments (screening, detailed in the next section); again, it was not known beforehand that such theories existed. Finally, some of the techniques and theories developed have found fruitful applications to other areas of cosmology \citep{Sakstein:2017lfm,Sakstein:2017nns,Sakstein:2018pfd}.


In cosmology, there is a fundamental degeneracy in describing physics beyond
GR with a cosmological constant, since any modification of the Einstein
equation (``modified gravity'') could be moved to the right-hand side
and be called a novel form of stress-energy (``dark energy''). In this review,
we will focus on theories that qualify as ``modified gravity'' under the
classification proposed in \citet{Joyce:2016vqv}: that is, they violate
the strong equivalence principle. In the vast majority of cases, the
phenomenology of these theories is characterized by a universally coupled,
scalar field-mediated fifth force (and because black holes do not have
scalar hair, this force violates the strong equivalence principle). We will
not consider theories that violate the \emph{weak} equivalence principle (WEP) at the level of the Lagrangian, such as theories involving a light scalar field which only couples to dark matter.

Thus, the main focus of this review will be on tests for the presence of fifth forces,  in cosmology and on cluster, galactic and stellar scales; an attempt to visually compare these experiments is given in \citet{2015ApJ...802...63B,Burrage:2017qrf,Burrage:2016bwy}.

\section{Screening mechanisms}
\label{sec:theory}




Most current theories for cosmic acceleration are theories of modified gravity, and hence a ubiquitous prediction is the presence of fifth forces on astrophysical scales. Existing solar system and laboratory constraints on these theories \citep{Burrage:2017qrf} require one to tune the new parameters to small values, essentially ruling them out as dark energy models. The idea behind screening mechanisms is to find theories that include a dynamical suppression of fifth forces i.e. they are naturally small on astrophysical scales as a consequence of their equation of motion rather than parameter tuning. With few exceptions (e.g. \citealt{Heckman:2019dsj}), the majority of viable dark energy models are either highly fine-tuned or include screening mechanisms. This is why the study of such mechanisms is so important.

After the bounds imposed by GW170817, a very general theory that is viable is given in equation \eqref{template}. Additionally, some sectors of DHOST theories (see e.g. \citealt{Crisostomi:2017lbg,Langlois:2017dyl,Dima:2017pwp}) and beyond Horndeski theories (see e.g. \citealt{PhysRevLett.119.251304,Ezquiaga:2018btd}) remain viable. The screening mechanisms discussed in this theory all fall into one or more of these theories. In particular, chameleon, K-mouflage, and Vainshtein screening all fit into \eqref{template} and Vainshtein breaking is exhibited by beyond Horndeski and DHOST theories.

\subsection{Principles of screening}
\label{sec:screening}

In order to motivate screening, we begin by considering what happens when we have a theory of gravity that does not screen. In the Newtonian, sub-horizon limit of GR, the dynamics of any system are described by
\begin{equation}
\dd s^2 = -(1+2\Phi)\dd t^2 + (1-2\Psi)\delta_{ij}\dd x^i\dd x^j, \label{eq:Phipsi}
\end{equation}
where $\Phi$ is the Newtonian potential and $\Psi+\Phi$ governs the motion of light. In GR, $\Phi=\Psi$ and the fields sourced by non-relativistic matter obey the Poisson equation $\nabla^2\Phi=4\pi G\rho$. The solution for a single source object of mass $M$ is $\Phi=\Psi=-GM/r$. The equations of motion give the gravitational acceleration of a body:
\begin{equation}\label{eq:fgrav}
\vec{a}_{\rm grav} = -\vec{\nabla}\Phi= -\frac{GM}{r^2}\hat{r}.
\end{equation}
Now let us modify GR by considering a scalar $\phi$ coupled to matter such that it mediates an additional or \emph{fifth} force. If the field is massless then one generically expects the Poisson equation\footnote{A massless scalar field has the Lagrangian $-\partial_\mu\phi\partial^\mu\phi/2$, which gives rise to the Laplacian operator in static situations. The $\alpha$ parameterization is based on the commonly studied Weyl coupling to matter via the metric $A^2(\phi)g_{\mu\nu}$. One has $\alpha(\phi)=\dd\ln A/\dd\phi$, which is not necessarily constant, although we take it to be so here for illustrative purposes.} $\nabla^2\phi = 8\pi\alpha G\rho$, which we have parameterized by a dimensionless $\mathcal{O}(1)$ parameter $\alpha$. This is solved by $\phi=-2\alpha GM/r$, i.e. a factor of $2\alpha$ larger than the GR solution. The scalar generates an additional acceleration
\begin{equation}
\vec{a}_5=-\alpha\vec{\nabla}\phi=-2\alpha^2\frac{GM}{r^2}\hat{r}.
\end{equation}
Thus the scalar mediates a force that is a factor $2\alpha^2$ larger than the force of gravity. This causes problems observationally. Since the field mediates a $1/r^2$ force between two bodies, the metric can be put into the parameterised post-Newtonian (PPN) form (note that signs differ from the conventional signs \citep{Will:2014kxa,Sakstein:2017pqi} due to our conventions in Eq.~\eqref{eq:Phipsi})
\begin{equation}
\dd s^2 = \left(1-2\frac{GM}{r}\right)\dd t^2 + \left(1-2\gamma\frac{GM}{r}\right)\delta_{ij}\dd x^i\dd x^j
\end{equation}
with Eddington light-bending parameter $\gamma = (1-2\alpha^2)/(1+2\alpha^2)$. This has been constrained to the $10^{-5}$ level by the Cassini satellite by means of the frequency shift of radio waves to and from the satellite as they passed near the sun \citep{Bertotti:2003rm}. This requires $2\alpha^2<10^{-5}$. To see that this forces the scalar into a cosmologically uninteresting region of parameter space, consider the Klein-Gordon equation $\ddot{\phi}+3H\dot{\phi}+\alpha G\rho = 0$, where $H$ is the Hubble parameter. The contribution from modified gravity is the final term, which must then be subdominant to the GR contribution by a factor $\mathcal{O}(10^{-3})$ and therefore any cosmic evolution of the scalar cannot be driven by modifications of gravity. Any cosmic acceleration is therefore due to the scalar field's potential and not modified gravity. One could add a cosmological constant (or quintessence) to drive the cosmic acceleration and look for deviations from GR on smaller scales, although in this case the acceleration would not be driven by modified gravity. At the level of linear cosmological perturbations the growth of dark matter is governed by the equation
\begin{equation}
    \ddot{\delta}_{\rm DM}+2\ddot{\delta}_{\rm DM}-\frac32\Omega_{\rm DM}(a)\left(1+2\alpha^2\right) = 0,
\end{equation}
where $\delta_\text{DM}$ is the dark matter overdensity, $\Omega_\text{DM}$ is the cosmic DM density relative to $\rho_\text{crit}$ and $a$ is the cosmic scale factor. Hence, even in this case the modifications are negligible once the Cassini bound is imposed.

Besides tuning $\alpha$ to small values, one could attempt to avoid this conclusion by introducing a mass for the scalar, so that it satisfies $\nabla^2\phi+m^2\phi=8\pi\alpha G \rho$, with solution $\phi=-2\alpha(GM/r) e^{-m r}$. This implies that $\phi$ mediates a Yukawa force with range $m^{-1}$. If this range is $\lsim 9$ AU (the Sun-Saturn distance) then the Cassini bound can be satisfied. Such forces are however heavily constrained by other means. For $\mathcal{O}(1)$ matter couplings, which are needed for cosmological relevance, lunar laser ranging (LLR) constrains the inverse-mass to be less than the Earth-Moon distance \citep{Murphy:2012rea,Merkowitz:2010kka}, and Earth-based torsion balance experiments, in particular the E\"{o}t-Wash experiment, constrains the range of the force to be sub-micron \citep{Adelberger:2003zx,Adelberger:2005vu,Kapner:2006si}. Cosmologically, however, the force range should be at least inverse-Hubble to play a role in the background evolution of the field. Indeed, the Klein-Gordon equation is now $\ddot{\phi}+3H\dot{\phi}+m^2\phi+\alpha G\rho = 0,$ and one thus requires $m\sim H$ for the field not to be over-damped at the present time. At the level of perturbations, the modifications of gravity are only relevant on scales smaller than $m^{-1}=\lambda_{\rm C}$ (i.e. inside the Compton wavelength; \citealt{Brax:2013yja}) and they are therefore irrelevant for structure formation too.

The argument above assumes that the dark energy scalar is coupled universally to all matter species. It is possible to couple only to dark matter and leave the visible sector untouched. Such models are often called \emph{coupled quintessence} or \emph{coupled dark energy} in general \citep{Amendola:1999er,Copeland2006}. In these cases, there are no fifth forces in the visible sector but there is between dark matter particles. Many of the tests described in this review do not apply to these models with the exception of tests of the equivalence principle between dark matter and visible matter. These models are less theoretically well-motivated because there is no symmetry that prevents the scalar coupling to visible matter and so the absence of any coupling is tantamount to fine-tuning. As mentioned above, we do not consider them here.


Tuning either $\alpha$ or $m$ fails to simultaneously produce interesting cosmology and satisfy solar system tests precisely because those parameters are universal. Screening mechanisms solve this problem by dynamically suppressing the modifications of GR in the Solar System without the need to tune any parameter to small values. This leaves them free to assume values with significance for cosmology. In particular, the issue with the approaches above stems from the fact that the equations of motion for the scalar are both linear and Poisson-like. Thus they are superfluous copies of the Poisson equation for the metric potentials of GR, which is sufficient by itself to explain solar system observations. The essence of screening is to alter the structure of the Poisson equation, either by introducing a nonlinear generalization of the Laplacian operator or by adding a nonlinear potential for the field. One can write a generalized Poisson equation
\begin{equation}
Z^{ij}(\phi_0)\partial_i\phi\partial_j\phi + m_{\rm eff}^2(\phi_0)\phi=8\pi\alpha(\phi_0)G\rho,
\end{equation}
where we have included a nonlinear kinetic term for $\phi$ and have allowed this, the effective mass $m_\text{eff}$, and the coupling to matter $\alpha$ to depend on the background field value $\phi_0$ i.e. we have expanded the total field as $\phi\rightarrow \phi_0+\phi$ with $\phi_0$ being the field sourced by the surrounding environment e.g. the cosmological or galactic scalar (precisely which depends on the situation being considered). The schematic solution is
\begin{equation}
\phi\sim\alpha(\phi_0)\frac{GM}{|Z(\phi_0)|r}e^{-m_{\rm eff}(\phi_0) r}.
\end{equation}
Screening works by adjusting $\phi_0$ so that one of the following three conditions is satisfied in the Solar System:
\begin{enumerate}
\item The effective mass for the field $m R\ll1$ so that the field is short-range.
\item The coupling to matter $\alpha(\phi_0)\ll1$ so that the fifth force is weak.
\item The kinetic factor $Z^{ij}\gg1$ so that the fifth force is suppressed.
\end{enumerate}
Importantly, since the background field may depend on environment, it is possible to satisfy any of these conditions without tuning any model parameter to small values. The density on Earth differs from that in the cosmological background by 29 orders of magnitude, and this makes it easy to construct screened theories that are relevant cosmologically but naturally suppressed in the Solar System. In the cases described above, the screening mechanism is called chameleon screening \citep{Khoury:2003aq,Khoury:2003rn}, symmetron \citep{Hinterbichler:2010es} and dilaton screening \citep{Brax_1}, and kinetic screening \citep{Vainshtein:1972sx,Nicolis:2008in} respectively. Chameleon, symmetron, and dilaton screening are qualitatively similar\footnote{Astrophysical tests are not particularly sensitive to the specific mechanism \citep{Sakstein:2015oqa}. Laboratory tests are more model-dependent \citep{Burrage:2017qrf}.}, so in this review we will refer to them collectively as \emph{thin-shell screening}. Similarly, \emph{kinetic screening} can be subdivided into Vainshtein and K-mouflage theories. We now describe each screening mechanism in turn.


\subsection{Thin-shell screening}

\subsubsection{Chameleon screening} 

\begin{figure}[ht]
{\includegraphics[width=0.45\textwidth]{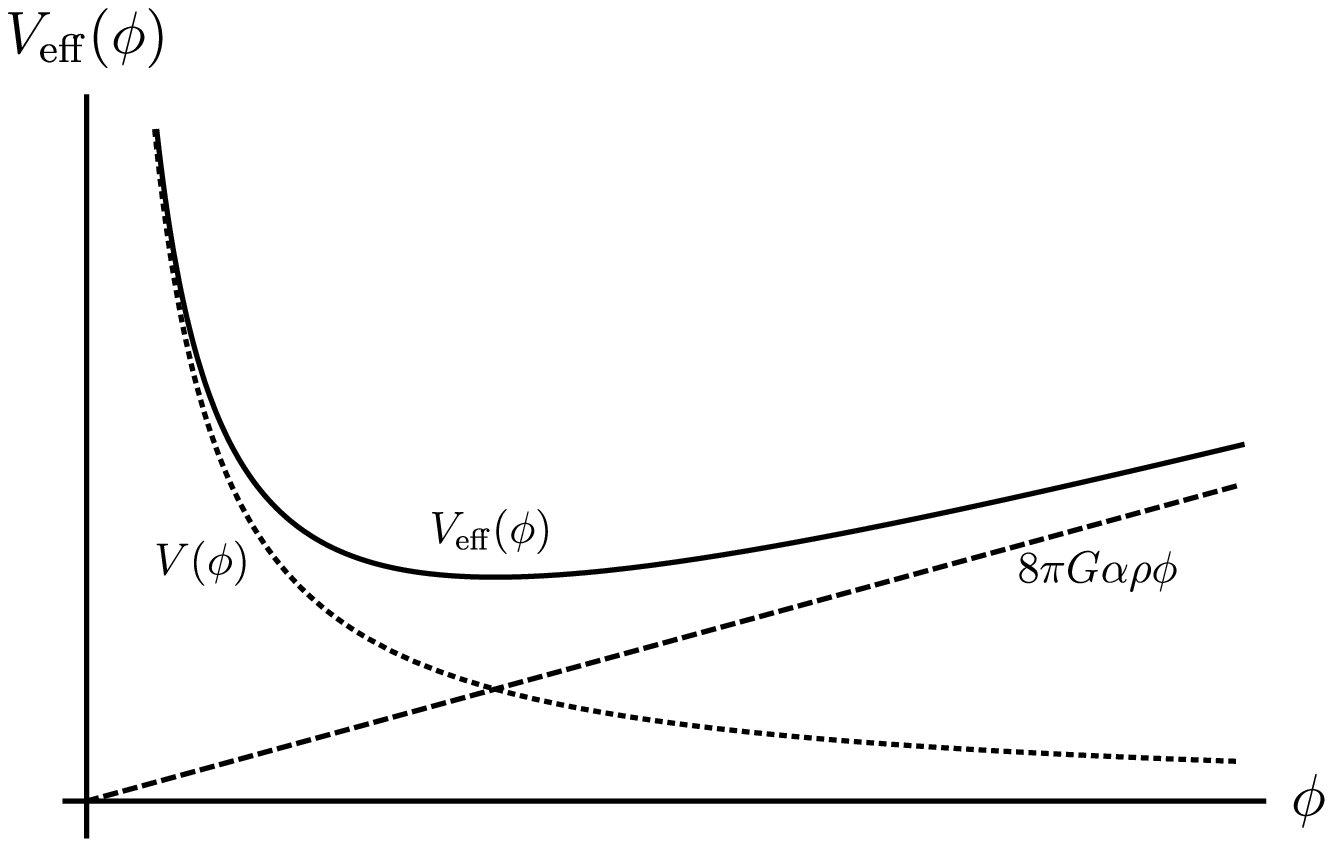}}
{\includegraphics[width=0.45\textwidth]{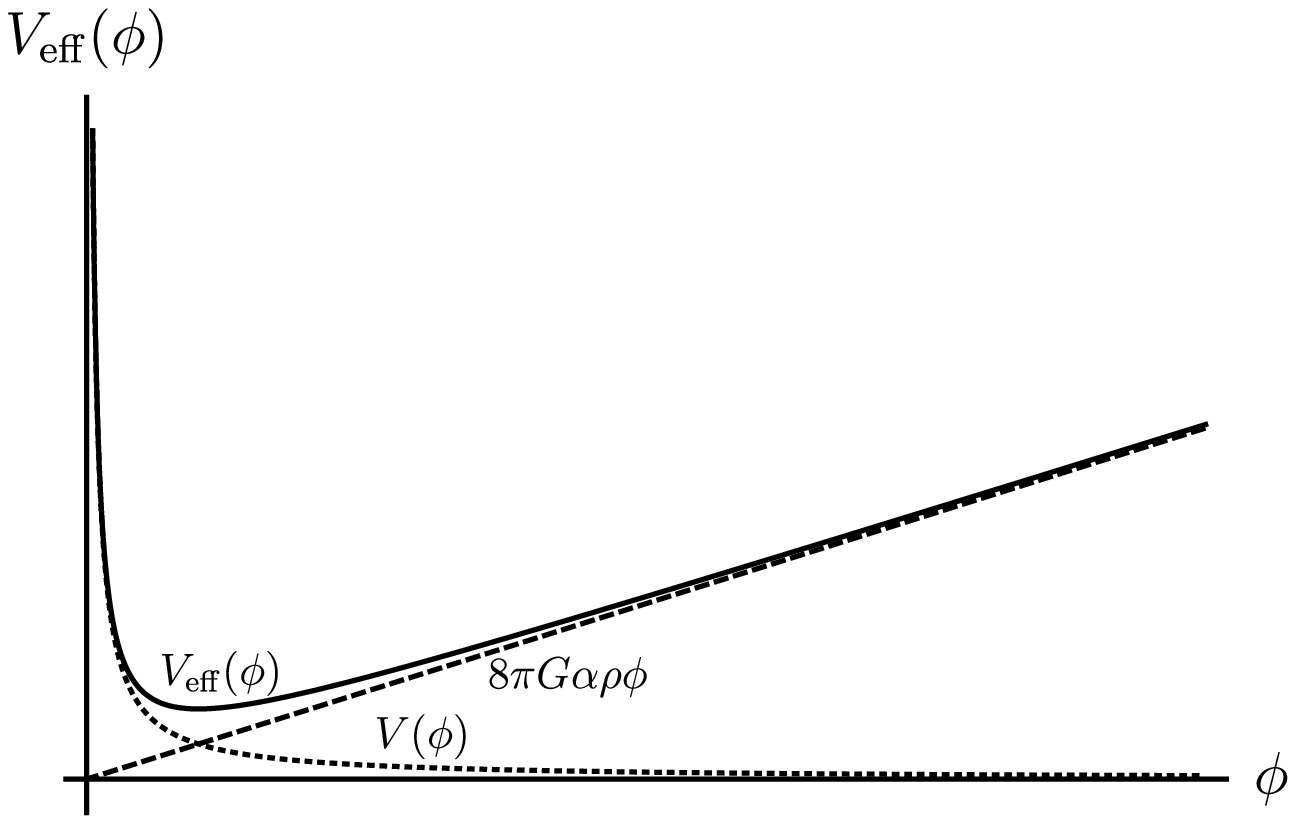}}
\caption{The chameleon effective potential for both low (left panel) and high (right panel) density environments. The bare potential $V(\phi)$ is shown by the dotted black line and the contribution from the matter coupling by the black dashed line. The effective potential is the sum of these contributions, and is shown in solid black.}\label{fig:cham_potentials}

\end{figure}

\noindent The equation of motion for the chameleon is \citep{Khoury:2003aq,Khoury:2003rn}
\begin{equation}\label{eq:chamEOM}
\nabla^2\phi= -\frac{n\Lambda^{4+n}}{\phi^{n+1}}+8\pi\alpha G\rho,
\end{equation}
for constant $n$, which describes a scalar coupled to matter with constant coupling $\alpha$ and a nonlinear scalar potential $V(\phi)=\Lambda^{4+n}\phi^{-n}$. The dynamics of the scalar can be thought of as arising from a density-dependent effective potential
\begin{equation}
V_{\rm eff}(\phi)=\frac{\Lambda^{4+n}}{\phi^n} + 8\pi\alpha\phi\rho,
\end{equation}
which is illustrated schematically for low and high densities in Fig.~\ref{fig:cham_potentials} for positive $n$. One can see that there is a density-dependent minimum given by
\begin{equation}
\phi_{\rm min}=\left(\frac{n\Lambda^{4+n}}{\alpha\mpl^n\rho}\right)^{\frac{1}{n+1}},
\end{equation}
and the effective mass for fluctuations about this minimum is 
\begin{equation}\label{eq:chammass}
m_{\rm eff} = V''_{\rm eff}(\phi)=n(n+1)\Lambda^{n+4}\left(\alpha\frac{\rho}{n\mpl\Lambda^{n+4}}\right)^{\frac{n+2}{n+1}}.
\end{equation}
It is both the density-dependent minimum and mass that make chameleon screening possible. One can see from equation \eqref{eq:chammass} or Fig.~\ref{fig:cham_potentials} that the mass is an increasing function of the density, so that the force can be made dynamically short-range within the Galaxy (high density) but long-range cosmologically (low density).

\begin{figure}[ht]
\centering
{\includegraphics[width=0.5\textwidth]{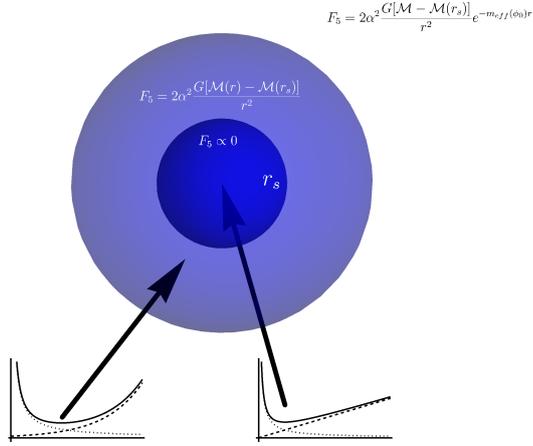}}
\caption{Thin shell screening. The field minimizes its effective potential inside the screening radius but begins to roll to the minimum in the low-density background outside. Only the mass outside the screening radius contributes to the fifth force.}\label{fig:thin_shell}
\end{figure}

The screening mechanism is illustrated qualitatively in Fig.~\ref{fig:thin_shell}. Consider a spherical object of high density (a star or dark matter halo for example) immersed in a larger medium of lower ambient density (a galaxy or the cosmic background for example). If the object is big enough (to be quantified shortly), the scalar will minimize its effective potential within the object and the equation of motion \eqref{eq:chamEOM} is $\nabla^2\phi=0$, i.e. the field is unsourced. Since the mass at the minimum is high, we expect this to remain the case as we move out from the center until the density falls to a point where the field can begin to roll to its asymptotic value $\phi_0$, which is the minimum of the effective potential in the background. We will refer to the radius where the field begins to roll as the \emph{screening radius} $\rs$. Outside this, the mass of the field is small ($m_{\rm eff} R\ll1$, where $R$ is the radius of the star), so the scalar field's motion is set by the density: $\nabla^2\phi=8\pi\alpha G\rho$. Integrating from $\rs$ to $R$ then yields a fifth-force acceleration
\begin{equation}\label{eq:f5chamshell}
a_5=2\alpha^2\frac{GM(r)}{r^2}\left[1-\frac{M(\rs)}{M(r)}\right]\quad \rs<r<R.
\end{equation}
Outside the object, one has $\nabla^2\phi+m_{\rm eff}^2(\phi_0)\phi=8\pi\alpha G\rho$, a massive Klein-Gordon equation with boundary condition at the object's surface altered by the screening radius. This gives
\begin{equation}\label{eq:f5chamshell2}
a_5=2\alpha^2\frac{GM}{r^2}\left[1-\frac{M(\rs)}{M}\right]e^{-m_{\rm eff}(\phi_0)r}\quad r>R.
\end{equation}
One can see that the force is suppressed by a factor $1-{M(\rs)}/{M}$ without the need to tune $\alpha$ to small values. The size of the screening radius determines whether or not the fifth force is screened. If $\rs\ll R$ then this factor is of $\mathcal{O}(1)$ and the force is unscreened, whereas if $\rs\approx R$ the field profile is sourced only by the mass inside a thin shell and the force is screened. The essence of chameleon screening, therefore, is that nonlinearities in the field conspire to remove the scalar charge of the source over much of the object's volume (this is often referred to as the \emph{thin shell effect}). For a spherical object, one can find $\rs$ by solving \citep{Davis:2011qf,Sakstein:2013pda,Burrage:2016bwy}
\begin{equation}\label{eq:rseq}
\chi\equiv\frac{\phi_0}{2\alpha\mpl}=4\pi G\int_\rs^R r'\rho(r').
\end{equation}
If this equation has no solutions then $\rs=0$ and the object is fully unscreened. This will be the case when
\be
\chi > \frac{GM}{R} = |\Phi| \quad\mbox{(unscreened).}
\ee
Hence, the Newtonian potential determines whether an isolated object is screened or not. For this reason, $\chi$ is often called the \emph{self-screening parameter}.

One popular model that exhibits the chameleon mechanism is Hu-Sawicki $f(R)$ gravity \citep{Hu:2007nk}, where one replaces the Ricci scalar in the Einstein-Hilbert action with
\begin{equation}
f(R)=R-a\frac{m^2}{1+(R/m^2)^{-b}}.
\end{equation}
This function is chosen such that in low curvature regimes such as the Solar System ($R\ll m$) one has $f(R)\approx R +\mathcal{O}(R^b/m^{2b})$ so that deviations from GR are suppressed, whereas in cosmological regimes $f(R)\approx R- am^2(1-({R}/{m^2})^{-b})$
so that the theory looks like a cosmological constant and small perturbations. Typically, one tunes $a$ and $m$ to match the $\Lambda$CDM background evolution and the additional terms are seen as causing deviations from $\Lambda$CDM at the linear and nonlinear levels. In terms of chameleons, one has $n=-b/(1+b)$ \citep{Burrage:2016bwy} so that $-1<n<-1/2$ and $\alpha=1/\sqrt{6}$. The only free parameter is the present-day cosmological background field value $f_{R0}=\dd f/\dd R|_{z=0}$. This controls the level of screening and sets $\chi$ via $f_{R0}=2\chi/3$.\par

\subsubsection{Symmetron screening}

\begin{figure}[ht]
\centering
{\includegraphics[width=0.5\textwidth]{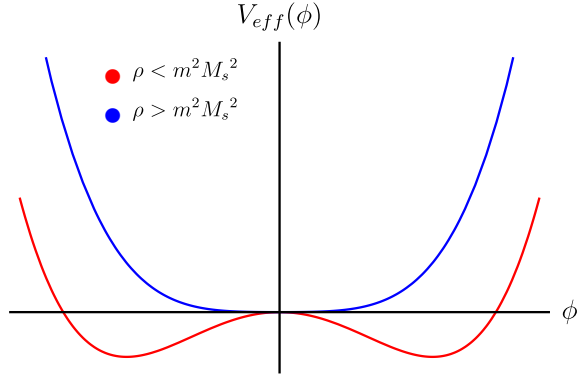}}
\caption{The symmetron effective potential. The red line is for $\rho<\mu^2\ms^2$ and the blue line is for $\rho>\mu^2\ms^2$.}\label{fig:symmetron}
\end{figure}

\noindent The symmetron \citep{Hinterbichler:2010es} screens in similar manner to chameleons. Its equation of motion is also the derivative of an effective potential
\begin{equation}
\nabla^2\phi=\frac{\dd V_{\rm eff}}{\dd\phi}\quad\textrm{with}\quad V_{\rm eff} = -\frac{\mu^2}{2}\left(1-\frac{\rho}{\ms^2\mu^2}\right)\phi^2+\frac{\lambda}{4}\phi^4,
\end{equation}
and the coupling to matter is $\alpha(\phi)=\mpl\phi/\ms^2$. There are then three free parameters, a quartic self-coupling $\lambda$ and two new mass scales $\mu$, the field's bare mass and $\ms$, which parameterizes the coupling to matter. This effective potential can have two different shapes depending on the density, as shown in Fig.~\ref{fig:symmetron}. When $\rho<\mu^2\ms^2$ there are two degenerate minima located at 
\begin{equation}
\phi_\pm\approx\frac{\mu}{\sqrt{\lambda}}
\end{equation}
so that the coupling to matter is $|\alpha(\phi_\pm)|=\mu\mpl/\lambda\ms^2$, which can be $\mathcal{O}(1)$. Conversely, when $\rho>\mu^2\ms^2$ there is a single minimum at $\phi=0$ so that $\alpha=0$ and the field does not couple to matter. Inside $\rs$ one has $\phi=0$ (provided the object is dense enough, $\rho>\mu^2\ms^2$) so that the field is unsourced. Outside $\rs$, the field begins to roll to $\phi_\pm$ where there is a non-zero matter coupling $\alpha(\phi_\pm)$. As in the chameleon case, the fifth force is then sourced by the mass inside the shell only. In particular, equations \eqref{eq:f5chamshell} and \eqref{eq:f5chamshell2} hold with $m_{\rm eff}(\phi_0)\rightarrow\mu$ and $\alpha\rightarrow\alpha(\phi_0)$. The main difference between the two mechanisms is that chameleons suppress the fifth force by having a large mass in dense environments and $\alpha\sim\mathcal{O}(1)$ on all scales, while symmetrons have a low mass on all scales and a small coupling to matter in dense environments. Another novel feature of the symmetron is the possibility of having domain wall solutions where the boundary conditions are such that the asymptotic field on different sides of a dense object can reside in different minima \citep{Pearson:2014vqa,Llinares:2014zxa}.

There are several variants of the symmetron including generalized symmetrons \citep{Brax:2011aw,Brax:2012gr} and radiatively stable symmetrons \citep{Burrage:2016xzz,Blinov:2018vgc}. Another variant, which is less-well studied, is the environment-dependent dilaton \citep{Brax_1}. These theories screen in a similar manner to symmetrons except that the suppression of the coupling to matter in dense environment is not due to a symmetry breaking transition. 

\vspace{3mm}


\subsection{Kinetic screening}
\label{sec:vainscreentheory}

\subsubsection{Vainshtein screening}

\begin{figure}[ht]
\centering
{\includegraphics[width=0.35\textwidth]{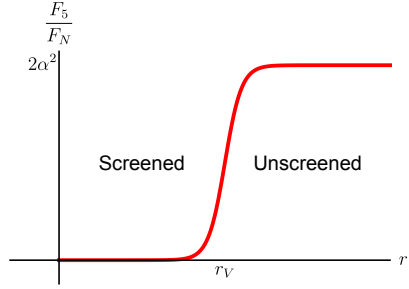}}
\caption{Vainshtein screening due to a point mass located at $r=0$. The red curve shows the ratio of the fifth- to Newtonian force outside the object.}\label{fig:vainshtein_screening}
\end{figure}

Vainshtein screening \citep{Vainshtein:1972sx} works in a qualitatively different way to thin-shell screening: instead of adding a scalar potential it changes the Laplacian structure of the Poisson equation. The Vainshtein mechanism is very generic in modified gravity theories \citep{Kimura:2011dc,Koyama:2013paa,Kobayashi:2014ida,Sakstein:2015aqx}, and arises in DGP models, generic Horndeski (and extensions) theories, and massive gravity (and extensions to multi-metric gravity). The quintessential example is the scalar field theory known as the ``galileon" because they are invariant under the galilean transformation $\phi \rightarrow \phi + c + b_\mu x^\mu$ \citep{Nicolis:2008in}.  The two most common examples of galileon theories are the cubic galileon, with equation of motion
\begin{equation}\label{eq:cubicgal}
\nabla^2\phi + \frac{r_c^2}{3}\left[(\nabla^2\phi)^2-\nabla_i\nabla_j\phi\nabla^i\nabla^j\phi\right] =8\pi\alpha G\rho,
\end{equation}
and the quartic galileon with equation of motion
\begin{align}\label{eq:quarticgal}
\nabla^2\phi + \frac{r_c^4}{4}\left[(\nabla^2\phi)^3-\nabla^2\phi\nabla_i\nabla_j\phi\nabla^i\nabla^j\phi+2\nabla_i\nabla_j\phi\nabla^j\nabla^k\phi\nabla_k\nabla^i\phi\right] =8\pi\alpha G\rho.
\end{align}
These contain familiar terms from the Poisson equation, the Laplacian and the matter sourcing, and also a new kinetic term parameterized by the \emph{crossover scale} $r_c$\footnote{Galileons have their roots in higher-dimensional brane world models where $r_c$ parameterises the scale at which higher-dimensional effects are important, hence its name.}. The screening is illustrated in Fig. \ref{fig:vainshtein_screening}. There are two regimes of interest. When the new kinetic terms are negligible, one is left with the Poisson equation and hence a fifth force that is a factor of $2\alpha^2$ larger than the Newtonian force. The difference arises when the Laplacian is negligible, in which case one finds that the fifth force is given by
\begin{align}\label{eq:galileonforce}
a_5=2\alpha^2\frac{GM}{r^2}\left(\frac{r}{\rv}\right)^q,
\end{align}
where $q=3/2$ for the cubic galileon and $q=2$ for the quartic. The new scale 
\begin{align}\label{eq:rv}
r_{\rm V}^3=
  \begin{cases}
            \frac{4}{3} \alpha GMr_c^2,  &\quad \textrm{cubic galileon}\\
   \sqrt{2}\alpha GMr_c^2, & \quad \textrm{quartic galileon}
  \end{cases}
\end{align}
is the \emph{Vainshtein radius}, which determines which of the two kinetic terms are dominant. If $r<\rv$ then the galileon terms dominate and the fifth force is therefore suppressed by a factor of $(r/\rv)^q$. The Vainshtein radius of the Sun (for theoretically interesting values of $r_c$) is $\mathcal{O}(100\textrm{ pc})$, showing that the region outside massive bodies is screened to large distance. Far beyond $\rv$, the fifth force is again a factor of $2\alpha^2$ larger than the Newtonian force and the theory can have cosmological consequences. 

The Vainshtein mechanism is not as efficient inside extended mass distributions. This is because equations \eqref{eq:cubicgal} and \eqref{eq:quarticgal} are total derivatives for spherically symmetric mass distributions and so one has a nonlinear generalization of Gauss' law. This means that only the mass inside the radius $r$ contributes to the field profile and so one has a radially-varying Vainshtein radius that is smaller than the Vainshtein radius found using the total mass \citep{Schmidt:2010jr}. For extended distributions, the galileon force profile is given by $a_5=2\alpha^2GM(r)/r^2g(r/r_*(r))$, where $r_*(r)^3=C\alpha r_c^2GM$ with $C=16/3$, $\sqrt{3/2}$ for the cubic and quartic galileon models respectively. The function $g$ is given by
\begin{equation}
g(\xi)=
  \begin{cases}
        2\xi^3\left(\sqrt{1+\xi^{-3}}-1\right) , & \quad \textrm{Cubic galileon}\\
    \xi^3\sinh\left[\frac{1}{3}\arcsinh\left(\frac{3}{\xi^3}\right)\right], &\quad \textrm{Quartic galileon},
  \end{cases}
\end{equation}
and is chosen so that the fifth force approaches the asymptotic solution $a_5=-2\alpha^2 GM/r^2$ at large distances.

\subsubsection{Vainshtein breaking: beyond Horndeski and DHOST}
\label{sec:VBintro}
Theories in the beyond Horndeski and DHOST classes exhibit a breaking of the Vainshtein mechanism: it operates perfectly outside of extended objects, but inside the Newtonian and lensing potentials are determined by modified equations of the form \citep{Kobayashi:2014ida,Koyama:2015oma,Crisostomi:2017lbg,Langlois:2017dyl,Dima:2017pwp}
\begin{align}
\frac{\dd\Psi}{\dd r}& =- \frac{GM}{r^2}-\frac{\Upsilon_1G}{4}\frac{\dd ^2 M(r)}{\dd r^2}\label{eq:VB1}\\
\frac{\dd\Phi}{\dd r}& =  -\frac{GM}{r^2} + \frac{5\Upsilon_2G}{4r}\frac{\dd M(r)}{\dd r}-\Upsilon_3G\frac{\dd ^2 M(r)}{\dd r^2},\label{eq:VB2}
\end{align}
where $M(r)$ is the mass enclosed within radius $r$. The dimensionless parameters $\Upsilon_i$ characterize the strength of the modifications and are very important from a cosmological point of view because they are related to the $\alpha_i$ parameters appearing in the effective description of dark energy \citep{Bellini:2014fua,Gleyzes:2014dya,Langlois:2017mxy}. In particular, one has
\begin{align}
\Upsilon_1&=4\frac{\left(\alpha_H+(1+\alpha_T)\beta_1\right)^2}{(1+\alpha_T)(1+\alpha_V-4\beta_1)-\alpha_H-1}\label{eq:BH1}\\
\Upsilon_2&=-\frac45\frac{\alpha_H\left(\alpha_H-\alpha_V+2\beta_1(2+\alpha_T)\right)+\beta_1\alpha_T\left(1+(1+\alpha_T)\beta_1\right)}{(1+\alpha_T)(1+\alpha_V-4\beta_1)-\alpha_H-1}\label{eq:BH2}\\
\Upsilon_3&=-\frac{\beta_1\left(\alpha_H+(1+\alpha_T)\beta_1\right)}{(1+\alpha_T)(1+\alpha_V-4\beta_1)-\alpha_H-1}\label{eq:BH3},
\end{align}
and so constraints on $\Upsilon_i$ directly constrain the cosmology of these theories\footnote{Note that this assumes that the effects of the local environment on the field profile are negligible. This is typically a good approximation because the radius of these objects is much smaller than the wavelength of the large scale galileon field sourced by the local dark matter e.g. halos and filaments implying that the galileon's field gradient is approximately linear. The galileon symmetry ($\phi\rightarrow\phi+b_\mu x^\mu +c$) ensures that adding a linear gradient to the solution leaves the physics unaltered and so the field due to the local environment does not diminish the sensitivity to the cosmological parameters.}. Since beyond Horndeski theories and DHOST theories in particular can survive the GW170817 bounds on $\alpha_T$ (depending on the choices of free functions) there has been a recent interest in constraining these parameters. Equations \eqref{eq:VB1} implies that constraints could be placed using non-relativistic objects (stars, galaxies, clusters) and equation \eqref{eq:VB2} implies that lensing can also be used. One can effectively set $\alpha_T=0$ in equations \eqref{eq:BH1}--\eqref{eq:BH3} to apply astrophysical bounds to constrain the remaining parameters \citep{Sakstein:2017xjx} or impose more relations between the parameters by demanding that $\alpha_T=0$ identically at the level of the functions appearing in the action \citep{Crisostomi:2017lbg,Langlois:2017dyl,Dima:2017pwp}.

It is possible to impose further restrictions on the class of DHOST theories to ensure that both gravitational waves travel luminally over cosmological distances, and that gravitons do not decay into dark energy (this requirement is highly constraining; \citealt{Creminelli:2018xsv,2019arXiv190607015C}). In these cases, and further restricting to theories that are stable and allow for the existence of Newtonian stars one finds a different kind of Vainshtein breaking to that described above \citep{Hirano:2019scf,Crisostomi:2019yfo}. Indeed, inside matter one has
\begin{equation}
    \label{eq:VBDHOSTIN}
    \frac{\dd\Psi}{\dd r}=-\frac{G_\Psi^{\rm in}M(r)}{r^2},\quad\frac{\dd\Phi}{\dd r}=-\frac{G_\Phi^{\rm in}M(r)}{r^2}
\end{equation}
while outside matter, they become
\begin{equation}
    \label{eq:VBDHOSTOUT}
    \frac{\dd\Psi}{\dd r}=-\frac{G_\Psi^{\rm out}M(r)}{r^2},\quad\frac{\dd\Phi}{\dd r}=-\frac{G_\Phi^{\rm out}M(r)}{r^2},
\end{equation}
where $G^{\rm in/out}_{\Psi/\Phi}$ are again related to $G$ and the parameters appearing in the effective description of dark energy. In these theories, the Vainshtein mechanism is broken both inside and outside of matter. Presently, the strongest bounds on the new parameters do not come from the astrophysical probes considered in this work, instead they come from either solar system tests of post-Newtonian gravity and the rate of the orbital decay of the Hulse-Taylor pulsar. For this reason, we will not discuss them in what follows. Devising novel astrophysical probes of these theories that could compete with other probes is certainly worthwhile, especially since, by design, they are able to evade the stringent bounds on DHOST theories.

\subsubsection{K-mouflage}

One variant on the Vainshtein mechanism that has attracted attention recently is K-mouflage models \citep{Babichev:2009ee,Brax:2012jr}. These are distinct from Vainshtein screening in that their equation of motion is a non-linear analog of Gauss' law. The simplest example is
\begin{equation}
    \label{eq:K-mouflage}
    \nabla_i\left(\nabla^i\phi+r_c\nabla_j\phi\nabla^j\phi\nabla^i\phi\right)=8\pi\alpha G\rho,
\end{equation}
 where $r_c$ is the analog of the crossover scale for Vainshtein screening. For a spherically-symmetric source, the screening is very similar to Vainshtein screening (Fig.~\ref{fig:vainshtein_screening}) in that there is a K-mouflage radius $r_K$ inside of which the fifth-force is suppressed and outside of which the fifth-force is unscreened and one has $a_5=-2\alpha^2 GM/r^2$. For example, inside the K-mouflage radius the model in Eq.~\eqref{eq:K-mouflage} gives \citep{Brax:2012jr}
 \begin{equation}
     a_5=2\alpha^2\frac{GM}{r^2}\left(\frac{r}{r_K}\right)^{\frac43};\quad r_K^2=2\alpha r_c GM.
 \end{equation}
One particularly interesting model that falls into the K-mouflage class is the DBIon \citep{Burrage:2014uwa}. 
 
K-mouflage screening models are less well explored than thin-shell and Vainshtein screening, although there has been some recent work over the last few years studying their viability \citep{Barreira:2015aea} and their cosmology \citep{Brax:2014wla,Brax:2018zvh}, as well as some steps towards identifying cosmological probes and placing constraints. These include linear probes \citep{Benevento:2018xcu}, galaxy clusters \citep{Brax:2015lra}, large scale structure \citep{Brax:2014yla}, and effects on non-linear scales \citep{Brax:2014gra}. We will not review novel probes of these models at the present time, but we anticipate that future revisions will include such discussions as the study of K-mouflage screening develops to the same stage as thin-shell and Vainshtein screening.

\subsection{Observational signatures}

\subsubsection{Equivalence principle violations}\label{sec:eqprinvio}

Perhaps the most important difference between Vainshtein and thin-shell screening is that they violate different equivalence principles. In particular, chameleons/symmetrons violate both the weak and strong equivalence principle, for \emph{macroscopic} objects, whereas Vainshtein-screened theories violate only the strong equivalence principle \citep{Hui:2009kc,Hui:2012qt,Hui:2012jb}. 

The weak equivalence principle (WEP) is the statement that all weakly gravitating bodies (whose mass does not receive significant contribution from gravitational binding energy) fall at the same rate in an externally applied field regardless of their internal structure and composition. Formally, one can consider a non-relativistic object at position $\vec{x}$ placed in an external gravitational field $\Phi^{\rm ext}$ and scalar field $\phi^{\rm ext}$. Its equation of motion is $M_I\ddot{\vec{x}}=-M_G\vec{\nabla}\Phi^{\rm ext}-QM_G\vec{\nabla}\phi^{\rm ext }$, where
$M_I$ is the object's inertial mass and $M_G$ its gravitational mass, which can be thought of as a ``gravitational charge'' that describes how it responds to an external gravitational field. Similarly, we have defined a scalar charge-to-mass ratio $Q$ that parameterizes the response of the object to an external scalar field. In GR $Q=0$ since the scalar is absent, and $M_I=M_G=M$. In scalar-tensor theories one still has $M_I=M_G$, but now $Q\ne0$. For the screening mechanisms above, one has, in the limit that the object is a test mass,
\begin{align}
Q=
  \begin{cases}
           \alpha\left[1-\frac{M(\rs)}{M}\right],  &\quad \textrm{thin-shell screening}\label{eq:Qcham}\\
   \alpha, & \quad \textrm{galileons, K-mouflage}
  \end{cases}.
\end{align}
Galileons and K-mouflage models therefore satisfy the WEP\footnote{Note that this is only the case for a single isolated test mass interacting with a galileon field whose wavelength is larger than the body's extent. Two-body systems can violate the WEP due to the highly nonlinear nature of the equation of motion \citep{Hiramatsu:2012xj,Andrews:2013qva,Kuntz:2019plo}. Interestingly,  \citet{Hiramatsu:2012xj} also find a 4\% enhancement of the galileon force compared with the one-body case. The WEP is also broken in more general Vainshtein-screened theories. Further investigation of these effects may yield new novel probes.} whereas chameleons/symmetrons violate it since the scalar charge depends on the screening radius, which in turn depends on the structure and composition of the object. It is worth emphasizing that all theories considered here preserve the WEP at the level of the action. The WEP violation in chameleon and symmetron theories arises because macroscopic screened objects lead to a strong distortion of the scalar field profile, despite being in the weak-gravity regime.

The strong equivalence principle (SEP) is the statement that any two objects will fall at the same rate in an externally applied gravitational field even if their self-gravity is considerable. There are very few theories apart from GR that satisfy the SEP and scalar-tensor theories are no exception. The violation of the SEP has its origin in the fact that the scalar couples to the trace of the energy-momentum tensor (rather than the energy-momentum pseudo-tensor) so that only non-relativistic matter contributes to the coupling and not the gravitational binding energy. Another way of seeing this is that there is a powerful no-hair theorem \citep{Hui:2012qt} (although see~\citealt{Sotiriou:2013qea,Sotiriou:2014pfa,Babichev:2016rlq} for exceptions) for black holes in generic scalar-tensor theories. The lack of scalar hair implies that the scalar charge is zero.

\subsubsection{Searching for screening}
\label{sec:searching}

Having elucidated the properties of and differences between the screening mechanisms, we now explain how to identify astrophysical objects in which to search for them. We begin with chameleon and symmetron screening and then move onto Vainshtein screening.

As the level of thin-shell self-screening is set by an object's Newtonian potential $|\Phi| = GM/R$ (the object is unscreened for self-screening parameter $\chi>GM/R$), identifying unscreened objects subject to fifth forces is tantamount to seeking those with low Newtonian potentials. Some commonly used object are listed in Table \ref{tab:newtonian_potential} along with their characteristic Newtonian potentials. On the face of it, the Earth and Moon should remain unscreened for very low background field values, making them excellent probes of modified gravity. It is important to note however that $\Phi$ receives an additional contribution from surrounding mass, and hence objects may be \emph{environmentally}- as well as self-screened. The Milky Way has a characteristic potential of $\mathcal{O}(10^{-6})$, and hence environmentally screens the Earth and Moon for typical field values.

Some of the most useful objects for testing thin-shell screening are post-main-sequence stars and dwarf galaxies. Post-main-sequence stars have masses of order their progenitor star's mass but radii 10 to 100 times larger, lowering their potential to $10^{-7}$ or less. In the case of galaxies, the virial theorem relates circular velocity to Newtonian potential:
\begin{equation}
v_c^2\sim\frac{GM}{R}.
\end{equation}
Spiral galaxies have $v_c\sim200$ km/s, giving ${GM}/{R}\sim 10^{-6}$, but dwarf galaxies have $v_c\sim50$ km/s so that ${GM}/{R}\sim 10^{-8}$. The strategy for searching for thin-shell screening effects is therefore to seek dwarf galaxies that are not environmentally screened, i.e. that reside in voids. These galaxies or their constituent post-main-sequence stars then serve as probes of screening. (We discuss the observational determination of environmental screening in Sec.~\ref{sec:screening_maps}.) No astrophysical objects have $\Phi \lesssim 10^{-8}$ (we are ignoring planets and smaller objects, which are typically screened by their environment and not useful for constraining these theories), and hence only laboratory tests can probe smaller values of $\chi$ \citep{Burrage:2016bwy}. 
Current bounds on $\chi$ imply that the these theories cannot act as dark energy \citep{Wang:2012kj} but they may have effects on smaller scales. Indeed the effective mass in the cosmological background is \citep{Brax:2012gr}
\begin{equation}
m_{\rm eff}^2\approx \frac{H_0^2}{\chi},
\end{equation}
so taking $\chi\lsim10^{-7}$ (commensurate with current bounds; see Sec.~\ref{sec:astro_tests}) corresponds to a Compton wavenumber $k \simeq (0.1\textrm{ Mpc})^{-1}$: the fifth force would only be operative on smaller scales i.e. scales smaller than $0.1 Mpc$.


\begin{table}[ht]
\centering
\begin{tabular}{c | c}\
Object & Newtonian Potential $\Phi$ \\\hline
Earth & $10^{-9}$\\
Moon & $10^{-11}$\\
Main-sequence star (Sun-like) & $10^{-6}$\\
Post-main-sequence star ($M=1$--$10M_\odot$, $R=10$--$100R_\odot$) & $10^{-7}$--$10^{-8}$\\
Spiral galaxy (Milky Way-like, $v_c\sim 200$ km/s) & $10^{-6}$\\
Dwarf galaxy ($v_c\sim 50$ km/s) & $10^{-8}$
\end{tabular}
\caption{Objects commonly considered as probes of thin-shell screening. The second column shows surface Newtonian potential $|\Phi| = GM/R$.}\label{tab:newtonian_potential}
\end{table}

Galileons are harder to test on astrophysical scales due to their highly efficient screening and nonlinear equations of motion. These make computing observables difficult. Indeed, the strongest bounds until recently came from lunar laser ranging (LLR), which restricts fractional deviations in the inverse-square law to $10^{-11}$ at the Earth--Moon distance \citep{Nordtvedt:2003pj,Merkowitz:2010kka,Murphy:2012rea,Murphy:2013qya}, allowing deviations of the form of Eq.~\eqref{eq:galileonforce} to be constrained directly \citep{Dvali:2002vf}. 

One promising test utilizes the violations of the strong equivalence principle discussed above, which lead to interesting novel effects detailed in Sec.~\ref{sec:SEPviolations}. Any system composed of both non-relativistic and strongly gravitating objects (for example, a galaxy comprising a central super-massive black hole as well as non-relativistic stars and gas) has the potential to exhibit violations of the SEP. In some theories, in particular, beyond Horndeski and DHOST (Sec.~\ref{sec:VBintro}), it is possible that the Vainshtein mechanism is broken inside objects \citep{Kobayashi:2014ida,Koyama:2015oma,Saito:2015fza}, which allows for additional tests \citep{Sakstein:2015aqx,Sakstein:2015zoa,Jain:2015edg,Sakstein:2015aac,Sakstein:2016ggl} that we discuss in Sec.~\ref{sec:vainshtein_breaking}.





\section{Surveys}
\label{sec:surveys}

\begin{figure}[ht]
\centering
{\includegraphics[width=0.7\textwidth]{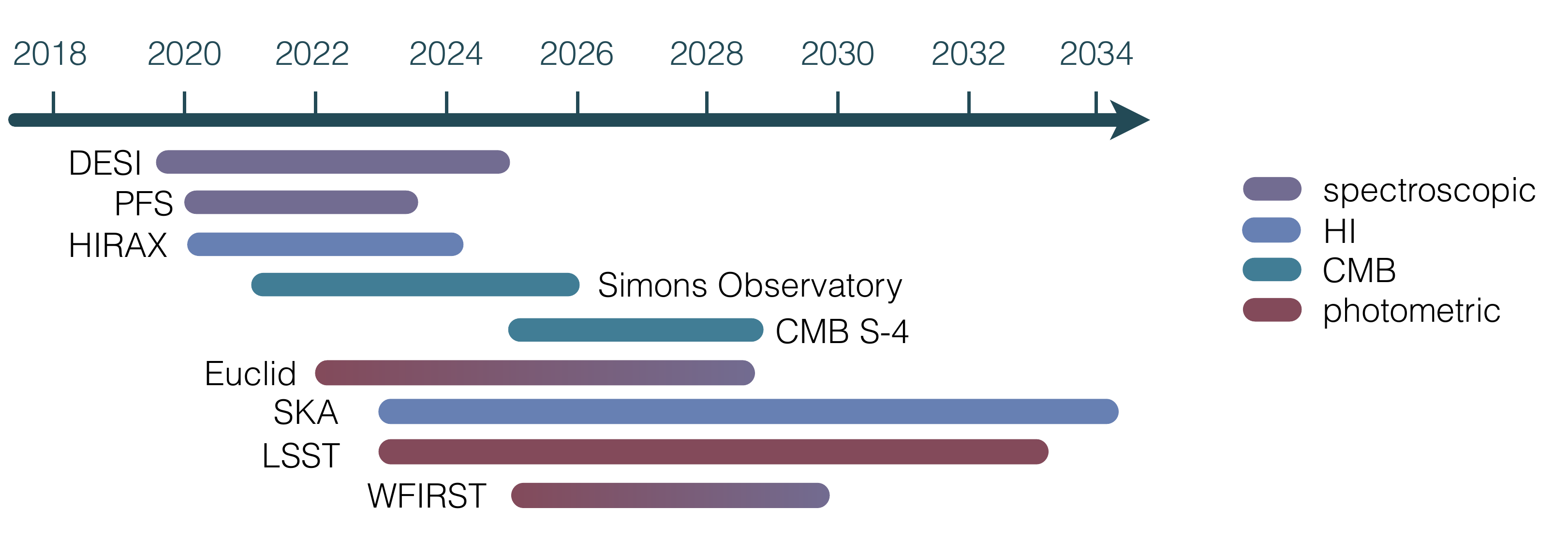}}
\caption{Timeline for future surveys.}\label{fig:surveys_timeline}
\end{figure}

In this section we describe the types of survey useful for constraining modified gravity, and list future surveys that will be particularly important in this regard. The surveys are summarised in Table~\ref{tab:survey_table}, while the timeline for upcoming surveys is shown in Fig.~\ref{fig:surveys_timeline}.

\subsection{Types of survey and available datasets}
\label{sec:surveys_current}

Several types of cosmological surveys are currently being carried out, which can roughly be divided into the following categories:

\begin{itemize}

\item{\tt spectroscopic galaxy redshift surveys:} Spectroscopic galaxy redshift surveys probe the three-dimensional matter density field by measuring angular galaxy positions and redshifts using spectroscopic methods. These redshifts are measured to high precision ($\sfrac{\Delta z}{z} \lesssim \mathcal{O}(10^{-3})$), as spectroscopy allows the identification of specific atomic transition lines in galaxy spectra. Assuming a cosmological model to relate redshifts to distances, these surveys can be used to measure the statistical properties of the galaxy density field. The main applications include inference of the distance-redshift relation through measurements of the Baryonic Acoustic Oscillations (BAOs) peak and measurement of the growth rate of structure through the anisotropy imprinted on the 2-point function by redshift space distortions (RSDs). BAOs are fluctuations in the matter density caused by acoustic waves in the pre-recombination plasma, which show up as an enhancement in galaxy clustering at a scale $\sim 150$ Mpc today and allow constraints to be placed on the components of the universe's density budget that determine its expansion history (see e.g. \citealt{Beutler:2016arn}). RSDs are discussed further in Sec.~\ref{sec:RSD}. Other applications of spectroscopic surveys include the study of cosmic voids or peculiar galaxy velocities. Current examples of these types of surveys include the Baryonic Oscillation Spectroscopic Survey (BOSS; \citealt{Alam:2015})\footnote{\url{http://www.sdss3.org/surveys/boss.php}} and its extension (eBOSS) \citep{Abolfathi:2017}\footnote{\url{http://www.sdss.org/surveys/eboss/}}, and WiggleZ \citep{Parkinson:2012}\footnote{\url{http://wigglez.swin.edu.au/site/}}. As measuring galaxy spectra is time-consuming, spectroscopic samples typically consist on the order of millions of objects.

\item{\tt photometric galaxy redshift surveys:}  Photometric surveys infer galaxy redshifts from measurements of their fluxes in several wavebands, resulting in samples of hundreds of millions of objects or more. Photometric redshift fitting codes usually rely on representative spectroscopic training samples, which allow the measured redshifts to reach accuracies of $\sfrac{\Delta z}{z} \gtrsim \mathcal{O}(10^{-2})$. Photometric galaxy redshift surveys can be used for galaxy clustering, weak lensing and cluster clustering measurements, amongst others. Due to the increased redshift uncertainties, these analyses are usually performed in 2D but some 3D information can be retrieved through tomographic techniques. Examples for current and completed photometric surveys include the Dark Energy Survey (DES; \citealt{Abbott:2018})\footnote{\url{http://www.darkenergysurvey.org}}, the Kilo Degree Survey (KiDS; \citealt{Kuijken:2015})\footnote{\url{http://kids.strw.leidenuniv.nl}}, the Canada-France-Hawaii Telescope Legacy Survey (CFHTLS)\footnote{\url{https://www.cfht.hawaii.edu/Science/CFHTLS/}; \url{http://www.cfhtlens.org/astronomers/content-suitable-astronomers}}, and surveys with the Hyper Suprime Cam (HSC; \citealt{Mandelbaum:2018, Aihara:2018})\footnote{\url{http://hsc.mtk.nao.ac.jp/ssp/}} on Subaru.

\item{\tt Cosmic microwave background surveys:} Cosmic microwave background (CMB) experiments measure the fluctuations in the temperature and polarization of the CMB. The ratio of CMB to foreground emission peaks at frequencies between 40 and 100 GHz but measurements are typically conducted in a range of frequencies between 10 and 300 GHz (or higher, in some cases) in order to separate the CMB from Galactic and extragalactic foregrounds. The primary CMB anisotropies probe the matter distribution at the last-scattering surface, but the observed anisotropies also receive contributions from the integrated Sachs-Wolfe (ISW) effect, gravitational lensing and the Sunyaev-Zel'dovich (SZ) effect which probe the low-redshift universe. Examples of current and completed CMB experiments include the Wilkinson Microwave Anisotropy Probe (WMAP; \citealt{Bennett:2013})\footnote{\url{https://wmap.gsfc.nasa.gov}}, Planck \citep{Planck-Collaboration:2016}\footnote{\url{https://www.cosmos.esa.int/web/planck}}, the Atacama Cosmology Telescope (ACT; \citealt{Louis:2017, De-Bernardis:2016})\footnote{\url{https://act.princeton.edu}}, the South Pole Telescope (SPT; \citealt{Henning:2018, Benson:2014})\footnote{\url{https://pole.uchicago.edu}}, POLARBEAR\footnote{\url{http://bolo.berkeley.edu/polarbear/}} and BICEP/Keck\footnote{\url{http://bicepkeck.org/}}.

\item{\tt Intensity mapping surveys:} Intensity mapping experiments forego identifying individual objects and, instead, measure the intensity of radiation  of a particular frequency as a function of angular position, typically with emphasis on a particular atomic line. These surveys are therefore sensitive to all sources of emission in some frequency range (galaxies, IGM, etc.) and by using the redshift of the line as a proxy for distance it is possible to trace the three-dimensional structure of the universe to great distance. Current efforts mainly focus on mapping the Hydrogen 21 cm line (HI) but other possibilities like CO or CII are also considered. For instance, the Canadian Hydrogen Intensity Mapping Experiment (CHIME)\footnote{\url{https://chime-experiment.ca}}, based at the Dominion Radio Astrophysical Observatory in British Columbia, Canada, is undertaking an HI intensity mapping survey in the frequency range $400-800$ MHz, corresponding to $1\le z \le 3$, covering approximately $25,000$ deg$^2$. The field of line-intensity mapping is in its infancy but recently HI, CO, CII and Lyman-$\alpha$ line-emission have been detected in cross-correlation using, amongst others, data from the Green Bank Telescope\footnote{\url{https://greenbankobservatory.org/}}.

\end{itemize}

\begin{table*}
\begin{center}
\begin{threeparttable}
\begin{tabular}{>{\centering}m{2.8cm}c>{\centering}m{1.2cm}>{\centering}m{1.7cm}>{\centering}m{3.5cm}>{\centering}m{3.cm}@{}m{0pt}@{}}
\hline\hline 
Survey & Duration & Area [sq.deg.] & $z$-range & Survey properties & Main probes & \\ \hline     

BOSS &  2008 - 2014 & $10,000$ & $0 - 0.7$ & $n_{\mathrm{obj}}\sim 1.5 \times 10^{6}$ & GC\tnote{a}, QSOs\tnote{b} & \\

eBOSS &  2014 - 2020 & 7500 & $0.6 - 3.5$ & $n_{\mathrm{obj}}\sim 1 \times 10^{6}$ & GC, QSOs, Ly-$\alpha$ & \\

WiggleZ &  2006 - 2011 & 1000 & $< 1.0$ & $n_{\mathrm{obj}}\sim 2 \times 10^{5}$ & GC & \\
 
DES &  2013 - 2019 & 5000 & $0 - 1.4$ & $m_{\mathrm{lim, r}} = 24$ \\ $n_{\mathrm{obj}}\sim 3 \times 10^{8}$ & WL\tnote{c}, GC, clusters, SNe Ia & \\

KiDS & 2013 - 2019 & 1500 & $\bar{z} \sim 0.7$ & $m_{\mathrm{lim, r}} = 24.9$\\ $n_{\mathrm{obj}}\sim 3 \times 10^{7} $ & WL & \\

HSC &  2014 - 2020 & 1400 & $\bar{z}\sim1$ & $m_{\mathrm{lim, r}} = 26.1$ \\ $n_{\mathrm{obj}}\sim 1 \times 10^{8}$ & WL, GC, clusters, SNe Ia,  & \\

WMAP &  2001 - 2010 & full sky & $1,100$ & res. $< 0.3^{\circ}$ \\ sens. $\sim 60 \mu$K arcmin & T\tnote{d}, P\tnote{d} & \\

Planck &  2009 - 2013 & full sky & $1,100$ & res. $< 10$ arcmin \\ sens. $\sim 45 \mu$K arcmin & T, P & \\

ACT/AdvACT &  2013 - 2019 & 1000/$18,000$ & $1,100$ & res. $\sim 1$ arcmin \\ sens. $\sim 30 \mu$K arcmin & T, P & \\

SPT/SPT3G &  2013 - 2023 & 500/2500 & $1,100$ & res. $\sim 1$ arcmin \\ sens. $\sim 17 \mu$K arcmin & T, P & \\ \hdashline

DESI &  2019 - 2024 & 14,000 & $0 - 3.5$ & $n_{\mathrm{obj}}\sim 3 \times 10^{7}$ & GC, QSOs, Ly-$\alpha$ & \\

PFS &  early 2020's & 2000 & $0.8 - 2.4$ & $n_{\mathrm{obj}}\sim 1 \times 10^{7}$ & near-field cosm., GC, Ly-$\alpha$ & \\

LSST &  2023 - 2033 & $20,000$ & $\bar{z} \sim 1.2$ & $m_{\mathrm{lim, r}} \sim 27$ \\ $n_{\mathrm{obj}}\sim 2 \times 10^{10}$ & WL, GC, SL\tnote{e}, clusters, SNe Ia & \\

WFIRST & mid-2020's  & 2000 & $1 - 3$ & $m_{\mathrm{lim, J}} \sim 26.9$ \\ $n_{\mathrm{obj}}\sim 5\times 10^{8}/$\\ $2 \times 10^{7}$ & WL, GC, SNe Ia & \\

Euclid & 2022 - 2028 & $15,000$ & $0.7 - 2.1$ & $n_{\mathrm{obj}}\sim 5\times 10^{7}/$\\ $1 \times 10^{9}$ & GC, WL & \\

Simons Observatory & 2021 - 2026 & $15,000$ & $1,100$ & res. $\sim 1.5$ arcmin \\ sens. $\sim 5\mu$ K & T, P & \\

CMB S-4 & mid-2020's & 8000 & $1,100$ & res. $\sim 3$ arcmin\tnote{f} \\ sens. $\sim 1\mu$ K arcmin\tnote{f} & T, P & \\

SKA & mid-2020's & $15,000$ & $<2$ & $n_{\mathrm{obj}}\sim 1 \times 10^{9}$ & HI IM\tnote{g}, HI GC & \\

HIRAX & 2020 - 2024 & $15,000$ & $0.8 - 2.5$ & res. $\sim 5-10$ arcmin & HI IM & \\

CHIME & 2017 - 2023 & $25,000$ & $1 - 3$ & res. $\sim 15-30$ arcmin & HI IM & \\


 \hline \hline
\end{tabular}
\begin{tablenotes}
\item[a] \footnotesize{GC: galaxy clustering.}
\item[b] \footnotesize{QSOs: quasars.}
\item[c] \footnotesize{WL: weak lensing.}
\item[d] \footnotesize{T: temperature, P: polarization (CMB).}
\item[e] \footnotesize{SL: strong lensing.}
\item[f] \footnotesize{According to Science Book specifications.}
\item[g] \footnotesize{IM: intensity mapping.}
\end{tablenotes}
\end{threeparttable}
\end{center}
\caption{Properties of a selection of past and current (above dashed line) and planned (below dashed line) surveys.}
\label{tab:survey_table}
\end{table*}



\subsection{Upcoming surveys}
\label{sec:surveys_future}



We are living in a golden age of survey science. An abundance of observational programs have been proposed that will substantially increase the amount and quality of data with which we can explore the various aspects of gravity discussed in this review. In what follows we will briefly summarize the key surveys which, hopefully, will be rolled out in the next decade or so.
\begin{itemize}
\item{\tt DESI:} The Dark Energy Spectroscopic Instrument \citep{DESI-Collaboration:2016}\footnote{\url{http://desi.lbl.gov}}, based at Kitt Peak in Arizona, will be used to perform a spectroscopic survey of over 30 million objects (luminous red galaxies, OII emitting galaxies, quasars and "bright" galaxies) out to a redshift of $z\sim 3.5$. The survey will cover about $14,000$ deg$^2$.
\item{\tt PFS:} The Subaru Prime Focus Spectrograph \citep{Takada:2014} will construct a spectroscopic redshift survey of emission line (OII) galaxies in the redshift range, $0.8\le z\le 2.4$ covering approximately $2,000$ deg$^2$ down to apparent magnitude $r\sim26$.\footnote{\url{http://pfs.ipmu.jp}}
\item{\tt LSST:} The Large Synoptic Survey Telescope \citep{Ivezic:2008, LSST-Science-Collaboration:2009}\footnote{\url{http://www.lsst.org}}, based in Chile, will undertake a deep (reaching an $r$-band magnitude limit of $r\sim27$) and wide ($20,000$ deg$^2$) imaging survey of the southern sky over 10 years. It will use photo-$z$ for radial information and will provide information on tomographic galaxy clustering and cosmic shear, strong lensing, galaxy cluster counts and type Ia supernovae. 
\item{\tt WFIRST:} The Wide-Field InfraRed Survey Telescope \citep{Spergel:2013}\footnote{\url{http://wfirst.gsfc.nasa.gov}} is a planned satellite mission that will carry out an imaging and a spectroscopic survey of $2,000$ deg$^2$ (reaching a $J$-band magnitude limit of $J\sim27$). This will result in an imaging catalogue of 500 million galaxies and spectra of 20 million galaxies in the redshift range $1\le z \le 3$.
\item{\tt Euclid:} Euclid \citep{Laureijs:2012}\footnote{\url{http://sci.esa.int/euclid/}} is a satellite mission to be launched in the early 2020s, which is going to cover $15,000$ deg$^2$ on the sky. It will determine the redshifts of $5 \times 10^7$ galaxies in the range $0.7 < z <1.8$ using an infrared spectrograph. It will further conduct a photometric survey of $10^9$ galaxies in the redshift range $0<z<2$. The spectroscopic data will be used mostly for galaxy clustering, baryon acoustic oscillation and redshift space distortion measurements, while the imaging data will be used to measure cosmic shear.
\item{\tt SKA:} The Square Kilometre Array \citep{Maartens:2015}\footnote{\url{https://www.skatelescope.org}} is a partially-funded radio facility which will be based in two sites -- the Karoo region in South Africa and the Murchinson region in Western Australia -- and will consist of three instruments: SKA1-MID, consisting of 254 single pixel dishes covering $350-1760$~MHz, SKA1-SUR, an array of 96 dishes with 36 beam-phased aperture arrays covering a similar frequency range, and SKA1-LOW, a set of 911 aperture array stations covering $50-350$~MHz. The SKA can be used for HI intensity mapping and for measuring spectroscopic redshifts and the galaxy continuum.
\item{\tt HIRAX:} The Hydrogen Intensity and Real-time Analysis eXperiment \citep{Newburgh:2016}\footnote{\url{https://www.acru.ukzn.ac.za/~hirax/}}, based in South Africa, is an HI intensity mapping survey covering a redshift range $0.8\le z \le 2.5$ with a sky coverage of $15,000$ deg$^2$.
\item{\tt Simons Observatory:} The Simons Observatory \citep{Ade:2018}\footnote{\url{https://simonsobservatory.org/index.php}} is a CMB experiment covering $15,000$ deg$^2$, at a resolution of $1-2$ arcmin with a sensitivity of approximately $5\mu K$. It will cover the multipole range $50<\ell<3000$ in temperature.
\item{\tt Stage 4 CMB observatory (S4):} Current ground-based CMB facilities will be superseded by a coordinated, multi-site experiment \citep{Abazajian:2016}. The aim is that the combined instruments will map $40\%$ of the sky with an rms noise sensitivity of $\sim 1\mu K$ arcmin in temperature and beams with a $\sim 3$ arcmin FWHM. This means that the S4 will effectively cover the multipole range $30<\ell<3000$ in temperature and $30<\ell<5000$ in polarization.\footnote{\url{https://cmb-s4.org}}

\end{itemize}

\newpage
\section{Nonlinear structure formation: (semi-)analytic approaches}
\label{sec:nonlinear}

\subsection{Nonlinear structure in GR vs. modified gravity}

As structures in the Universe grow, the density fluctuations become larger than unity on small scales and the structure formation enters in the non-linear regime. At large scales, density perturbations remain small and linear perturbation theory is applicable. However, these scales are prone to cosmic variance and the bulk of information is available on non-linear scales in future surveys. Thus, it is important to describe non-linear structure formation accurately in order to distinguish modified gravity models from GR. Although cosmic variance becomes small on small scales, there are increasing uncertainties from baryonic physics. Understanding of baryonic effects is even more important in modified gravity models as they can be degenerate with the effects of modified gravity. Screening mechanisms also play a role at small scales and it make it harder to distinguish modified gravity models from GR. 

In this section we summarise analytic and semi-analytic methods for describing the nonlinear regime of structure in GR, and their extensions to modified gravity, discuss the impact of baryonic physics on small scale structure and its degeneracy with the behaviour of gravity, and describe estimators that have recently been developed for maximising sensitivity to deviations from GR.


\subsubsection{(Semi-)Analytic approaches in GR}

A key ingredient for predicting several important cosmological observables is the nonlinear matter power spectrum. On quasi-linear scales, Eulerian Standard Perturbation Theory (SPT) provides a way to predict the onset of nonlinearity  (see \citealt{Bernardeau:2001qr} for details and references therein). The dark matter particle number density in phase space obeys the Vlasov equation, which describes phase-space conservation of the number density. In Eulerian perturbation theory, the Vlasov equation is first approximated by the continuity and Euler equations by taking the first two moments of the phase space density of dark matter particles and neglecting the stress tensor. These equations 
describe the evolution of the density perturbation and velocity divergence. In SPT, these nonlinear equations are solved perturbatively assuming smallness of these quantities. In order to obtain the leading order correction to the linear power spectrum, the 1-loop power spectrum, solutions for these quantities need to be obtained up to the third order in perturbations. The maximum wavenumber $k_{\rm max}$ below which the SPT prediction for the matter power spectrum at 1-loop order agrees with N-body simulation results is empirically determined by the following formula \citep{2009PASJ...61..321N}:
\begin{equation}
\frac{k_{\rm max}^2}{6 \pi^2} \int^{k_{\rm max}}_0 P_{L}(q, z) dq < C,
\end{equation}
where $P_L(q,z)$ is the linear power spectrum at redshift $z$. The constant value $C$ was calibrated from N-body simulations in a $\Lambda$CDM model as $C=0.18$ by imposing that the 1-loop SPT prediction agrees with N-body results within $1\%$ at $k < k_{\rm max}$ \citep{2009PASJ...61..321N} (note that this value depends on cosmological parameters, in particular on the amplitude of the linear power spectrum). Although this condition was calibrated in GR simulations, it gives a useful indication for the validity of the perturbation theory even in modified gravity models such as $f(R)$ and DGP models \citep{Koyama:2009me}.

SPT is known to suffer from several problems, which can be traced back to the fact that it does not consistently capture the effect of small-scale nonlinear perturbations on large-scale perturbations.
\comment{
Given the limited range of the convergence regime of 1-loop results, the 2-loop power spectrum is widely used now in GR \citep{Beutler:2016arn}.  
Firstly, the contribution from higher order loop corrections is not suppressed even at low wavenumbers at low redshifts. For example, it was shown that three loop corrections become larger than lower order loops at $z=0$ \citep{Blas:2013aba}. Secondly, the SPT over-predicts the impact of the small scale nonlinearity on large scale modes \citep{Nishimichi:2014rra}. N-body simulations show that the statistical properties of the large-scale structure of the universe are much more insensitive to the details of the small-scale physics. Finally, in configuration space the 1-loop power spectrum fails to describe the BAO bump in the correlation function \citep{Baldauf:2015xfa}.   
}%
To overcome these problems, various improvements have been proposed. The 1-loop power spectrum has the form 
\begin{equation}
P(k) = G(k, z)^2 P_i(k) + P_{{\rm MC}} (k, z), 
\end{equation}
where $P_i$ is the initial power spectrum, $G(k, z)$ is the propagator, which reduces to the growth function $D(z)$ at linear order, and the second term describes the mode coupling between different $k$ modes. The higher order loop corrections to the propagator can be resummed and the propagator is modified to  
\begin{equation}
G(z, k) = \exp \left(- \frac{k^2 D^2 \sigma_v^2}{2} \right) D(z), 
\end{equation}
where $\sigma_v$ is the linear velocity dispersion. Renormalised Perturbation Theory (RPT) approaches are based on this resummation of the propagator \citep{Crocce:2005xy}. Recently there has been a debate on the validity of this resummation. It was pointed out that this exponential damping disappears if a similar resummation is performed for the mode coupling term \citep{2012JCAP...04..013T,sugiyama}. Also the resummation of the propagator breaks the Galilean symmetry of the original equations \citep{Peloso:2016qdr}. The Effective Field Theory (EFT) of Large Scale Structure takes a different approach \citep{baumann/etal:2012,Carrasco:2012cv}. In this approach, only the effect of large-scale modes on the BAO feature is resummed. This leads to the damping of the BAO feature, leaving the smooth part of the power spectrum untouched. The effect of ultraviolet (UV) modes is included in the form of the counter term and this counter term needs to be calibrated using simulations or observations.    

On fully nonlinear scales, the halo model \citep{Cooray:2002dia} gives an intuitive understanding of how the nonlinear power spectrum should look like.
It assumes all matter in the Universe to be located in virialized structures, or halos.
This allows one to compute the statistics of the matter density field such as the power spectrum from the spatial distribution and density profiles of halos.
In order to fit simulations, however, the halo model approach needs to be tweaked by introducing several free parameters. The most widely used approach is the one proposed by \citet{Mead}. In this approach, there are seven parameters that need to be calibrated by simulations. The key input in this approach is the variance of the linear density perturbations smoothed on a comoving scale $R$. Another approach is to provide a mapping between the linear power spectrum $P_L(k)$ and the nonlinear power spectrum. The halofit model provides fitting formulae for this mapping calibrated from a suite of N-body simulations \citep{Smith:2002dz}. The fitting formula was revised by \citet{Takahashi:2012em} and this is now widely used to predict the nonlinear power spectrum for a given linear power spectrum computed by the linear Einstein-Boltzmann code such as CAMB and CLASS. Finally, there are attempts to create emulators for nonlinear power spectra using a carefully chosen sample set of cosmological simulations and provide accurate predictions over the wide parameter space \citep{Lawrence:2017ost,Knabenhans:2018cng}.   

\begin{figure}[ht]
\centering
	{\includegraphics[angle=270,width=13cm]{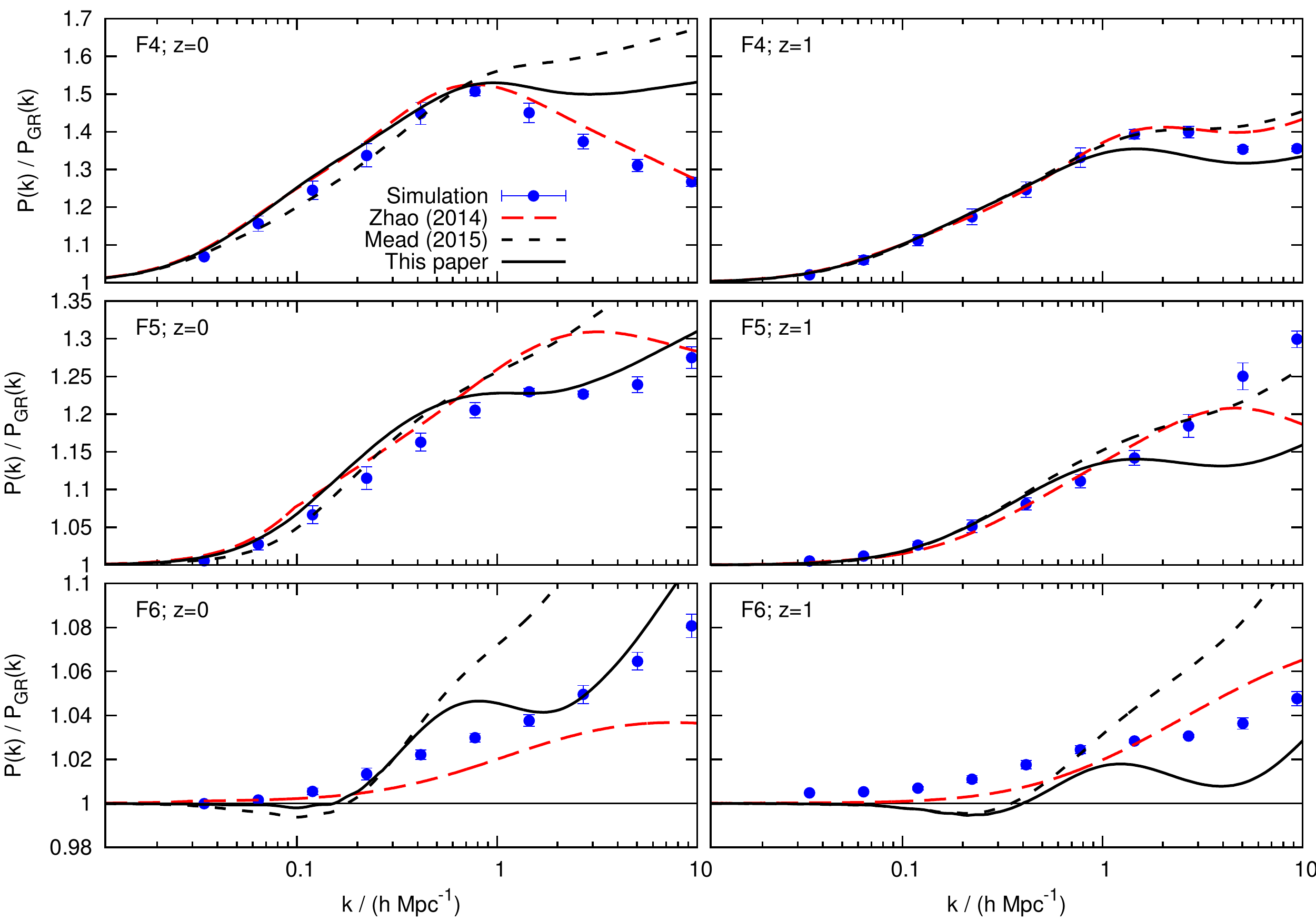}}
	\caption{
	A comparison of the ratios of power spectra for $f(R)$ models compared to an equivalent LCDM model for $|f_{R0}| = 10^{-4}$ (top), $10^{-5}$ (middle) and $10^{-6}$ (bottom) at $z =0$ (left-hand column) and 1 (right-hand column). We show the power spectrum from the simulations of Li et al. (\citealt{2013MNRAS.428..743L}; blue points) together with that from two versions of the halo model; that of Mead et al. (\citealt{Mead}, short-dashed; black) and Mead et al.(\citealt{Mead_2} solid; black). The former does not take into account chameleon screening and thus over-estimates the deviation from GR. We also show the MG-HALOFIT model of  Zhao (\citealt{Zhao:2013dza}; long-dashed; red) that was fitted to the same simulation data as shown (amongst others) and provides a better fit. From \citet{Mead_2}. 
}\label{fig:nonlinearpk}
\end{figure}

\subsubsection{(Semi-)Analytic approaches in modified gravity}
\label{sec:nonlinear_semiMG}

The approaches developed for GR can be extended to compute nonlinear power spectra in modified gravity. The main complication comes from the screening mechanism. Screening modifies the Poisson equation so that the relation between the Newton potential and density becomes nonlinear. In the perturbation-theory approach, this nonlinear Poisson equation can be expanded in terms of the density field and the 1-loop power spectrum can be computed \citep{Koyama:2009me, Cusin:2017wjg}. Another complication arises in chameleon screening. In this case, due to the mass of the scalar field, the linear growth function becomes scale dependent and the higher order solutions in SPT need to be computed numerically \citep{Taruya:2014faa, Bose:2016qun}. The regime of validity of perturbation theory is then expected to be a function of the field mass. An EFT approach has been applied to DGP models in \citet{2018JCAP...04..063B}.  

The halo model or halofit model predicts the nonlinear power spectrum for a given linear power spectrum. Since these models were developed in GR, screening mechanisms are not taken into account. As a consequence, a naive use of these approaches by just replacing the linear power spectrum by the one in modified gravity over-predicts deviations from GR because it does not account for the suppression of the fifth force due to screening \citep{2008PhRvD..78l3523O,Zhao:2010qy}. Thus these approaches need to be modified in the presence of screening. For an $f(R)$ gravity model, an extension of halofit was developed in \citet{Zhao:2013dza}. The halo model has also been extended to include the effect of screening through the modification of spherical collapse model parameters \citep{schmidt:08,schmidt:10,Lombriser:2013eza}. Fig.~\ref{fig:nonlinearpk} shows the comparison of these approaches for the ratios of power spectra for $f(R)$ models compared to $\Lambda$CDM \citep{Mead_2}. \citet{2019MNRAS.tmp.1778C} proposed a method to use the reaction of a $\Lambda$CDM matter power spectrum to the physics of an extended cosmological parameter space by adopting the halo model and nonlinear perturbation theory. Emulators of the deviation from $\Lambda$CDM matter power spectrum in Hu-Sawicki $f(R)$ models were presented in \citet{2019arXiv190308798W}. 

On linear scales, there is a well-defined connection between the statistics of galaxies and that of matter (see \citealt{biasreview} for a review), known as bias. It is important to take into account that, if the growth in the model is scale-dependent, then the bias relation also becomes scale-dependent \citep{parfrey/hui/sheth:2011}. A particularly interesting target is the cross-correlation of two galaxy populations differing in bias or screening properties, which leads to parity breaking in the relativistic correlation function that enhances sensitivity to fifth forces \citep{Bonvin:2013ogt}. Note however that in many viable models, these effects only appear on scales that are already moderately nonlinear (see Fig.~\ref{fig:nonlinearpk}), so that nonlinear effects (including nonlinear galaxy bias and velocity bias) present in GR need to be taken into account carefully. The modification to the Euler equation leads to the appearance of new terms in the spherical harmonic expansion of the correlation function, which may be observable with present and upcoming spectroscopic surveys such as DESI \citep{Bonvin_MG,Gaztanaga:2015jrs}. The octopole in particular may provide a relatively clean probe of screening per se (especially at higher $z$), as opposed simply to the modified growth rate in the cosmological background which shows up predominantly in the dipole \citep{Kodwani}.

Finally, it is worth noting that statistics beyond the power spectrum may be useful for breaking degeneracies between the parameters of $\Lambda$CDM and modified gravity. For example, the convergence power spectrum is degenerate between $f_{R0}$ and $\sigma_8$ and $\Omega_\text{m}$. This degeneracy can however be broken with information from the bispectrum and/or clustering statistics based on peak counts \citep{Shirasaki}.

\subsection{Baryonic effects and small scale structure}
\label{sec:baryons}

\noindent Any effect that impacts nonlinear scales is a potential systematic for tests of modified gravity. A notable example is the effect of baryons, which is able to alter appreciably the distribution of total matter on small scales, and consequently lead to biased cosmological constraints if not properly accounted for (e.g. \citealt{Hearin_Zentner, 2011MNRAS.417.2020S}). We begin this section with a general discussion of the effects of baryons on the clustering of matter, before going on to discuss methods to incorporate them in modified gravity predictions, and some degeneracies that arise.

The fact that baryons are subject to pressure forces which become relevant below their Jeans scale immediately implies modifications to the distribution of total matter, compared to a case in which structure formation takes place only under the influence of gravity. For example, gas loses energy via radiative cooling as it falls into gravitational potential wells, which makes it easier to trigger the formation of high-density structures such as gas and stellar disks. As baryons fall towards the centre of potential wells they drag dark matter gravitationally, via a process known as adiabatic contraction \citep{Blumenthal,Gnedin_1,Gnedin_2}. Both these effects enhance the clustering power on scales $k > {\rm few}\times h/{\rm Mpc}$ ($\ell \gtrsim 3000 - 5000$ in weak-lensing spectra; \citealt{White04, Guillet}).

\begin{figure*}
  \subfigure[]
  {
    \includegraphics[width=0.5\textwidth]{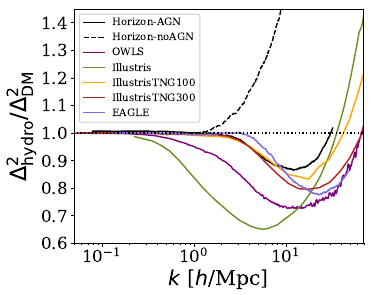}
  }
  \subfigure[]
  {
    \includegraphics[width=0.5\textwidth]{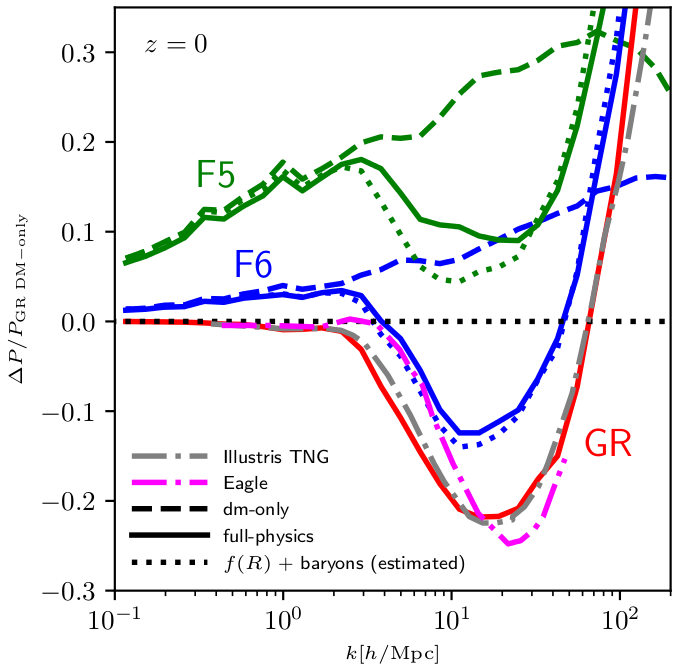}
  }
  \caption{(a) The change in power spectrum due to baryonic effects as a function of scale, compared between the Horizon, OWLS, Illustris and EAGLE simulations. Reproduced from \citet{Chisari}. (b) Competition between baryonic and modified gravity effects: on scales $k \gtrsim 2\ h/{\rm Mpc}$, baryonic effects can noticeably suppress (solid green and blue) the enhancement in clustering power due to the fifth force (dashed green and blue) in $f(R)$ models. Reproduced from \citet{Arnold:2019vpg}.}
  \label{fig:PS_baryons}
\end{figure*}

There are however also baryonic processes that work to suppress the amplitude of the matter power spectrum. First, beyond galactic scales, the hot gas found in the intra-cluster medium (ICM) is more diffuse than dark matter in clusters, which reduces power relative to N-body predictions on scales below a few ${\rm Mpc}$ \citep{Zhan}. Second, and perhaps most importantly for weak-lensing applications, the violent ejections of matter from both supernova (SNe) and active galactic nuclei (AGN) can redistribute matter out to scales of the halo virial radius. SNe are the primary cause of mass loss below the knee in the galaxy stellar mass function ($M_* \approx 10^11 M_\odot$), and AGN the primary cause above. For example, the OWLS simulation suite \citep{OWLS} includes nine baryonic physics models, differing in the description of AGN, SNe, gas cooling and stellar initial mass function (IMF). These models were intended to span a range of a priori plausible possibilities for baryonic physics, making them a useful reference point for the impact of baryons as well as a testbench for empirical and semi-analytic parametrisations. Current state-of-the-art cosmological hydrodynamical simulations, such as EAGLE \citep{Schaye, Crain}, Illustris and IllustrisTNG \citep{Illustris, 2018MNRAS.473.4077P, 2018MNRAS.475..676S}, MassiveBlack \citep{MassiveBlack} and Horizon \citep{HorizonAGN, HorizonAGN_2}, all include prescriptions for AGN feedback which is thought to be a crucial ingredient to bring the results of these simulations closer to observations. The baryonic impact on the power spectrum measured from these simulations is shown in the left panel of Fig.~\ref{fig:PS_baryons}. Current simulations agree qualitatively on the main changes to the power spectrum, namely a suppression of power on scales $k \gtrsim 1 h/ {\rm Mpc}$ due to AGN feedback and a turnaround on scales $k \approx 10\ h/{\rm Mpc}$ due to adiabatic cooling. A point to stress however is that these simulations disagree on the exact magnitude of the effects, which can be traced back to differences in the physical implementation of baryonic physical processes on scales below the resolution scale of these simulations (the so-called {\it sub-grid physics}). This uncertainty is a considerable hurdle for constraining additional parameters beyond $\Lambda$CDM, e.g. those of modified gravity.

\citet{Arnold:2019vpg,Arnold:2019zup,Leo:2019ada} have recently made progress on understanding the interplay between modified gravity and baryonic effects by presenting the results of a first set of full modified gravity ($f(R)$) galaxy formation simulations (see Sec.~\ref{sec:simulations} below for a discussion on $N$-body methods in modified gravity). These are shown to produce a number of baryonic observables that are in broad agreement with observations, including the formation of disk galaxies similar to our Milky Way. One of the key findings is that for weaker modified gravity (e.g. Hu-Sawicki $|f_{R0}|=10^{-6}$), the effects of baryons and modified gravity are separable and additive, while for stronger models such as $|f_{R0}|=10^{-5}$ the two effects are coupled which require full simulations to reproduce. Fig.~\ref{fig:PS_baryons}b shows the impact on the total matter power spectrum; the competing effects between the fifth force and AGN feedback are apparent, but the scale-dependence is different so the degeneracy is not perfect. The work of \citet{Arnold:2019vpg,Arnold:2019zup,Leo:2019ada} also shows that the properties of baryonic matter, such as stars, gas, neutral hydrogen and black holes, can be significantly affected by modified gravity in a way that is not captured in detail by simple analytic models.
While simulations are required to track the full impact of baryons on matter clustering, they are computationally expensive to run. This makes it infeasible to marginalise over baryonic effects to constrain physics such as modified gravity by simulating the full range of possible baryonic and gravitational models. Effort is therefore being devoted to the development of empirical or semi-analytic models that capture the effects of baryons with some free parameters that can be marginalized over in real data analyses. A convenient framework for this is the halo model, in which the impact of baryons can be captured by modifications to the assumed density profile of dark matter haloes \citep{Rudd, Zentner, Zentner_2, Hearin_Zentner, Mead}. For example, \citet{Mead} proposed an augmented halo model that includes two parameters that govern the inner halo structure and which is able to describe the power spectrum in each of the OWLS simulation variants at the $\lesssim 5\%$ level. In \citet{Mead_2} this halo model is further extended to account for the effects of chameleon and Vainshtein screening, enabling the degeneracies between modified gravity and baryonic effects to be investigated. They find that while baryons mostly affect internal halo structure, modified gravity effects impact also the halo distribution, thereby making their effects potentially distinguishable (this is in line with Fig.~\ref{fig:PS_baryons}b from modified gravity simulations of galaxy formation).

Other methods besides halo model modifications have been developed for the purpose of determining and accounting for the effect of baryons on the power spectrum. One is based on principal component analyses (PCA), where one possible goal is to identify and discard the PCA components of the data vectors that are most contaminated by baryonic effects, which yields less degraded parameter constraints compared to simply discarding the small-scale elements of the data vectors \citep{Eifler, Mohammed, 2019MNRAS.488.1652H}. Other approaches are built on perturbation theory \citep{Lewandowski}, or more straightforwardly on fitting empirical functions to hydrodynamical simulation results \citep{Harnois}. Absent precise knowledge of baryonic effects, a combination of approaches will likely be necessary to maximise the scientific return of future surveys aimed at extending lensing power spectra to quasi- and non-linear scales. This is especially important for the case of tests of gravity given the degeneracies that may arise.

\subsection{Novel estimators for the nonlinear regime}

The screening effects of modified gravity models suppress deviations from general relativity in high-density regions.
These regions, however, contribute most to the matter power power spectrum in the nonlinear regime on small scales, which renders a detection of a potential modification of gravity more difficult.
In \citet{Lombriser:2015axa,White:2016yhs,Llinares:2017ykn,Valogiannis:2017yxm,Hernandez-Aguayo:2018yrp,Armijo:2018urs}, it was therefore proposed that the up-weighting of low-density, unscreened, regions in the calculation of the 2-point statistics would enhance the detectability of a modification.
Moreover, by considering different weights, one could more readily discriminate gravitational modifications from baryonic effects or new fundamental physics that can yield degenerate contributions in the power spectrum.
Density-weighting has been studied for marked correlation functions~\citep{White:2016yhs}, 
density transformations~\citep{Neyrinck:2009fs}, and clipping~\citep{Simpson:2011vn,Simpson:2013nja}. 

The truncated density field in clipping~\citep{Simpson:2011vn,Simpson:2013nja} is given by
\begin{equation}
 \tilde{\delta} = \delta_c(\vec{x}) = \left\{
 \begin{array}{ll}
  \delta_0 \,, & \delta(\vec{x})>\delta_0 \,, \\
  \delta(\vec{x}) \,, & \delta(\vec{x})\leq\delta_0 \,,
 \end{array}
 \right. \label{eq:clipping}
\end{equation}
for an overdensity $\delta(\vec{x})$ at position $\vec{x}$. The transformation was applied in \citet{Lombriser:2015axa,Valogiannis:2017yxm} to $f(R)$ and more general chameleon models as well as nDGP gravity.
A percent-level measurement of the clipped power spectrum at $k<0.3~\textrm{h/Mpc}$ would improve constraints on $f(R)$ gravity to $|f_{R0}|\lesssim 2\times10^{-7}$~\citep{Lombriser:2015axa}.

Motivated by the observation that the nonlinearly evolved density field is well described by a lognormal distribution even though the initial density field is nearly Gaussian, \citet{Neyrinck:2009fs} proposed the logarithmic transformation of the density field
\begin{equation}
 \tilde{\delta} = \ln(\delta+1) \,.\label{eq:logdensity}
\end{equation}
The transformation in Eq.~(\ref{eq:logdensity}) also reduces the relative weights of high-density regions in the calculation of 2-point statistics. The transformation was applied to $f(R)$ gravity in \citet{Lombriser:2015axa,Llinares:2017ykn,Valogiannis:2017yxm} and symmetron models in \citet{Llinares:2017ykn,Valogiannis:2017yxm}.
A generalization and optimization for testing modifications of gravity with Eq.~\eqref{eq:logdensity} was proposed in \citet{Llinares:2017ykn}. 
%
Finally, \citet{White:2016yhs} proposed marked correlation functions~\citep{White:2008ii} to re-weight the density field.
An analytic function of this transformation is given by~\citet{Valogiannis:2017yxm}
\begin{equation}
 \tilde{\delta} = \left( \frac{\rho_* + 1}{\rho_*+\bar{\rho}(\delta+1)} \right)^p \,, \label{eq:marked}
\end{equation}
with free parameters $\rho_*$ and $p$.


\citet{Valogiannis:2017yxm} compared the performance of the density transformations, Eqs.~(\ref{eq:clipping}), (\ref{eq:logdensity}), and (\ref{eq:marked}), in enhancing the signal-to-noise, finding that the marked transformation performs best for scales $k<2$~h/Mpc whereas clipping performs better when including smaller scales $k>2$~h/Mpc.

The above studies considered the ideal cases of upweighting certain regions of the dark matter density field, which is not directly observable. In observations, weak lensing tomography can be used to reconstruct the 3D large-scale distribution of matter, but the current data quality is still not sufficient for applying the above density weighting schemes to test gravity. An alternative way is to apply similar weighting schemes but to the 2D cosmic shear or weak lensing peak fields instead of the 3D matter density field. Finally, the density-weighting schemes can be applied to 3D tracer fields of the large-scale structure, such as galaxies, galaxy clusters or quasars, of which the most interesting one is the galaxy field, for which one can have high tracer number density and probe low redshift (where modified gravity effects are stronger). The practical challenge, however, is the lack of a reliable model to predict how galaxies populate the dark matter field from first principles, in particular in modified gravity models. In particular, the (unknown) physics of galaxy formation could have degenerate effects on the 2-point statistics with those of density weighting (see Sec.~\ref{sec:baryons}).  No matter which gravity model is the correct one it must produce a galaxy distribution that is in agreement with real observations. Following this logic, \citet{Hernandez-Aguayo:2018yrp,Armijo:2018urs} produced mock galaxy catalogs for different $f(R)$ scenarios, by tuning an empirical galaxy populating model in each of them individually, which have the same projected 2-point correlation function. They found that this greatly reduces the difference in the marked correlation functions of these models for various marks including the one described in Eq.~(\ref{eq:marked}). This is hardly surprising given that the marks are defined using the galaxy field itself, which has been fixed against observations for the different gravity models. \citet{Hernandez-Aguayo:2018yrp,Armijo:2018urs} proposed to include additional information in the marks, e.g., the masses or gravitational potentials of the host haloes of the galaxies, and found that this increases the differences in the marked correlation functions. The challenge then becomes how to find such additional information that can be reliably measured from observations (see also Sec.~\ref{sec:screening_maps}).


\section{Cosmological simulations}
\label{sec:simulations}


N-body simulation codes are currently the most reliable tool to predict the distribution and evolution of matter in the nonlinear regime of structure formation.  This is true for any theory of gravity, but for the case of theories with screening the importance of N-body simulations is even stronger.  This is because they are currently the only means to accurately investigate the types of signatures that the screening mechanisms -- which are themselves nonlinear phenomena -- leave on large scale structure. Being numerically expensive, these simulations cannot (at least yet) be used in thorough explorations of theory space, nor for running Monte Carlo constraints on the parameter space of specific theories. However, by focusing on a small number of representative models, N-body simulations of modified gravity have taught us a great deal about the types of signatures predicted for the nonlinear regime of structure formation and what observational tests can be designed to detect those signatures.

The first simulations of the Hu-Sawicki $f(R)$ model were performed in \citet{2008PhRvD..78l3523O} on a fixed-resolution (i.e., no mesh refinement) grid using the Newton Gauss-Seidel relaxation method. This code was subsequently adapted for simulations of the DGP model in \citet{2009PhRvD..80d3001S, 2009PhRvD..80l3003S}.  The latter model was also simulated using a fixed grid, but with a different (the so-called FFT-relaxation) algorithm \citep{2009PhRvD..80j4005C}. Shortly afterwards, adaptive mesh refinement (AMR) simulations for a range of modified gravity models \citep{Li:2009sy,Li:2010re3,Li:2010re,Li:2010re2,Zhao:2010qy,Brax:2011ja,Davis:2011pj} were developed by modifying the N-body code {\sc mlapm} \citep{Knebe:2001av}; these were however serial simulations with limited efficiency.

The field of modified gravity simulations took a significant step forward with the development of parallel AMR codes, which allow for computationally affordable investigations of matter clustering on scales that would be difficult to be resolved with fixed grid or serial AMR codes. These include the {\sc ecosmog} code \citep{2012JCAP...01..051L}, which is built on {\sc ramses} \citep{2002A&A...385..337T}; the {\sc mg-gadget} code \citep{2013MNRAS.436..348P}, which is a modified version of {\sc Gadget3} \citep{2005MNRAS.364.1105S};  the {\sc isis} code \citep{2014A&A...562A..78L}, also a modified version of {\sc ramses}; and a modified version \citep{Arnold:2019vpg} of the moving mesh N-body and hydro code {\sc arepo} \citep{arepo}. Some of these codes (together with the DGP code of \citet{2009PhRvD..80d3001S, 2009PhRvD..80l3003S}) were compared in a code comparison paper \citep{codecomp}. At the time of writing, these codes are not all able to simulate the same classes of models\footnote{Also, none of these codes is currently publicly available, though interested readers could write to their authors to request copies of some of them.}.

In this section, we briefly review the main features of modified gravity N-body algorithms and recent developments in the validation and optimization of these methods\footnote{In accordance with the scope of this review, in this section we only focus on cosmological simulations that probe phenomena on scales that are relevant to cosmological/astrophysical tests, and therefore do not review the numerical simulation methods for strong field applications such as stellar collapse, black-hole mergers, etc. (see, e.g., \citealt{2015arXiv150206853C} for a review). Furthermore, a remark should made that all the modified gravity simulation codes mentioned here work for the Newtonian limit; indeed, there has been recent progress towards simulating modified gravity in the general relativistic regime, e.g. with the advent of the {\tt fRevolution} code \citep{Reverberi:2019bov}.}. The code comparison paper \citep{codecomp} is also a useful first read about these simulations, and a more thorough discussion of simulation techniques in modified gravity models is given in \citet{modsimbook_Li}.

\subsection{The algorithm: relaxations with multigrid acceleration}

In gravity-only N-body simulations, the nonlinear evolution of the total matter density in the universe is followed by sampling it with N-body tracer particles, with positions and velocities determined by the total force they experience at a given time step. In modified gravity simulations, under the so-called weak-field (WFA) and quasi-static (QSA) approximations (see Sec.~\ref{sec:qsa} below), this force is given by $-\nabla\Phi$, with the gravitational potential $\Phi$ obeying a modified Poisson equation of the form
\bq\label{eq:poissongen}
\nabla^2\Phi = 4\pi G\delta \rho + f(\phi, \nabla\phi, \nabla^2\phi, \cdots),
\eq
where the density perturbation $\delta \rho$ is determined by the particle distribution in a given time step and $f$ is some model-specific function of a scalar field and/or its (usually spatial) derivatives. The first term in the RHS of Eq.~(\ref{eq:poissongen}) is given by the standard Poisson equation in Newtonian gravity. The scalar field $\phi$ obeys a nonlinear Klein-Gordon equation, which in the weak-field and quasi-static approximations can be cast in the following generic form:
\bq\label{eq:eomgen}
\mathcal{L}\left[\phi; \delta\rho\right] = \mathcal{S}\left(\delta\rho\right),
\eq
in which $\mathcal{L}$ is a model-specific nonlinear derivative operator acting on $\phi$ (which can depend on the density perturbation $\delta\rho$) and $\mathcal{S}$ is a function of $\delta\rho$ that sources the scalar field. The simulation particles are evolved according to
\bq\label{eq:geodesic}
\ddot{\bf x} + 2H\dot{\bf x} = -\vec{\nabla}\Phi,
\eq
which is as in standard Newtonian simulations\footnote{Throughout this section, we use the terminology `Newtonian simulations' rather than `GR simulations' for usual simulations of the $\Lambda$CDM model, to avoid potential confusion with `general relativistic simulations', which aim to solve Einstein equations beyond the Newtonian-type Poisson equation.}, just with a modified dynamical potential, $\Phi$. The main objective of the modified gravity algorithms is to solve Eq.~(\ref{eq:eomgen}) to then be able to construct the extra term $f(\cdots)$ in the modified Poisson equation Eq.~(\ref{eq:poissongen}). Once this correction is found, the N-body calculation proceeds as in standard Newtonian simulation codes.

Since the scalar field $\phi$ obeys a nonlinear field equation, it is in general not possible to solve for the modified forces using pairwise force summation, as done in the tree algorithm for Newtonian N-body simulations. Instead, Eq.~(\ref{eq:eomgen}) is solved using the finite-difference method on a grid. The current state-of-the-art modified gravity N-body codes solve Eq.~(\ref{eq:eomgen}) with AMR, which makes use of a suite of grids that refine in high matter density. This ensures sufficiently high force resolution where it is needed, while saving computational time in regions of fewer particles, where the force resolution can be lower. The algorithm consists of taking a discretized version of Eq.~(\ref{eq:eomgen}) defined on the AMR grid and updating the values of $\phi$ in grid cells using some relaxation arrangement until convergence is reached. Effectively, the algorithm solves the equation 
\bq\label{eq:eomgen-disc}
\mathcal{T}_{ijk}^\ell \equiv \mathcal{L}\left[\phi_{ijk}^{\ell}\right] - S_{ijk}^{\ell} = 0,
\eq
where $\left\{ijk\right\}$ is the cell index and $\ell$ is the refinement level. The iterations can be carried out with the {Newton-Gauss-Seidel} (NGS) method, in which the value of the scalar field is updated as
\bq\label{eq:newton}
\phi_{ijk}^{{\rm new},\ell} = \phi_{ijk}^{{\rm old},\ell} - \frac{\mathcal{T}_{ijk}^\ell}{\partial \mathcal{T}_{ijk}^\ell/\partial\phi_{ijk}^{{\rm old},\ell}},
\eq
in which the evaluation of $\mathcal{T}^\ell_{ijk}$ often involves the values of $\phi$ in neighbouring cells. The iterations proceed until a sufficiently good solution to Eq.~(\ref{eq:eomgen-disc}) is obtained which, in practice, is achieved by checking if certain statistics of $\mathcal{T}_{ijk}^\ell$ over the whole AMR structure, $s(\mathcal{T})$ (e.g., the root-mean-squared, the mean of the modules or the maximum of the modules of $\mathcal{T}_{ijk}$ in all cells) drops below some pre-specified threshold, or if the estimated error in the solution is already much smaller than the truncation error from discretizating the continuous equations on a grid.

During the first few Gauss-Seidel iterations, the value of $s(\mathcal{T})$ usually decays quickly, showing that the numerical solution is converging towards the true solution. But the convergence becomes slower afterwards, because the Fourier modes of the error with wavelengths larger than the grid size are slow to reduce. To circumvent this problem, and hence improve the performance of the algorithm, most codes make use of {\it multigrid acceleration} (see, e.g., \citealt{Briggs} for an introduction), in which a hierarchy of coarser grids is used to help reduce the long-wavelength Fourier models of the error and speed up the convergence. In practice, the operation goes as follows. Once the convergence becomes too slow on a level $\ell$, the equation is interpolated to a coarser grid labelled as $\ell-1$, where the larger grid cell size helps to reduce longer-wavelength modes of the error. This process can continue for several coarser levels $\ell-2$, $\ell-3$, $\cdots$.
The coarser-level solutions are subsequently interpolated back to level $\ell$ where the solutions to $\phi^\ell_{ijk}$ are corrected. If convergence has still not been reached (i.e., $s(\mathcal{T})$ is still not small enough), then the process -- called a multigrid cycle -- is repeated. There are different ways to arrange the multigrid cycle, such as V-cycles and W-cycles. 

Eq.~(\ref{eq:newton}) is analogous to using the Newton Raphson method to solving a nonlinear algebraic equation. This `Newton' part is an approximation to the nonlinear algebraic equation which causes additional error in the solution $\phi^\ell_{ijk}$; this additional error actually accounts for a substantial fraction of the time spent on NGS relaxations, because the highly nonlinear nature of the equation makes it hard to reduce. However, in certain cases it is possible to make a field redefinition, $\phi_{ijk}\rightarrow u_{ijk}$, so that $u_{ijk}$ satisfies a different nonlinear algebraic equation, usually up to cubic or quartic order, that can be solved analytically. This enables one to solve Eq.~(\ref{eq:newton}) directly without having to resort to the Newton approximation, therefore greatly improving the performance of the algorithm. An application of this trick to the Hu-Sawicki $f(R)$ model is given in \citet{2017JCAP...02..050B,Arnold:2019vpg}, but it should be noted that although this method can be applied to a wide variety of modified gravity models of interest, it does not work in general cases. Still, a lesson from this is that, if a field redefinition can make the equation less nonlinear, then the performance can be greatly improved.

Another way to improve the performance of these simulations is to first manipulate the target equations -- in particular those involving nonlinear terms of the partial derivatives of $\phi$ --  analytically and try to recast them into a more numerically stable form. For instance, the first simulations of DGP gravity \citep{2009PhRvD..80d3001S, 2009PhRvD..80l3003S} iterate an equation which, due to  $\phi_{ijk}$ appearing in not only the operator $\mathcal{L}$ but also the source term $\mathcal{S}$ in Eq.~(\ref{eq:eomgen}), was numerically unstable and suffered from slow convergence. \citet{2013JCAP...05..023L} subsequently demonstrated that, after using an {operator-splitting trick} \citep{2009PhRvD..80j4005C} to rewrite Eq.~(\ref{eq:eomgen}) and get rid of $\phi_{ijk}$ from $\mathcal{S}$, it was possible to make the relaxation iterations stable and fast to converge. The operator splitting method was generalized and implemented in simulations of more complicated Vainshtein screening models, like Quartic Galileon, in \citet{2013JCAP...11..012L}. 

The general algorithm outlined above applies to virtually all simulations of modified gravity models with screening. The details of the discretisation and implementation of the field equations, however, differ in models with different screening mechanisms.

\subsection{The validity of the quasi-static approximation}\label{sec:qsa}

To date, the majority of simulations of modified gravity models have been performed in the weak field and quasi-static approximations. The WFA essentially states that the amplitudes of the scalar field perturbations ($\delta\phi/M_{\rm Pl}$) and gravitational potentials are much smaller than the speed of light squared, so that we are far away from relativistic strong-field regimes, which is true on cosmological scales. On the other hand, the QSA amounts to a special treatment of terms involving time derivatives of the scalar field perturbation ($\dot{\delta\phi} = \dot{\phi}-\dot{\bar{\phi}}$), by assuming that it can be neglected {\it when compared} with terms which involve spatial gradients of $\phi$, i.e.,
\be
\ddot{\delta\phi}\sim H(\dot{\delta\phi}) \sim H^2 \phi \ll\nabla^2\phi = \nabla^2(\delta\phi)\,.
\ee

A number of studies exist which addressed the validity of the QSA. For example, focusing on $f(R)$ models and working in linear perturbation theory, \citet{2014PhRvD..89b3521N} confirmed that the use of the quasi-static approximation on sub-horizon scales is not a worry if certain observational viability conditions are met (which are the case for the models that would be simulated anyway). \citet{2015JCAP...02..034B} included time derivatives in $N$-body simulations of the Hu-Sawicki $f(R)$ using an implicit method (i.e., the time derivative of the scalar is evaluated with a backward finite difference using the value at the previous time step), and found virtually the same probability distribution function of the density field and power spectrum of density and velocity fields, compared to simulations run under the quasi-static approximation. For Vainshtein screening models, the linear small-scale limit was found to be independent of the time derivatives in the Galileon models \citep{2012PhRvD..86l4016B} and for Horndeski theories in general \citep{Lombriser:2015cla}. \citet{2009PhRvD..80d3001S} demonstrated the validity of the quasi-static approximation in simulations of DGP gravity with self-consistency tests, and \citet{2015PhRvD..92f4005W} demonstrated explicitly that the quasi-static approximation works well by studying the evolution of spherically symmetric structures in the DGP and Cubic Galileon models (see also \citealt{2014PhRvD..90l4035B}). To our knowledge, there has been no explicit check of the validity of the QSA in simulations of more complicated theories such as Quartic \& Quintic Galileon and beyond Horndeski models to date. These can be particularly interesting since their field equations can contain terms such as $\nabla^2\dot{\phi}=\nabla^2\dot{\delta\phi}\sim H\nabla^2\delta\phi$, that are not necessarily negligible compared with other terms, and must therefore be treated carefully. 

There are, on the other hand, models for which we do know the QSA is not a good approximation. One example is the symmetron model \citep{Hinterbichler:2010es}, in which the scalar field $\phi$ stays at its true vacuum with $\phi=0$ in high-density regions and early times, but develops two possible vacuua with $\pm|\phi_{\ast}|$ (where $\phi_\ast$ depends on the model parameters) in low-density regions and late times. Disconnected spatial regions can relax to either of the two vacuua, and a domain wall forms at their boundaries. Consider, for example, a bubble where $\phi=-|\phi_\ast|$ that is surrounded by much larger regions where $\phi=|\phi_\ast|$: as structure formation proceeds, it may be energetically more favourable for the bubble to take $\phi=|\phi_\ast|$ as well, and there can be a fast transition of $\phi=-|\phi_\ast|\rightarrow|\phi_\ast|$ which means that the time derivative of $\delta\phi$ can be non-negligible. A {\it non-quasi-static} code that explicitly evolves the scalar field in time with a leap-frog method was developed in \citet{2014A&A...562A..78L, 2013PhRvL.110p1101L}. Using this code, in \citet{2014PhRvD..89h4023L}, the authors ran N-body simulations of the symmetron model in and without the QSA to conclude that the impact on standard matter statistics is negligible, despite the formation of interesting domain wall effects in the distribution of the scalar field in the non-quasi-static cases. Furthermore, \citet{2017PhRvL.118j1301H, 2018JCAP...06..035I} studied the impact that propagating scalar waves in the symmetron model (which arise from the full non-QSA equations) have on the screening efficiency in the Solar System.  The conclusion was that for realistic directions of propagation of the incident scalar waves, the effects of scalar waves are negligible, i.e., the QSA remains a valid approximation. 

In general, the QSA approximation is expected to fail on scales comparable or larger than the scalar field horizon, which is set by the propagation speed of its fluctuations \citep{2015PhRvD..92h4061S}. However, most N-body simulations of modified gravity to date probe scales that are well within the corresponding scalar field horizons, which is the justification behind the adoption of the QSA. Various recent works have thus concluded that the QSA for the evolution of $\phi$ is not a source of concern in N-body simulations of modified gravity. We note however that the tests have so far been focused on existing classes of screened modified-gravity theories (such as $f(R)$, Vainshtein and symmetron models), and that similar tests will need to be performed as new theories are developed.

\subsection{Approximate speed-up methods}

The simulations of modified gravity are notoriously slow compared to their standard gravity counterparts even with parallelization. This is because of the numerically demanding relaxation iterations that take place when solving Eq.~(\ref{eq:eomgen}). On the other hand, the exploitation of the data from future surveys would benefit greatly if analysis pipelines can be validated and calibrated using mock catalogues constructed from simulations of different theories of gravity. The construction of these mocks with the resolution and volumes required for these surveys therefore motivated efforts to improve the performance of modified gravity simulation algorithms. We have mentioned above that field redefinition to make the equations less nonlinear is one way to achieve this, but there are various other possibilities, mostly involving some approximations (see \citealt{Li:2019re} for a recent review). Being approximate methods, these are less accurate than full simulations; nevertheless, the idea is that they can still be tremendously helpful in certain applications and/or regimes, provided that we understand their limitations and possible implications.

An example of such efforts is that undertaken by \citet{2015PhRvD..91l3507W} (inspired by \citealt{2009PhRvD..80f4023K}), in which instead of solving Eq.~(\ref{eq:eomgen}) numerically, one solves a linearised version of it with an analytic {\it screening factor} derived by assuming spherical symmetry. Concretely, Eq.~(\ref{eq:eomgen}) becomes $\nabla^2\phi = \mathcal{S}_{screen}\left(\delta\rho\right)$ with $\mathcal{S}_{screen}$ being  some nonlinear function of the density that takes screening into account and that recovers the linear theory result if $\delta\rho \ll \bar{\rho}$. This way, the scalar field equation can be solved with the same fast methods used for the standard Poisson equation, effectively only doubling the computational cost of the gravity calculation in the N-body code. In their simulations, the authors demonstrate that they can recover the power spectrum of the full simulations to better than $3\%$ accuracy up to $k \lesssim 1\ h/{\rm Mpc}$ (the agreement improves with decreasing $k$). The predicted mass functions also agree relatively well with the results from full simulations, but it would be interesting to extend the comparison to other statistics, namely those associated with the lowest density regions like void counts and void profiles. 

In the standard algorithm, the iterations of Eq.~(\ref{eq:eomgen}) need to be performed for (i) every simulation time step; (ii) all cells on a given AMR level; and (iii) for every AMR level. Every time the code enters a new refinement level, the time step is halved and the number of time steps doubled to ensure numerical accuracy. Therefore, while a simulation generally consists of a few hundred coarse time steps on the non-AMR level, there can be tens of thousands of fine time steps and the scalar field needs to be solved at each of them.  For the case of Vainshtein screening models, there is a way to speed up the algorithm significantly by relaxing condition (iii), for a negligible sacrifice in accuracy. As explained in \citet{2015JCAP...12..059B}, the speed-up trick is implemented by not iterating the scalar field explicitly above a given refinement level $l_{\rm iter.}$, and instead just taking its value by interpolation from some lower refinement level where the iterations took place. Naturally, skipping the iterations of Eq.~(\ref{eq:eomgen}) on levels $l > l_{\rm iter.}$ speeds up the code, but at the cost of a large error on the calculation of the fifth force. The key point to note here, however, is that in highly refined regions, the fifth force is a small fraction of the total force, and hence, an error on its evaluation constitutes only a small and affordable error on the much larger (GR dominated) total force. In \citet{2015JCAP...12..059B}, with simulations of the nDGP model, the authors have demonstrated that it is possible to speed up the performance of the algorithm by factors of $10$, with very little loss in accuracy in the matter power spectrum for $k \lesssim 5\ h/{\rm Mpc}$; halo properties such as their abundance, mass, density profiles and peculiar velocity were also virtually unaffected by the speed-up trick.  Note that this method relies on the correlation between highly-refined regions and screening efficiency, which exists in Vainshtein screening models but not in chameleon models. This hinders the applicability of the same idea to simulations of chameleon models, as demonstrated in the appendix of \citet{2017JCAP...02..050B}. One can also relax condition (i), as done in \citet{Arnold:2016arp}, by choosing not to solve the scalar field on every fine time step, in the expectation that the field values do not change substantially between neighbouring fine time steps; a more detailed test of how this affects the accuracy is yet to be carried out for general setups.


It is also worth noting the recent generalisations of the COLA (COmoving Lagrangian Acceleration) method \citep{2013JCAP...06..036T} to theories of modified gravity with screening and scale-dependent growth \citep{2017PhRvD..95j3515V, 2017JCAP...08..006W}. This approximate method can prove very useful in the fast generation of a large number of halo/galaxy catalogues for various theories of modified gravity, even if the detailed mass distribution on very small scales is less accurate due to the approximations made in the method.

To summarize, over the last couple of years, a variety of developments took place that brought the performance of N-body simulations of modified gravity to a level nearly comparable to standard simulations of $\Lambda$CDM. The simulation codes developed are finally able to reach the resolution and volume specifications required for the planning of future large scale structure surveys.

\section{Cosmological tests}
\label{sec:cosmo_constraints}

This section provides an overview on modified gravity constraints which come from probes of the expansion history of the universe and observations of the structure on very large scales. Naturally, there is a smooth transition of these latter probes to the astrophysical tests described in the next section. Our goal is not to provide a precise division here, but to work from the largest scales towards successively smaller scales. 


\subsection{Parametrized vs. model-by-model approaches}
\label{subsec:paramvsmodel}
\begin{figure*}
  \subfigure[]
  {
    \includegraphics[width=0.5\textwidth]{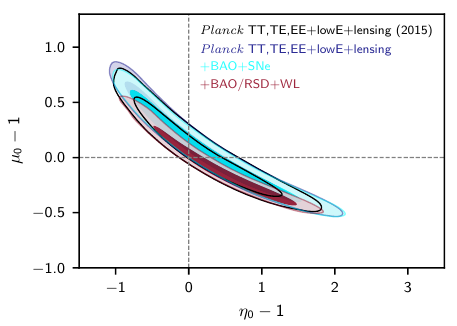}
  }
  \subfigure[]
  {
    \includegraphics[width=0.5\textwidth]{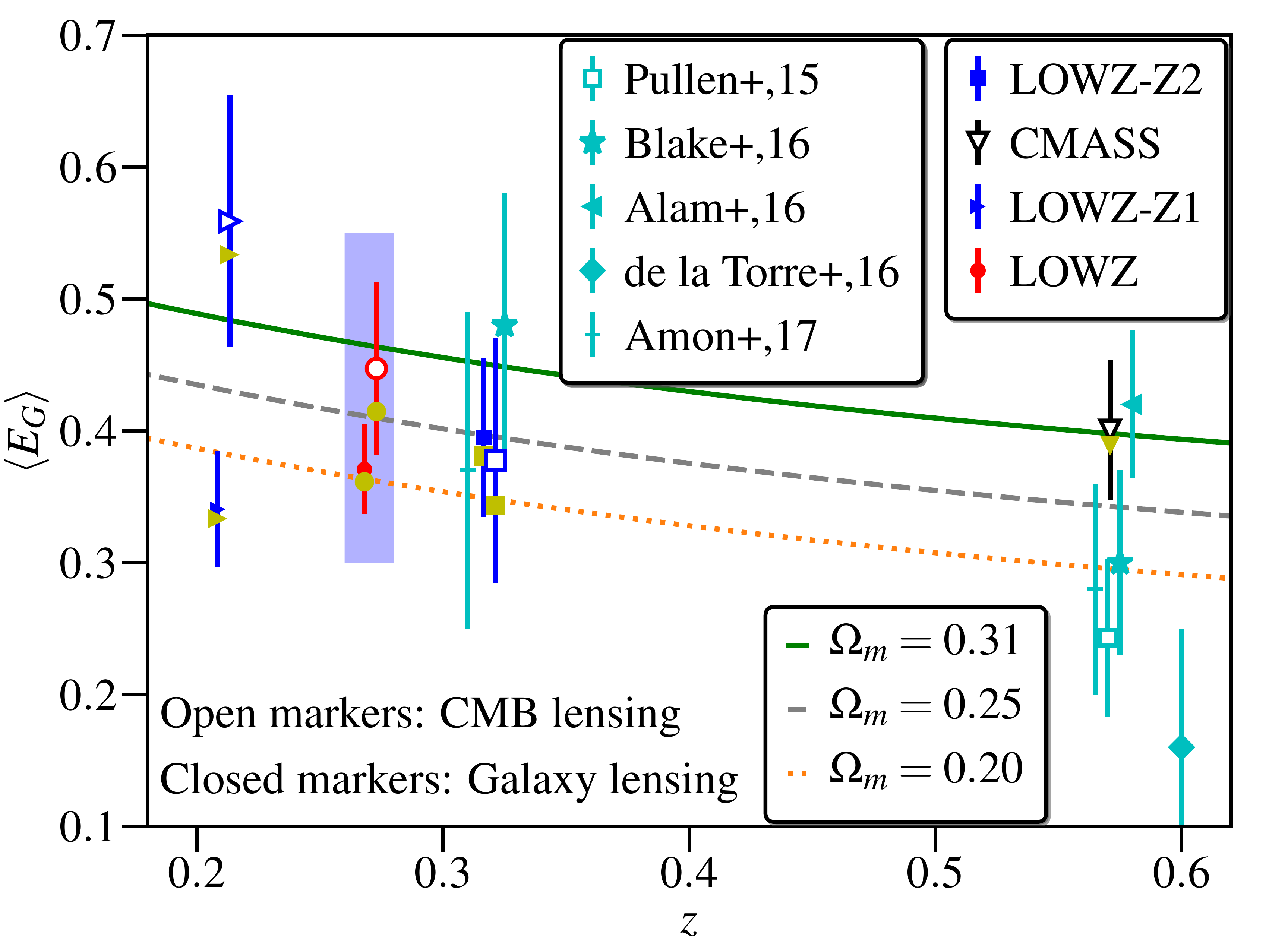}
  }
  \caption{(a) Constraints on parameters $\mu$ and $\eta$ (Eq.~\ref{eq:mu}) describing deviations of the gravitational potentials from their expected values in GR. The $\Lambda$CDM prediction is at the intersection of the dashed lines. Reproduced from \citet{Planck_cosmological_parameters}. (b) Constraints on $E_G$ (Eq.~\ref{eq:EG}) from CMB and galaxy-galaxy lensing compared to $\Lambda$CDM expectations of varying $\Omega_m$. Reproduced from \citet{Singh}.}
  \label{fig:mu_eta_EG}
\end{figure*}

Cosmological constraints studies on modified gravity can be broadly divided into constraints on parametrized frameworks and constraints on concrete theories. The parametrized approach \citep{2007PhRvD..76j4043H,2011PhRvD..84l4018B, 2013PhRvD..87b4015B} introduces a few parameters or functions that describe departures from the fiducial GR case, and which can be constrained with large-scale structure observations. For example, on sub-horizon scales, the potentials $\Phi$ and $\Psi$ appearing in the line element (where $\tau$ is conformal time)
\begin{align}
ds^2&=a^2\left[-\left(1+2\Psi\right)d\tau^2+\left(1-2\Phi\right)dx^2\right]  
\label{lineel}
\end{align}
can be parametrized by introducing two independent free functions of time and wavenumber:
\begin{align}\label{eq:mu}
-k^2 \Phi &\equiv 4 \pi G a^2 Q(a,k) \rho,  \\ \nonumber 
-k^2 \Psi &\equiv 4 \pi G a^2 \mu(a,k) \rho,  \\ \nonumber
-k^2 (\Phi +\Psi) &\equiv 8 \pi G a^2 \Sigma(a,k) \rho, \\ \nonumber
\eta(a,k) &\equiv \Phi/\Psi.
\end{align}
For example, the function $Q$ describes an effective gravitational constant that non-relativistic particles are sensitive to. Relativistic particles such as photons in lensing observations would instead be sensitive to the function $\Sigma = \left(Q + \mu\right)/2$; note that only two of the four functions $Q, \mu, \Sigma, \eta$ are strictly needed/independent. In GR, they are all equal to unity, and hence, if the data prefers a departure from this result (either in time or in space), then this would signal a need to go beyond GR. The current constraints from \textit{Planck}+BAO+RSD+SNe+WL on $\mu$ and $\eta$ (at $z=0$ and assumed scale-independent) are shown in the left-hand panel of Fig.~\ref{fig:mu_eta_EG}.

Another popular way to parametrize modified gravity effects is via the so-called Effective Field Theory of Dark Energy (EFTofDE) \citep{2013JCAP...02..032G, 2013JCAP...08..010B, 2013JCAP...08..025G, 2017PhRvD..95b3518L, 2017PhRvD..96f3516G}, in which the parametrization is done at the level of the linearized action of Horndeski (and beyond) models. In the EFTofDE approach, the evolution of linear perturbations is completely encoded by five functions of time, plus a specification of the Hubble rate $H(a)$.

An appealing aspect of parametrized frameworks is that they allow for systematic and fairly model-independent constraints on modified gravity (e.g.~\citealt{2016JCAP...02..053B, 2017JCAP...02..043B}). A main disadvantage lies in its limited regime of applicability, which encompasses only linear scales. The EFTofDE cannot be used to study nonlinear structure formation because the formalism builds on top of a linearized Lagrangian (see however \citealt{Cusin:2017wjg, 2018JCAP...04..061C} for recent work on how to go beyond this limitation). In other parametrized frameworks, the corresponding free functions would become too general to be satisfactorily constrained by the data (see, however, \citealt{Brax:2011aw,Brax:2012gr, 2012JCAP...10..002B,Brax:2013mua} for a few attempts to parameterize chameleon/symmetron/dilaton theories and screening mechanisms more generally~\citep{2016JCAP...11..039L} in the nonlinear regime). One can also translate model-indepdendent parameter constraints onto any theory that falls within the parameterized framework, though the results are generally weaker than if one had directly analyzed the model of interest.

In the model-by-model approach, by focusing only on one model at a time, one can afford detailed investigations in the nonlinear regime of structure formation, typically based on N-body simulations. A disadvantage of this approach is that it is harder to generalize the conclusions obtained from one model onto the rest of the theory space. To make progress nonetheless, the philosophy adopted by the community has been that of ``electing'' a model that is representative of some type of phenomenology (e.g. $f(R)$ for chameleon or DGP/Galileons for Vainshtein), which is then used to place benchmark constraints on the size of the deviations from GR that are supported by a given dataset. The body of work developed for models such as $f(R)$, DGP and Galileon gravity has already taught the community a great deal about the most promising ways to test gravity in cosmology. The analysis pipelines that have been developed for such models should in principle  be  adaptable to other theories such that, as new models are developed, work on constraining them can take place straightaway.

\subsection{Main cosmological datasets and observational signatures}\label{sec:cosmo2}

We now summarize the main cosmological observables that can be used to test modified gravity. The line that separates astrophysical from cosmological datasets is not always clear;  in what follows, we limit ourselves to  probes that are sensitive to the expansion rate of the universe or to structure formation on linear to quasi-linear scales (with the exception perhaps of weak-lensing cosmological analyses that  extend into the nonlinear regime).  It is also worth noting that the datasets outlined below are most powerful when used in combination, rather than on their own.

\begin{itemize}

\item \underline{Expansion history from CMB, BAO and SNIa}

Data from CMB, BAO and SNIa are amongst the most robust datasets in cosmology to probe the expansion rate $H(z)$ of the universe. The agreement between GR-based $\Lambda{\rm CDM}$ and these data is sometimes misinterpreted as a requirement on modified gravity models to possess a $\Lambda{\rm CDM}$ background limit, i.e., some choice of parameters that very closely reproduces a cosmological constant. This is a confusion that is worth clarifying: a model in which $H(z)$ differs from $\Lambda{\rm CDM}$ can still stand a chance at being compatible with CMB, BAO and SNIa data. Such compatibility, if it exists, would occur for cosmological parameter values (e.g.~$H_0$, $\Omega_m$, $\Omega_K$, neutrino masses, etc.) that are different from those obtained in constraint analyses that assume $\Lambda{\rm CDM}$. The assessment of observational viability rests however on whether or not the full model yields acceptable fits. The adoption of $\Lambda{\rm CDM}$-inspired cosmological parameters in analyses of self-accelerating models without a $\Lambda{\rm CDM}$ limit may thus result in biased conclusions. An illustrative case is that of the Galileon model which displays nearly the same goodness-of-fit to CMB data as $\Lambda{\rm CDM}$, albeit with very different $H_0$ and $\Sigma m_\nu$ values \citep{2017A&A...600A..40N, Barreira:2014jha} (note, however, that the Galileon is ruled out when ISW data (see ISW discussion below) and constraints on the propagation speed of gravitational waves (cf.~Sec.~\ref{sec:GWs}) are considered).

In addition to the sensitivity to the expansion history, the CMB temperature power spectrum is also sensitive to late-time  structure formation via the ISW effect, which affects the low-$\ell$ part of the spectrum. Despite being limited by cosmic variance, this probe is nonetheless stringent enough to rule out drastic scenarios such as the self-accelerating branch of the DGP model \citep{Fang:2008kc} and a large portion of the Galileon model parameter space \citep{2012PhRvD..86l4016B, 2013PhRvD..87j3511B, Barreira:2014jha}. The power spectrum of the reconstructed CMB lensing potential is sensitive to late-time structure formation over a wider range of scales. CMB lensing measurements from Planck and other surveys have become a useful dataset for tests of modified gravity as well \citep{2016A&A...594A..14P}. We refer the interested reader to \citet{2018PhRvD..97b3520B} for a Boltzmann-Einstein code comparison project in various modified gravity cosmologies, with comparisons made at the level of the CMB temperature, polarization and lensing power spectra, as well as the linear matter power spectrum.

\item \underline{ISW-galaxy cross-correlation}

The cross-spectrum of CMB temperature maps and foreground galaxy distributions can be written as
\bq\label{eq:Ctg}
C^{Tg}(\ell) = 4\pi \int \frac{{\rm d}k}{k} \Delta^{\rm ISW}(\ell, k) \Delta^{\rm g}(\ell, k) P_\mathcal{R}(k)
\eq
where $P_\mathcal{R}(k)$ is the power spectrum of curvature fluctuations and $\Delta^{\rm g}(\ell, k)$ is a galaxy distribution kernel that depends on the bias and redshift distribution of the galaxy sample at hand. The ISW kernel is given by
\bq\label{eq:ISWkernel}
\Delta^{\rm ISW}(\ell, k) = \int_{\tau_{\rm rec}}^{\tau_0} {\rm d}\tau \frac{{\rm d}\left(\Phi(k, \tau)+\Psi(k, \tau)\right)}{{\rm d}\tau} j_\ell(k(\tau-\tau_0)),
\eq
which shows that these data \citep{2008PhRvD..77l3520G, 2016A&A...594A..21P, 2015PhRvD..91h3533F} can be used to constraint $\Sigma$ in Eqs.~(\ref{eq:mu}). The fact that $C^{Tg}(\ell)$ is sensitive to whether the potentials become deeper or shallower with time (i.e., the sign of ${\rm d}\left(\Phi + \Psi\right)/{\rm d\tau}$, where $\tau$ is the conformal time) plays a crucial role in cosmological constraints of Galileon gravity. For instance, the best-fitting Cubic Galileon model to the CMB data displays only a modest increase in the amplitude of the low-$\ell$ part of the CMB temperature power spectrum, but its predicted cross-correlation with galaxies from the WISE survey ($z \approx 0.3$) is negative (dashed red line in the left panel of Fig.~\ref{fig:cosmo}), indicative that the potentials are getting deeper (more negative) with time. Data from the ISW effect (both from CMB temperature \citep{2012PhRvD..86l4016B, 2013PhRvD..87j3511B, Barreira:2014jha} and its cross-correlation with galaxies \citep{Renk:2017rzu}) is in fact what sets some of the tightest constraints on the Covariant Galileon model because of its modifications to $\Sigma(a,k)$; the few corners of the parameter space that survive the ISW tests end up being ruled out by their anomalous speed of gravitational waves.

Examples of other models that have been constrained with ISW-galaxy cross-correlations include the DGP model \citep{Lombriser:2009xg}, $f(R)$ \citep{2007PhRvD..76f3517S,Lombriser:2010mp}, massive gravity \citep{2015PhRvD..91h4046E} and a kinetic gravity braiding toy-model that interpolates between $\Lambda{\rm CDM}$ and Cubic Galileon limits \citep{Kimura:2011td}. Analyses of the ISW effect are typically restricted to the largest observable scales where linear theory is valid, and hence screening mechanisms are not a source of complication. A complication that arises is that the bias of the foreground galaxies is degenerate with the amplitude of the signal. The bias can be estimated by cross-correlating CMB lensing maps with the galaxy sample (e.g. \citealt{2015PhRvD..91h3533F}); the resulting values are in general different in different models by virtue of different dark matter clustering \citep{Renk:2017rzu}.


\begin{figure*}
  \subfigure[]
  {
    \includegraphics[width=0.5\textwidth]{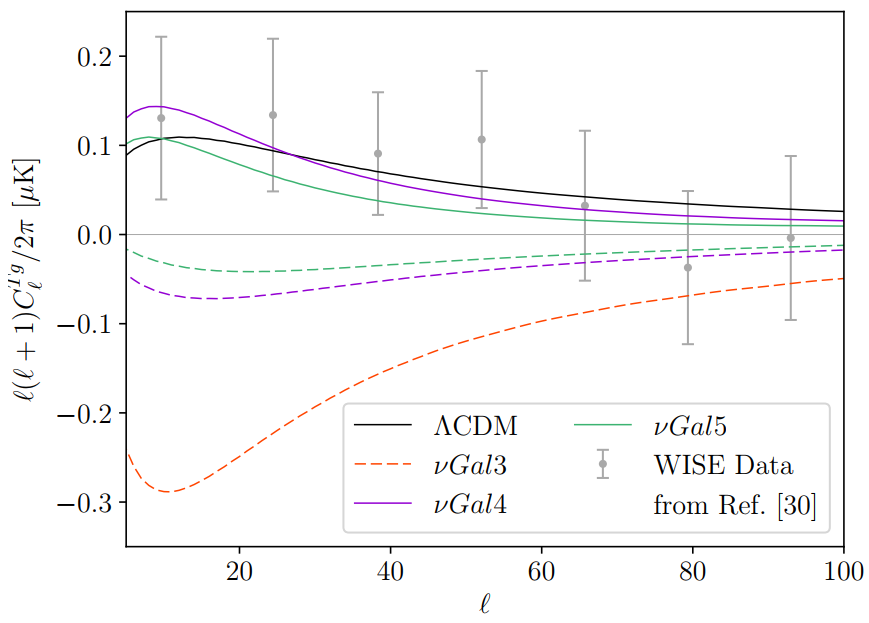}
  }
  \subfigure[]
  {
    \includegraphics[width=0.5\textwidth]{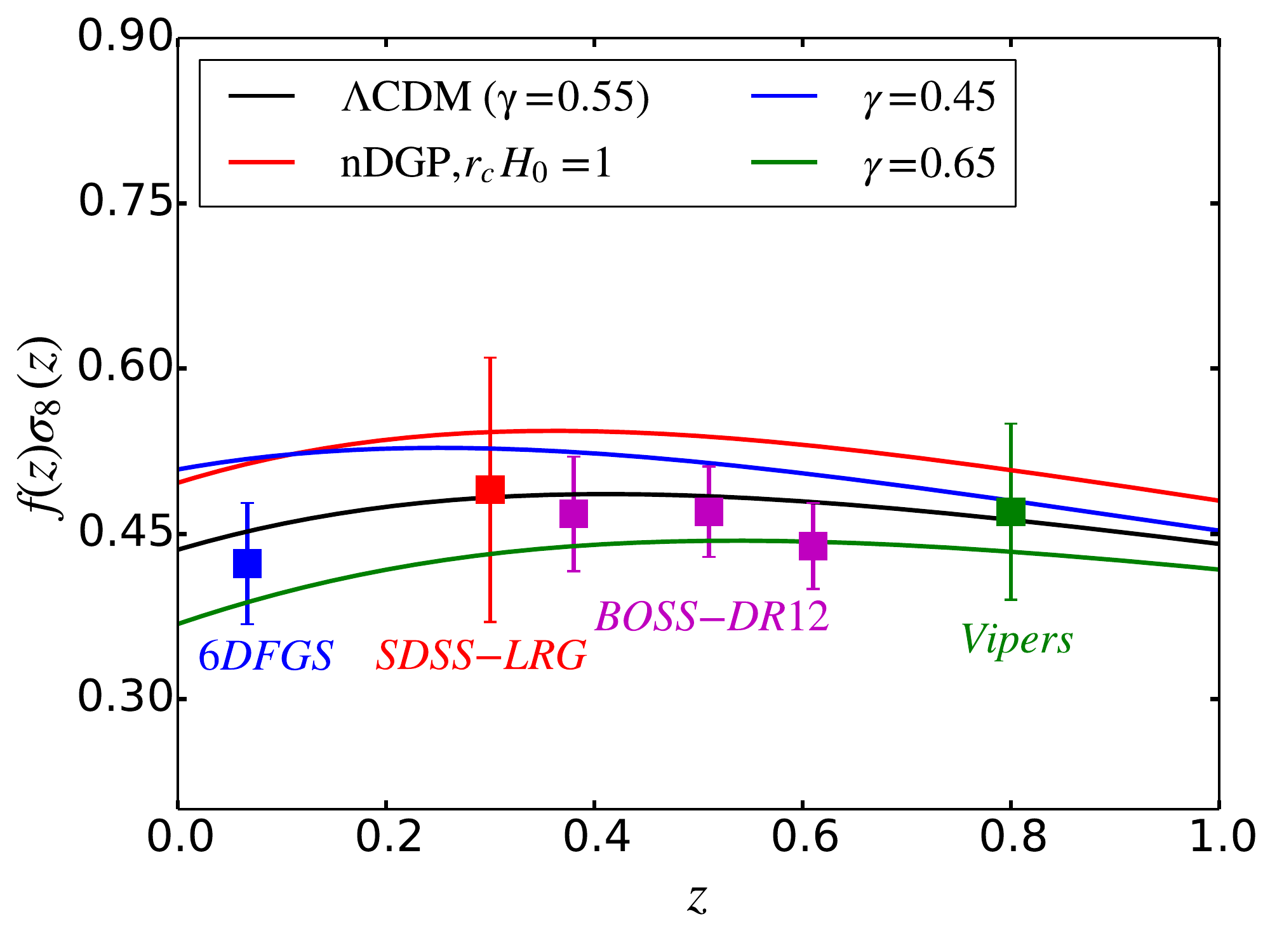}
  }
  \caption{(a) ISW-galaxy cross-correlation from the WISE survey \citep{2015PhRvD..91h3533F} together with predictions from a number of parameter choices in the Covariant Galileon model. The dashed-red curve is the best-fitting Cubic Galileon model to the data, which illustrates the high degree of tension plaguing this model. The figure in panel (a) is reproduced from \citet{Renk:2017rzu}. (b) Time-evolution of $f\sigma_8$ for $\Lambda{\rm CDM}$, nDGP and two values of growth-index $\gamma$. From left to right, the data points shown correspond, respectively, to the analyses of the 6DFGS survey \citep{2012MNRAS.423.3430B}, Luminous Red Galaxy sample from SDSS-DR7 \citep{2014MNRAS.439.2515O}, BOSS DR12 \citep{2017MNRAS.470.2617A} and Vipers surveys \citep{2013A&A...557A..54D}, as labeled.}
  \label{fig:cosmo}
\end{figure*}

\item \underline{\it Redshift Space Distortions and Growth rate}

The rate at which structure grows in the universe offers a natural and powerful way to test gravity in cosmology (see also Sec.~\ref{sec:RSD} on RSD below). This growth rate is usually quoted in terms of the parameter $f\sigma_8$, where $f$ is the logarithmic derivative of the growth factor w.r.t.~the scale factor: $f = {\rm dln}D/{\rm dln}a$ and $\sigma_8$ the root-mean-squared of the linear density field at redshift $z=0$ on scales of $8\ {\rm Mpc}/h$. The parameter $\sigma_8$ is a measure of the amplitude of the linear matter power spectrum, and the combination $f\sigma_8$ it what can be constrained with RSD data \citep{2009JCAP...10..004S}. The growth factor is governed by the equation (dots are physical time derivatives)
\bq\label{eq:growthfactor}
\ddot{D} + 2H\dot{D} - \frac{3}{2} Q(a, k) \Omega_m H_0^2 a^{-3} D = 0,
\eq
where $Q(a,k)$ is defined in Eqs.~(\ref{eq:mu}).  In GR, the growth rate is to a good approximation given by $f(a) = \Omega_m(a)^\gamma$, with $\gamma \approx 0.55$ ($\gamma$ is called the growth-index \citep{2005PhRvD..72d3529L, 2007APh....28..481L}). This has made it popular to use $f\sigma_8$ data to place constraints on $\gamma$ to look for eventual deviations from its GR value.

Observationally, $f\sigma_8$ can be inferred from the anisotropic clustering of galaxies on large scales in redshift space (cf.~Sec.~\ref{sec:RSD}; see also \citealt{2017MNRAS.470.2617A} for the latest analysis of the BOSS survey). Specifically, $f\sigma_8$ enters as one of the parameters of theoretical models of the galaxy power spectrum in redshift space that are fitted to the data. The ingredients of these models include modeling of RSD, galaxy bias \citep{2016arXiv161109787D}, as well as a prescription for the nonlinear clustering of matter (galaxy clustering studies are restricted to large enough scales on which baryonic feedback effects on the power spectrum are not a concern).  Carrying out such analyses for galaxy samples at different redshifts leads to constraints on  $f\sigma_8$ as a function of time. Different theories of gravity in general make different predictions for the time-dependence of $f\sigma_8$, thereby allowing to distinguish between competing models (cf.~right panel of Fig.~\ref{fig:cosmo}).

The validity of the galaxy clustering models must be checked first using mock galaxy samples constructed using N-body simulations. These validation steps have been carried out almost entirely for mocks in $\Lambda{\rm CDM}$ cosmologies \citep{2013MNRAS.433.1202S, 2017MNRAS.470.2617A}; only a few such validations analyses exist in modified gravity. For example, \citet{2016PhRvD..94h4022B} validated the use of the BOSS clustering wedges analyses pipelines \citep{2013MNRAS.433.1202S} using nDGP mocks (see also~\citealt{2017PhRvD..96b3519B} for a similar study), and~\citet{Hernandez-Aguayo:2018oxg} conducted a comparative study for both nDGP and $f(R)$ models. For chameleon screening models with scale-dependent linear growth \citep{2012MNRAS.425.2128J}, \citet{2014PhRvD..89d3509T} found that GR-based models of galaxy clustering can fail to recover the correct value of $f\sigma_8$ when applied to Hu-Sawicki $f(R)$ simulations. For future data with higher statistical precision, this modeling systematic must be addressed with detailed simulations.

The tightest cosmological constraints on the nDGP model (with a $\Lambda{\rm CDM}$ background) to date come  from the $f\sigma_8$ determinations of the BOSS DR12 release (performed with appropriate validation steps \citep{2016PhRvD..94h4022B}), which demonstrates the central role these data can play in tests of gravity. \citet{2016arXiv161200812M} uses a compilation of $f\sigma_8$ data to constrain the growth index $\gamma$, as well as the free functions in Eqs.~(\ref{eq:mu}). See also \citet{2016JCAP...12..032T} for a forecast study of the constraining power of $f\sigma_8$ data from SKA and Euclid.

The $E_G$ statistic described below in Sec.~\ref{sec:RSD} has also been estimated using RSD and lensing data \citep{zhang/etal:2007, pullen/etal:2016, 2017arXiv171110999A, Singh}; the right panel of Fig.~\ref{fig:mu_eta_EG} displays a recent compilation. None of the data published to date revealed any need for modifications of GR, but the current level of precision of this statistic is relatively poor compared to other probes of gravity.


\item \underline{\it Weak-lensing}

Weak gravitational lensing (see e.g. \citealt{2001PhR...340..291B, 2008ARNPS..58...99H} for reviews) provides a direct probe of the distribution of matter in the late-time universe. That is, it is not subject to galaxy bias or RSD, although intrinsic alignments are a systematic contaminant that needs to be carefully controlled for. Lensing thus offers complementary information to test gravity in cosmology \citep{Schmidt:2008hc, 2013MNRAS.429.2249S, 2015PhRvD..91h3504L, 2017arXiv171201846F}. The angular power spectrum of the lensing convergence $\kappa$ can be written as (adopting the Limber approximation and a single lensing source redshift at comoving distance $\chi_S$)
\bq\label{eq:Clkappa}
C_{\kappa}(\ell) = \int _0^{\chi_S} \left(\frac{g(\chi)}{\chi}\right)^2 \left[\Sigma\left(z(\chi), \frac{\ell}{\chi}\right)\right]^2 P_m\left(\frac{\ell}{\chi}, z(\chi)\right),
\eq
where $g(\chi) = (3H_0^2\Omega_m/2/c)(1+z)(\chi_S-\chi)\chi/\chi_S$ is called the lensing kernel. If modified gravity enhances structure formation ($Q > 1$ in Eqs.~(\ref{eq:mu})), then this boosts the amplitude of the three-dimensional matter power spectrum $P_m$, and consequently, $C_{\kappa}(\ell)$. Modified gravity models that have $\Sigma \neq 1$ will display additional effects.
 
Weak lensing analyses can be carried out to smaller length scales compared to galaxy clustering analysis, for which modeling uncertainties associated with galaxy bias preclude the use of data on scales $k \gtrsim 0.2\ {h / {\rm Mpc}}$. On these smaller scales, however, baryonic feedback effects can alter appreciably the amplitude of the total clustering of matter \citep{2016MNRAS.461L..11H, 2018MNRAS.475..676S}. These effects are relatively poorly understood and remain a source of theoretical systematics in real constraint analyses (see Sec.~\ref{sec:baryons}). Similarly, massive neutrinos can also have a non-negligible impact on the small-scale matter power spectrum, and with degenerate effects with modified gravity. These are all complications that are present in {\it standard} GR analyses, but that gain further importance in modified gravity because of the degeneracies that can be at play between these effects and the effects of fifth forces \citep{2015MNRAS.454.2722H, 2014MNRAS.440...75B, 2019MNRAS.486.3927H}. Finally, on such small scales, it becomes imperative to appropriately take into account screening effects. The difficulty of parameterizing the impact of screening on the small-scale matter power spectrum is currently hindering the use of small-scale weak-lensing shear data in tests of gravity.  For instance, \citet{2017MNRAS.471.1259J, 2018MNRAS.474.4894J} place constraints on modified gravity parametrizations using lensing data from KiDS, but with small-scale data removed (see e.g.~Fig.~13 of \citealt{2017MNRAS.471.1259J} to appreciate the loss in constraining power). In constraints of the Galileon model, \citet{2017arXiv171104760P} also considers only sufficiently large scales in their lensing dataset. The development of tools to accurately predict small-scale clustering in modified gravity cosmologies (cf.~Sec.~\ref{sec:nonlinear_semiMG}) will therefore prove extremely valuable to the analyses of future weak-lensing surveys (see e.g. \citealt{2018arXiv180104251S} for cosmic shear forecasts for Euclid in the context of modified gravity). We note also that the adequacy of current models of intrinsic alignment should also be re-evaluated in the context of modified gravity models.

\end{itemize}

To conclude this discussion, we comment on which types of theories are susceptible to be most tightly constrained by cosmological probes, compared to astrophysical ones. For self-accelerating models with non-$\Lambda{\rm CDM}$ expansion histories, cosmological constraints using CMB, BAO and SNIa data should represent the first line of testing to determine which cosmological parameters yield acceptable fits to these data (if any) and which are therefore worthy of more dedicated and lengthier studies (e.g., involving N-body simulations). The usefulness of cosmological data in tests of gravity is also dependent on the screening mechanism. For instance, we will see in the next section that astrophysical tests have a stronger potential to constrain chameleon-like theories due to the environmental dependence of the screening efficiency, compared to cosmological ones. On the other hand, Vainshtein screening models tend to have stronger signatures on cosmological observables: the tight constraints on the nDGP model using $f\sigma_8$ data \citep{2016PhRvD..94h4022B} and on Galileons using ISW-galaxy cross-correlations \citep{Renk:2016olm} are illustrative of this point.

\section{Astrophysical tests}
\label{sec:astro_tests}

We now turn to the main focus of the review: astrophysical probes of modified gravity. We begin by discussing tests that are applicable to modified gravity in general without dependence on the screening mechanism (associated with galaxy velocities \ref{sec:RSD}, dark matter halos \ref{sec:universal_tests} and voids \ref{sec:universal_tests2}), then we present probes of thin-shell and Vainshtein screening specifically. We preface this with a discussion of the ``screening maps'' that are necessary to identify regions of the universe, and hence galaxies, that are unscreened. Where such results exist, we provide quantitative detail on the constraints achieved, and those achievable using future data with the characteristics we describe. The reader interested in more in-depth discussions and technical derivations on some of the topics presented here are are referred to \citet{Sakstein:2018fwz}.

\subsection{Galaxy velocities and redshift space distortions}
\label{sec:RSD}


The matter velocity field is particularly sensitive to modifications of gravity. This is because velocities are determined by a single time integral over the acceleration, and hence they typically display stronger signatures of modified gravitational forces, compared to the matter density field (which is determined by two time integrals). A way to probe these statistics directly is via galaxy velocities, which are an unbiased tracer of the matter velocity field on large scales, by way of the equivalence principle. 

On quasi-linear scales, the matter velocity field is governed by the Euler equation
\be
\frac{\partial}{\partial\tau} \vec{v} + (\vec{v}\cdot\vec{\nabla}) \vec{v} + \mathcal{H} \vec{v} 
= -\vec{\nabla} \Phi - \mathcal{Q} \vec{\nabla}\phi\,.
\label{eq:euler}
\ee
The quantity $\mathcal{Q}$ is the cosmological analog of the scalar charge-to-mass ratio $Q$ introduced in Sec. \ref{sec:eqprinvio} (with factors of $\alpha$ suitably scaled) and represents the strength of the coupling of a given test object to the scalar field.  Any modifications to gravity which affect the time-time component of the metric will contribute directly to changes in the velocity field. In GR, the equivalence principle guarantees that any difference between the velocity field $\v{v}_g$ of galaxies and matter has to be due to non-gravitational forces such as baryonic pressure. On scales much larger than the Jeans length of the gas, this pressure is negligible, and hence there is no velocity bias. Note however that in frameworks such as the Effective Field Theory of Large Scale Structure the Euler equation is expected to be only a first order approximation to the behaviour of dark matter. Higher-order corrections may be partially degenerate with with modified-gravity effects if the gradient of the scalar field
$\vec{\nabla}\phi$ is suppressed on large scales, for example due to a finite mass of the field.

In modified gravity theories, however, there can be violations of the equivalence principle, which can be used in observational tests (cf.~Sec.~\ref{sec:eqprinvio}). For example, in thin-shell screening models the scalar charge of a screened object, say some sufficiently massive galaxy, is strongly reduced: $\mathcal{Q} \ll 1$. The velocity of this galaxy will effectively only be sourced by $\vn\Phi$, and would thus fall at a slower rate compared to a less massive, unscreened galaxy with $\mathcal{Q} = 1$ \citep{Hui:2009kc}. Similar lines of reasoning hold for black holes in Vainshtein screening theories, which carry no scalar charge $\mathcal{Q}=0$, and thus fall at different rates than the gas and stars in the same galaxy (\citealt{Hui:2012jb}; cf.~Sec.~\ref{sec:SEPviolations}).


On large scales, in a frame comoving with the galaxy velocities, the clustering pattern of galaxies would be isotropic. In reality, however, peculiar velocities of galaxies perturb the measured redshift via Doppler effect (and consequently the inferred line-of-sight distance), which induces an anisotropy in the galaxy distribution that is proportional to the galaxy velocities. This effect is referred to as redshift space distortions (RSD). Specifically, the inferred three-dimensional position of a galaxy $\v{x}_\text{obs}$ differs from the true unobserved one $\v{x}$ by \citep{kaiser:1987}
\be
\v{x}_\text{obs} = \v{x} + \mathcal{H}^{-1} (\hat{n}\cdot \v{v}_g) \hat{n}\,,
\label{eq:RSD}
\ee
where $\hat{n}$ is the angular position of the galaxy on the sky. The observed {\it redshift-space} fractional galaxy density perturbation $\d_g^s$ is related to the rest-frame one $\d_g$ through
\be
\d_g^s(\v{x}_\text{obs}) = \frac{\d_g(\v{x})}{1 + \mathcal{H}^{-1} \hat n^i \hat n^j \partial_i v_{g,j}(\v{x})} \Big|_{\v{x} = \v{x}_\text{obs} - \mathcal{H}^{-1} (\hat{n}\cdot \v{v}_g) \hat{n}} \,.
\label{eq:dgs}
\ee
Working to linear order in $\d_g$ and $\v{v}_g$, the power spectrum of galaxies on large scales becomes \citep{kaiser:1987}
\be
P_g^s(\v{k}) = (b_1 + f \mu^2)^2 P_m(k)\,,\quad \mbox{where}\quad \mu = \frac{\v{k}\cdot\hat{n}}{k}\,,
\label{eq:Pkkaiser}
\ee
which is clearly anisotropic as it depends explicitly on the angle between the Fourier modes and the line-of-sight direction $\hat{n}$. The degree of anisotropy is proportional to the linear growth rate $f = d\ln D/d\ln a$, which can be therefore extracted from galaxy surveys \citep{2012MNRAS.423.3430B, 2014MNRAS.439.2515O, 2017MNRAS.470.2617A, 2013A&A...557A..54D} (cf.~right panel of Fig.~\ref{fig:cosmo}). Different theories of gravity make different predictions for the growth rate, which makes redshift space distortions a powerful tool to test gravity (cf.~Sec.~\ref{sec:cosmo2}).

Another interesting probe of gravity is the so-called $E_G$ statistic \citep{reyes/etal:2010}. This is obtained by combining the quadrupole moment of the anisotropic galaxy power spectrum $P_g^s(\v{k})$ (which is proportional to the cross-correlation between galaxies and the velocity divergence field $\theta = \nabla\v{v}$, $P_{g\theta}$) with the cross-correlation between lensing maps (either CMB lensing or cosmic shear) and galaxy positions, $P_{g\nabla^2\Phi_{\rm lens}}$ (where we write $\Phi_{\rm lens} = (\Phi + \Psi)/2$ for short):
\be
 E_G \equiv \frac{P_{g(\nabla^2 \Phi_{\rm lens})}}{P_{g\theta}}  \stackrel{\text{in GR}}{=}      \frac{\Omega_{m0}}{f(z)} = \frac{\Omega_{m0}}{\Omega_m(z)^{0.55}}\,,
\label{eq:EG}
\ee
where the last two equalities assume large linear scales and GR. The numerator of the $E_G$ statistic is sensitive to modifications to the lensing potential ($\Sigma$ in Eqs.~(\ref{eq:mu})), whereas the denominator to changes to the dynamical potential ($Q$ in Eqs.~(\ref{eq:mu})). The ratio between these two potentials is 1 in GR, and so this statistic directly targets modifications to gravity \citep{zhang/etal:2007, pullen/etal:2016, 2017arXiv171110999A, Singh}. Importantly as well is the fact that this statistic is constructed to cancel the effects of galaxy bias at linear order.

The kinematic Sunyaev-Zel'dovich (kSZ) effect, which describes the shifts in the temperature of CMB photons caused by the bulk momentum of hot ionized gas inside clusters, offers another probe of the cosmic velocity field. Observationally \citep{2012PhRvL.109d1101H, 2016MNRAS.461.3172S, 2017JCAP...03..008D}, what has been detected is the so-called pairwise kSZ signal whose amplitude can be written as
\bq\label{eq:pKSZ}
\frac{T_{pKSZ}(r)}{T_{\rm CMB}} = \tau_e \frac{v_{12}(r)}{c},
\eq
where $v_{12}(r)$ is the mean pairwise velocity of clusters separated by $r$ and $\tau_e$ is their mean optical depth, which is given as a line-of-sight integral of the free electron number density $n_e$ inside the clusters, $\tau_e = \sigma_T \int {\rm d}\chi n_e$. Modified gravity can leave signatures on the pairwise kSZ effect via the modifications to the pairwise velocity of clusters $v_{12}(r)$ \citep{keisler/schmidt,2014PhRvL.112v1102H}, which quantifies the mean velocity at which galaxy clusters approach one another due to the influence of gravity. In linear theory, and assuming linear cluster bias $b$ and unbiased cluster velocities, the pairwise velocity is given by (see e.g. \citealt{2001MNRAS.326..463S, 2015ApJ...808...47M})
\bq\label{eq:v12}
v_{12}(r) = -\frac{2}{3}arHf \frac{b\bar{\xi}(r)}{1 + b^2\xi(r)},
\eq
where $\xi(r)$ is the matter correlation function and $\bar{\xi}(r) = (3/r^3) \int {\rm d}r'r'^2/\xi(r')$. Different modified gravity models can thus display distinct $T_{pKSZ}(r)$ predictions via modifications to the growth rate $f$. A main complication in using current kSZ data to constrain modified gravity (see \citealt{2015ApJ...808...47M} for a forecast study) lies in the uncertain value of $\tau_e$, which is degenerate with the amplitude of $v_{12}(r)$ and could well vary from one modified gravity model to another. In fact, other sectors of the cosmological/astrophysical community have an interest on the kSZ effect as a probe of $\tau_e$, not the theory of gravity. One way to make progress is to use the shape of $v_{12}(r)$ \citep{keisler/schmidt}, another is to calibrate $\tau_e$ using hydrodynamical simulations of modified gravity.

Beyond the linear regime described by \refeq{Pkkaiser}, there is also strong motivation to use galaxy statistics on smaller scales. Unfortunately, disentangling galaxy velocities from their nontrivial clustering in real space becomes complicated on small scales. A compromise was proposed by \citet{lam/etal:2012}, who considered the cross-correlation between massive galaxy clusters and field galaxies. This essentially corresponds to a specific projection of the galaxy phasespace. Due to the presence of the massive halo hosting the galaxy cluster, the velocities of galaxies within a separation of $\sim 5-20h^{-1} \text{Mpc}$ are dominated by coherent infall motion on the massive cluster. This considerably simplifies the modeling of this observable \citep{zu/weinberg:2013,lam/etal:2013}.

\subsection{Galaxy clusters}
\label{sec:universal_tests}

The enhanced growth of structure in the presence of a fifth force leaves an imprint on the abundance (\citealt{schmidt:08}; Sec.~\ref{sec:abundance}) and profiles (\citealt{lombriser:11b}; Sec.~\ref{sec:profiles}) of galaxy clusters.
For chameleon models, one of the most prominent effects is an increase of the abundance of massive clusters for large scalar field values.
In the case of small field values, this enhancement is counteracted by the screening effect as well as by the Yukawa suppression \citep{Hui:2007zh,Martino:2008ae,2011PhRvD..83f3511P} beyond the Compton wavelength of the background field.
Both contribute to a recovery of Newtonian gravity and of cluster abundance in agreement with $\Lambda$CDM at high masses, restricting observable effects to smaller masses.
In addition to the mass dependence, chameleon screening introduces a dependence on the environment of the clusters.
The overall shape of the matter density profiles within the cluster is not strongly affected by the scalar field, but an increase in the halo concentration \citep{lombriser:12,Lombriser:2013eza,Shi:2015aya,Mitchell:2019qke} and effects on splashback in the outer regions of clusters provide observable signature (Sec.~\ref{sec:profiles}).
An additional test of gravity can be performed by comparing the distribution, temperature, and pressure of gas with the dark matter profile in the interior of the cluster (Sec.~\ref{sec:profiles}).

\subsubsection{Cluster abundance} \label{sec:abundance}




The statistics of virialized clusters is well described by excursion set theory, where the collapsed structures are associated with regions where the smoothed initial matter densities exceed the linear collapse density threshold $\delta_c$.
The variance of the density field $\sigma^2$ characterizes the size of such a region.
Variation of the variance (or the smoothing window size) causes incremental changes 
in the smoothed initial overdensity field that are independent of previous values for uncorrelated wavenumbers. This describes a Brownian motion of the smoothed matter density field, where the increment is a Gaussian field with zero mean.
The distribution $f$ of the Brownian motion trajectories that first cross a flat barrier $\delta_c$ at a given variance was described in \citet{press:74, 1991ApJ...379..440B}.
Relaxing the assumption of sphericity of the halo, the barrier, however, is no longer flat.
The first-crossing distribution based on excursion set results for ellipsoidal collapse was described by Sheth and Tormen~\citep{sheth:99, sheth:99b, sheth:01},
\bq
 \nu \, f(\nu) = \mathcal{N} \sqrt{\frac{2}{\pi} q \, \nu^2} \left[ 1 + \left(q \, \nu^2\right)^{-p} \right] e^{-q\,\nu^2/2} \label{eq:st}
\eq
with peak-threshold $\nu\equiv\delta_c/\sigma$, normalisation $\mathcal{N}$ such that $\int d\nu \, f(\nu) = 1$, as well as $p=0.3$ and $q=0.707$.
Hereby, $q$ was set to match the halo mass function 
\bq
 n_{\ln M_{vir}} \equiv \frac{d n}{d\ln M_{vir}} = \frac{\bar{\rho}}{M_{vir}} f(\nu) \frac{d \nu}{d\ln M_{vir}} \label{eq:hmf}
\eq
measured with $\Lambda$CDM $N$-body simulations.

Most tests of modified gravity with clusters have been performed within the chameleon-screened $f(R)$ paradigm. The halo mass function defined by Eqs.~(\ref{eq:st}) and (\ref{eq:hmf}) with the $\Lambda$CDM value of $\delta_c$ has been shown to provide a good description to $N$-body simulations of $f(R)$ gravity for large field values, while a modified collapse threshold $\delta_c$ derived from a collapse calculation with enhanced forces provides a conservative lower limit on the effects for small field values~\citep{schmidt:08}. 
The latter case was used to infer constraints on the Hu-Sawicki and designer models from the cluster abundance measured with Chandra X-ray~\citep{schmidt:09,ferraro:10}, SDSS MaxBCG~\citep{Lombriser:2010mp}, and ROSAT BCS, REFLEX, and Bright MACS data~\citep{Cataneo:2014kaa} at the level of $|f_{R0}|\lesssim10^{-5}-10^{-4}$.
\citet{Cataneo:2016iav} argued that future cluster surveys will improve this bound to a level comparable with solar system constraints at $|f_{R0}|\lesssim10^{-6}$.
Note that $f(R)$ designer models are constructed to exactly reproduce a $\Lambda$CDM expansion history whereas for Hu-Sawicki models, the Hubble functions differ at $\mathcal{O}(f_{R0})$. 
Since very small, this correction is usually neglected for the Hu-Sawicki model, but as a consequence the $f(R)$ functions differ between the two models.
Another useful proxy for halo abundance is the high signal-to-noise ratio peaks of weak lensing convergence field, which are believed to correspond to the largest dark matter halos; \citet{Liu:2016xes} used the peak counts from CFHTLenS data to derive a constraint at $|f_{R0}|\lesssim10^{-5.2}$, and this bound is expected to become much stronger with future larger lensing surveys.
These, however, require a more accurate modelling of the halo mass function that accounts for the chameleon screening effect.
An improved description of the halo mass function in the small-field regime is obtained by adopting the spherical collapse critical density $\delta_c$ determined by the mass and environment dependent spherical collapse model for chameleon and $f(R)$ gravity~\citep{li:11a,Lombriser:2013wta,Lombriser:2013eza} in Eq.~(\ref{eq:st}).
A fitting function for the halo mass function of the Hu-Sawicki $f(R)$ model was developed from these computations in \citet{Cataneo:2016iav}, yielding 5\% accuracy in the relative enhancement of the modified cluster abundance.

Alternatively to adopting the Sheth--Tormen halo mass function, excursion set theory approaches to computing the first-crossing distribution with the moving barrier defined by the linear chameleon collapse density have been pursued in \citet{Li:2012ll,Lam:2012ll} based on Lagrangian and Eulerian definitions of the environment, performing numerical integrations and Monte Carlo simulations.
The simpler Sheth--Tormen prescription combined with the mass and environment dependent spherical collapse model and a subsequent averaging over the probability distribution of the Eulerian environment, however, was found to show the better agreement with $f(R)$ $N$-body simulations~\citep{Lombriser:2013wta}.
Alternatively to the top-hat approximation implemented in the chameleon spherical collapse computations of \citet{li:11a,Lombriser:2013wta,Lombriser:2013eza}, the $f(R)$ evolution of an initial density profile was considered in \citet{Borisov:2011fu} and applied in \citet{Kopp:2013lea} to develop an analytic halo mass function function based on excursion set theory with a drifting and diffusing barrier.
In \citet{2011PhRvD..84h4033L}, the chameleon screening effect was incorporated in the Sheth--Tormen halo mass function with a phenomenological transition in the variance that interpolates between the linearized and suppressed regimes that is calibrated~\citep{2011PhRvD..84h4033L,Lombriser:2013wta} to $N$-body simulations of the Hu-Sawicki $f(R)$ model.
A comparison of these different approaches can be found in \citet{Lombriser:2014dua}.

The enhancement of the halo mass function in Vainsthein-type modified gravity models shows qualitatively different features from those observed in chameleon type models (in the small field regime), e.g. \citet{2009PhRvD..80l3003S,Barreira:2013xea,Barreira:2014zza}. The Vainshtein screening which is very efficient near/inside massive objects does not prevent more massive halos from forming in these models, possibly because the long-range fifth force has managed to accrete more matter towards the surroundings of these halos, creating a larger reservoir of raw material for their growth. Analytic results based on the excursion set theory agree qualitatively with predictions of simulation \citep{schmidt:10}, although a re-calibrated Sheth--Tormen formula was found to work much better for the Cubic and Quartic Galileon models \citep{Barreira:2014zza}. Hence, cluster abundance is also expected to be useful to constrain this type of model. \citet{2019arXiv190706657D}, for example, used weak lensing peaks as a proxy of massive dark matter halos and found a strong constraining potential on the nDGP model. Quantitative constraints do not yet exist for other screening models such as symmetron and K-mouflage, although modifications to spherical collapse and hence the halo mass function have been explored \citep{Taddei:2013faa,Taddei:2013bsk, Davis:2011pj, Brax:2015lra, Brax:2014yla}.

\subsubsection{Cluster profiles and splashback}
\label{sec:profiles}


Apart from their abundance, the internal structure of dark matter halos can
also be used as a probe. Well within the virial radius of halos,
the spherically-averaged dark matter distribution formed in  $f(R)$ $N$-body simulations is  well described~\citep{lombriser:12,Shi:2015aya,Mitchell:2019qke} by the Navarro-Frenk-White (NFW; \citealt{Navarro:1995iw}) profile
\begin{equation}
 \rho(r) = \frac{\rho_{\rm s}}{\frac{r}{r_{\rm s}} \left( 1+\frac{r}{r_{\rm s}} \right)^2 } \,, \label{eq:nfw}
\end{equation}
which were originally proposed to fit halos formed in $\Lambda$CDM and other models.
The characteristic density $\rho_{\rm s}$ and scale $r_{\rm s}$ can also be written as functions of the virial halo mass $M_{\rm vir}$, defined by the virial overdensity $\Delta_{\rm vir}$, and virial halo concentration $c_{\rm vir}\equiv r_{\rm vir}/r_{\rm s}$.
The halo concentration for clusters formed in chameleon $f(R)$ gravity was measured in $N$-body simulations in \citet{lombriser:12,Shi:2015aya} and found to be enhanced with respect to the concentration of $\Lambda$CDM halos.
This also causes an enhancement in $\rho_{\rm s}$ and a decrease of $r_{\rm s}$ compared to their $\Lambda$CDM counterparts.
However, for small-field values, chameleon screening suppresses the enhancement in the concentration, recovering $\Lambda$CDM values for high-mass clusters.
A mass and environment-dependent modeling of the chameleon halo concentration based on the spherical collapse calculations that captures these effects was introduced in \citet{Lombriser:2013eza}. In \citet{Mitchell:2019qke} a detailed analysis based on a large suite of simulations with varying box sizes and resolutions was conducted, and it was found that the effect of $f(R)$ gravity on halo concentration can be well described by a universal fitting formula that depends only on the combination $\bar{f}_R(z)/(1+z)$ -- where $\bar{f}_R(z)$ is the background scalar field at redshift $z$ -- which works for a wide range of model parameters, halo masses and redshifts. Observationally, miscentering and other issues can complicate the determination of halo concentration. 

At a few virial radii, in the infall region of the cluster, the halo density profile, or halo-matter correlation function, exhibits an enhancement in chameleon models, relative to GR, caused by the late-time gravitational forces~\citep{schmidt:08, lombriser:11b,zu:13}.
The signature can be well described by the halo model.
Through stacking of galaxy clusters it has been used to derive a constraint of $|f_{R0}|\lesssim10^{-3}$~\citep{lombriser:11b}.

The splashback feature in galaxy clusters has recently been introduced as a dynamical boundary of cluster halos \citep{Diemer:2014xya}. More specifically, after its initial infall stage, material being accreted by dark matter halos experiences a turn around. The location at which it does so defines the boundary of the multistreaming region of the halo \citep{Adhikari:2014lna,Shi:2016lwp} and corresponds to the outermost caustic. The definition of a turn-around radius, however, may be complicated by the surroundings of the halo being a collection of filaments, sheets and voids~\citep{Hansen:2019juz}. At this point, the logarithmic slope of the density profile is predicted to drop significantly below the NFW value (between -2 and -3), before rising again to the two-halo term value. Hence this \emph{splashback radius} provides a clear observational signature in both the galaxy number density (which has been measured in redMaPPer-selected clusters \citep{Rykoff:2013ovv} from SDSS and DES Y1  \citep{More:2016vgs,Baxter:2017csy}), and weak lensing \citep{Chang:2017hjt}.

\begin{figure}[t]
\centering
{\includegraphics[width=0.95\textwidth]{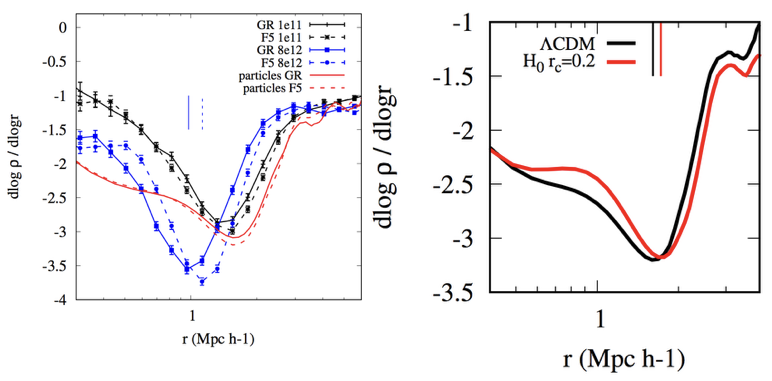}}
\caption{\emph{Left panel}: Comparison of the splashback radius (defined as the minimum of each curve) for GR and chameleon theories. F5 models correspond to $\chi=10^{-5}$. The red lines correspond to particles and the blue and black to subhalos with masses indicated in the figure. \emph{Right panel}: The density slope for dark matter particles in GR (black) and a Vainshtein-screened theory with $r_c=600$ Mpc (red). The splashback radius is the minimum of each curve. Reproduced from \citet{Adhikari:2018izo}.}\label{fig:splashback}
\end{figure}

A theoretical study dedicated to the splashback feature for both chameleon and Vainshtein screening has been conducted in \citet{Adhikari:2018izo}.
The reasons for its sensitivity to modified gravity are twofold.
First, for model parameters where the fifth forces are important on galactic scales, the splashback radius is located around the transition from the screened to the unscreened regime. Second, accreted material began its life well outside the screening/Vainshtein radius and its subsequent dynamics therefore has some memory of the unscreened fifth force. 
Using a combination of analytic approximations for the dynamics of accreted shells and N-body simulations, \citet{Adhikari:2018izo} found that the splashback radius for dark matter particles in Vainshtein-screened theories with $r_c\sim \mathcal{O}(500)$ Mpc (corresponding to theories that are relevant on galaxy scales and not excluded by GW170817) is significantly larger than GR. 
For chameleon theories the signature is on galaxy scale subhalos within clusters. Subhalos experience dynamical friction as they pass through their host, which results in their splashback radius being smaller than particles. The effect is mass dependent with subhalos having mass ratio $M_{\rm sub}/M_{\rm host}>0.01$ exhibiting significantly smaller splashback radii. For the chameleon theories studied, the reduced dynamical friction  resulted in smaller subhalo splashback radius than in GR.
Some examples of these effects are shown in Fig.~\ref{fig:splashback}. 
Recently, a semi-analytic study has revealed that for certain parameters, the symmetron can produce deviations of $\mathcal{O}(10\%)$ \citep{Contigiani:2018hbn}.

\subsubsection{Dynamical vs lensing masses}
\label{sec:dynmass}




The mass distribution of galaxies and clusters determines the gravitational potential that lenses the photon trajectories and govern the kinematic properties of the cluster.
%
A generic feature of modified gravity theories is an inequality between the two metric potentials.
Photons respond to the sum of the metric potentials while nonrelativistic tracers such as stars and galaxies, which typically move with speeds of 100-1000 km/s, respond to the time-time potential. 
Thus inferences made about the gravitational potential or the mass distribution using photons (i.e. lensing or the ISW effect) can be discrepant with dynamical masses inferred from stars and galaxy motions. This is often cast in terms of the PPN parameter $\gamma$, and provides a powerful test of gravity on a variety of scales:

\begin{compactitem}
\item On the smallest scales, $\mathcal{O}$(1 kpc), strong lensing and stellar dynamical mass estimates within lens galaxies can be compared \citep{Collett:2018gpf,Schwab:2009nz}.

\item A similar comparison of strong lensing by galaxy clusters with galaxy velocity dispersions or X-Ray masses can be made (length scales $\mathcal{O}$(100 kpc)) \citep{Schmidt:2010jr,Pizzuti:2016ouw}

\item Weak lensing mass profiles of galaxies and clusters provide mass estimates out to the virial radii of their halos and beyond. For clusters, these can be compared with SZ masses (that are calibrated using GR-based simulations) that extend to the virial radius. On slightly larger scales the lensing masses can be compared with dynamical estimates of infall motions from redshift space surveys \citep{lam/etal:2012, zu/etal, Terukina:2013eqa, Sakstein:2016ggl, Wilcox:2015kna}. On similar scales, $\sim 0.1 - 10\ {\rm Mpc}/h$, \citet{2017MNRAS.467.3024L} reported an interesting discrepancy between the measured galaxy-galaxy lensing signal of BOSS CMASS galaxies in CFHTLenS and CS82 fields and the prediction from a number of mock catalogues with CMASS-like clustering and stellar mass functions.

\item On even larger scales, $\sim 10-100$ Mpc, assuming linear bias for the galaxies, weak lensing and redshift space power spectra have been used to estimate the $E_G$ parameter \citep{zhang/etal:2007, Blake:2015vea,  pullen/etal:2016, Singh:2018flu} which is reliably known for GR. 

\end{compactitem}

Further information to be compared with lensing comes from cluster gas temperature, density and pressure profiles measured with X-ray and SZ data. The gas density 
$\rho_{\rm gas}$ and pressure $P$ relate to the dynamical mass profile $M_{\rm dyn}(r)$ as
\begin{equation}
\frac{1}{\rho_{\rm gas}(r)}\frac{d P(r)}{d r} = - \frac{G\,M_{\rm \rm dyn}(r)}{r^2} \,. \label{eq:hydroeq}
\end{equation}
The dynamical mass $M_{\rm dyn}$ differs from the lensing mass due to the
fifth force that can be interpreted as arising from the gradient of the scalar field profile. This can be modelled following \citet{Schmidt:2010jr,pourhasan:11,lombriser:12,Lombriser:2014nfa}.
%

A simple approximation for chameleon gravity is given by~\citep{Lombriser:2014dua}
\begin{equation}
 M_{\rm dyn}(r) \simeq \left\{ 1 + \frac{\Theta(r-r_{\rm c})}{3+2\omega}\left[ 1 - \frac{M(r_{\rm c})}{M(r)} \right] \right\} M(r)_{\rm lens}, \label{eq:dynamicalmass}
\end{equation}
where $\Theta$ is the Heaviside step function, $M(r)_{\rm lens}$ is the lensing mass, and the chameleon screening scale is
\begin{equation}
 r_{\rm c} \simeq \frac{8\pi G\rho_{\rm s}r_{\rm s}^3}{3+2\omega} \frac{1}{1-\phi_{\rm env}} - r_{\rm s}
\end{equation}
with the scalar field value $\phi_{\rm env}$ in the environment ($\phi=1+f_R$ for $f(R)$ gravity). A fitting function for the relation between the dynamical and true masses for $f(R)$ gravity, calibrated using a large suite of N-body simulations with varying resolutions and box sizes, was given in \citet{Mitchell:2018qrg}:
\begin{equation}
\frac{M_{\rm dyn}}{M_{\rm lens}} = \frac{7}{6}-\frac{1}{6}\tanh\left[p_1\left(\log_{10}(M_{\rm lens}/M_\odot)-p_2\right)\right],
\end{equation}
where $p_1$ and $p_2$ are given by
\begin{eqnarray}
p_2 &=& 1.503\log_{10}\left[\frac{|f_R(z)|}{1+z}\right]+21.64,\\
p_1 &=& 2.21,
\end{eqnarray}
works very well for all models with $|f_{R0}|\in\left[10^{-6.5},10^{-4}\right]$ in a wide range of redshifts $z\in[0,1]$. In particular, the constancy of $p_1$ and the slope of $p_2$ (1.503) are very close to the predictions of thin-shell modelling (which gives a slope 1.5 for $p_2$).

Assuming no non-thermal pressure, the gas pressure, density and temperature are related by $P=P_{\rm thermal}\propto\rho_{\rm gas}T_{\rm gas}$. Hence, in hydrostatic equilibrium, the lensing, X-ray surface brightness, X-ray temperature, and SZ observations are uniquely determined from any combination of two profiles adopted for either $P_{\rm thermal}$, $\rho_{\rm gas}$, $T_{\rm gas}$, and $M$.
A combination of the four measurements therefore breaks degeneracies among the profiles and yields a powerful test of gravity~\citep{2014JCAP...04..013T}.
Combining weak lensing measurements with gas observations from the X-ray surface brightness and temperature as well as the Sunyaev-Zel'dovich effect from the Coma cluster, \citet{2014JCAP...04..013T} inferred constraints on chameleon models that correspond to $|f_{R0}|\lesssim6\times10^{-5}$ when cast in terms of Hu-Sawicki $f(R)$. The same constraint was obtained in \citet{Wilcox:2015kna} from the stacked profiles of 58 clusters with combined XMM Cluster Survey X-ray and CFHTLenS weak lensing measurements but with no SZ data. This test has also been conducted for Galileon gravity \citep{Terukina:2015jua} and beyond Horndeski theories \citep{Sakstein:2016ggl}.



The method of comparing the hydrostatic and lensing masses of galaxy clusters has also been applied to Vainshtein breaking theories (see Sec.~\ref{sec:vainshtein_breaking}). Unlike thin-shell screening theories where the Newtonian potential is altered but not the lensing potential, Vainshtein breaking alters both potentials (see Eqs.~\eqref{eq:VB1} and \eqref{eq:VB2}). For an NFW halo, the masses are
\begin{align}
M(r)_{\rm dyn}&=M_{\rm NFW}+\pi\Upsilon_1r_s^3\rho_s\left(1-\frac{r_s}{r}\right)\left(1+\frac{r_s}{r}\right)^{-3}\\
M(r)_{\rm lens}&=M_{\rm NFW}+\frac{\pi r_s^3\rho_s}{2}\left[\left(\Upsilon_1+5\Upsilon_2+4\Upsilon_3\right)-\left(\Upsilon_1+5\Upsilon_2+4\Upsilon_3\right)\frac{r_s}{r}\right].
\end{align}
Assuming $\Upsilon_3 = 0$ (corresponding to beyond Horndeski theories with no DHOST terms), \citet{Sakstein:2016ggl} constrained $\Upsilon_1$ and $\Upsilon_2$ by comparing $M_{\rm hydrostatic}$ and $M_{\rm lens}$ for a sample of 58 X-ray selected clusters for which lensing data from CFHTLenS and X-ray data from XMM-Newton was available. 
The hydrostatic mass was found using the X-ray surface brightness temperature and the lensing mass was found by stacking the profiles. The $2\sigma$ bounds $\Upsilon_1=-0.11^{0.93}_{-0.67}$ and $\Upsilon_2=-0.22^{1.22}_{-1.19}$ were obtained.

The above method requires lensing data to be available for X-ray or SZ-selected clusters. An alternative way to constrain chameleon-type models that uses the fact that the dynamical and true masses of clusters can be different, without needing lensing data, is to consider the gas fractions of clusters. The largest galaxy clusters form from regions whose initial sizes are over 10 Mpc, and it is expected that the mass ratio between baryonic (dominated by hot gas) and dark matter components inside them is close to the cosmic average, $\Omega_b/\Omega_m$ \citep{White:1993wm}, making clusters `standard buckets'. The cluster gas fraction profile, $f_{\rm gas}(r)\equiv M_{\rm gas}(<r)/M_{\rm tot}(<r)$, can be estimated by measuring the mass profile ($M_{\rm tot}(<r)$) and hot gas ($M_{\rm gas}(<r)$)  profile. These can be obtained by respectively observing the X-ray temperature and luminosity profiles. 
In chameleon models, 
the measured $M_{\rm tot}(<r)$ is the cluster dynamical mass, 
which implies that for unscreened halos the observationally inferred value of $f_{\rm gas}(r)$, $f^{\rm obs}_{\rm gas}(r)$, is related to the true value, $f^{\rm true}_{\rm gas}(r)$, by
\begin{equation}
f^{\rm obs}_{\rm gas}(r) = \frac{M_{\rm gas}(<r)}{M_{\rm tot,dyn}(<r)} = \frac{M_{\rm tot,lens}(<r)}{M_{\rm tot,dyn}(<r)}\frac{M_{\rm gas}(<r)}{M_{\rm tot,lens}(<r)} \equiv \eta(r)f^{\rm true}_{\rm gas}(r),
\end{equation}
where $\eta(r)\in[3/4,1]$ encodes the screening effect. Because it is $f^{\rm true}_{\rm gas}(r)$ that is directly related to $\Omega_b/\Omega_m$, this suggests that in chameleon models, depending on the screening, the observed gas fraction can have a systematic difference from the expected value from constraints on $\Omega_b, \Omega_m$ by other observations such as the CMB. Since the screening depends on the background scalar field value and redshift, $\eta(r)$ can evolve with redshift, which can also be a signature of departures from $\Lambda$CDM. \citet{Li:2015rva} estimates that Hu-Sawicki $n=1$ $f(R)$ model with $|f_{R0}|=3\sim5\times10^{-5}$ is in tension with the gas fraction data of the 42 clusters analysed by \citet{Allen:2007ue}.

The distinction between dynamical and lensing masses in cluster observations is also important for the modelling of cluster scaling relations -- the relations that is often used to infer a cluster's mass from some observational proxy. For example, in $f(R)$ gravity, an interesting observation is that the gas density profile of a halo with dynamical mass $M_{\rm dyn}^{f(R)}=M_\ast$ from a simulation with baryon density parameter $\Omega_b$ is very similar to that of $\Lambda$CDM halos whose lensing (or true) mass is given by $M_{\rm lens}^{\Lambda{\rm CDM}}=M_\ast$ but from a simulation with a baryon density parameter $\Omega_bM_{\rm lens}^{f(R)}/M_{\rm dyn}^{f(R)}$, where $M_{\rm lens}^{f(R)}$ is the true mass of the $f(R)$ halo whose dynamical mass is equal to $M_\ast$. This is easy to understand: dynamically the $f(R)$ halo whose dynamical mass is $M_\ast$ and the $\Lambda$CDM halo whose true mass is $M_\ast$ are indistinguishable, but assuming that all haloes have the same ratio of baryon-to-true-halo mass, the $f(R)$ halo, which has a smaller true mass, would have a smaller baryon mass. \citet{He:2015mva} shows this using non-radiative hydrodynamical simulations: 
\begin{equation}
\rho_{\rm gas}^{f(R)}(r) = \frac{M^{f(R)}_{\rm lens}(<r)}{M^{f(R)}_{\rm dyn}(<r)} \: \rho_{\rm gas}^{\Lambda{\rm CDM}}(r),
\end{equation}
where $\rho^{f(R)}_{\rm gas}$, $\rho^{\Lambda{\rm CDM}}_{\rm gas}$ are respectively the hot gas density profiles of the $f(R)$ halo whose dynamical mass is $M_\ast$ and the $\Lambda$CDM halo whose true mass is $M_\ast$. Cluster observables, such as the SZ Compton-y parameter $Y_{\rm SZ}$, its X-ray counterpart $Y_{\rm X}$ and the X-ray luminosity $L_{\rm X}$, are usually integrated quantities that can be schematically written as
\begin{equation}
Y = \int^r_0{\rm d}r'4\pi r'^2\left(\rho_{\rm gas}(r')\right)^a\left(T_{\rm gas}(r')\right)^b,
\end{equation}
where $a,b$ are power indices. Therefore, from the above equation, $Y^{f(R)}$ and $Y^{\Lambda\rm CDM}$ satisfy
\begin{eqnarray}\label{eq:scaling_rescaling}
Y^{f(R)} &=& \int^r_0{\rm d}r'4\pi r'^2\left(\rho^{f(R)}_{\rm gas}(r')\right)^a\left(T^{f(R)}_{\rm gas}(r')\right)^b\nonumber\\ 
&=& \left[\frac{M^{f(R)}_{\rm lens}}{M^{f(R)}_{\rm dyn}}\right]^a\int^r_0{\rm d}r'4\pi r'^2\left(\rho^{\Lambda{\rm CDM}}_{\rm gas}(r')\right)^a\left(T^{\Lambda{\rm CDM}}_{\rm gas}(r')\right)^b = \left[\frac{M^{f(R)}_{\rm true}}{M^{f(R)}_{\rm dyn}}\right]^aY^{\Lambda{\rm CDM}},
\end{eqnarray}
where we have used the fact that for a $f(R)$ halo with dynamical mass equal to $M_\ast$ and a $\Lambda$CDM halo with true mass equal to $M_\ast$ the temperature profiles $T_{\rm gas}(r)$ should be identical. Eq.~(\ref{eq:scaling_rescaling}) is useful because using it one can directly obtain the $f(R)$ cluster scaling relation, $Y^{f(R)}\left(M^{f(R)}_{\rm dyn}\right)$ from the corresponding $\Lambda$CDM scaling relation $Y^{\Lambda{\rm CDM}}\left(M=M_{\rm dyn}^{f(R)}\right)$ (which are usually better known or easier to obtain), without having to run large suites of hydro simulations in $f(R)$ gravity \citep{He:2015mva}. \citet{Mitchell:2018qrg} proposes a framework to use these `rescaled' $f(R)$ cluster scaling relations to relate cluster observables from X-ray and SZ surveys to cluster masses, and put constraints using the abundance of clusters.

\subsection{Voids in galaxy surveys}
\label{sec:universal_tests2}

The term `cosmic void' is broadly used to refer to large (typically from $5-100\ {\rm Mpc}$) underdense regions of the universe, characterized by mass outflow from their centers onto the higher-density mass-accreting filaments and walls that define their boundaries \citep{2004MNRAS.350..517S, 2012ApJ...761...44S, 2017MNRAS.465..746S,  2016MNRAS.456.4425C}. Being the parts of the cosmic web with the lowest density, they are regions where the screening mechanisms are expected to be the least efficient; this gives void-related observations a promising potential to test gravity on astrophysical/cosmological scales. 

\begin{figure*}
  \subfigure[]
  {
    \includegraphics[width=0.5\textwidth]{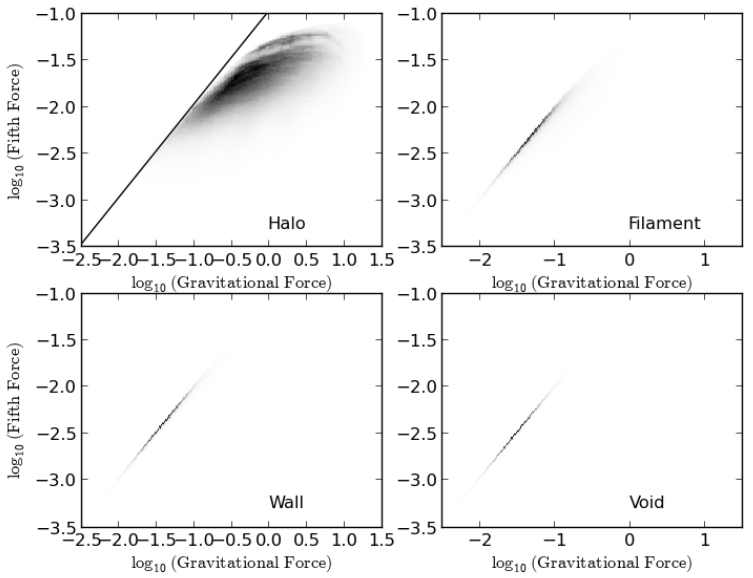}
  }
  \subfigure[]
  {
    \includegraphics[width=0.5\textwidth]{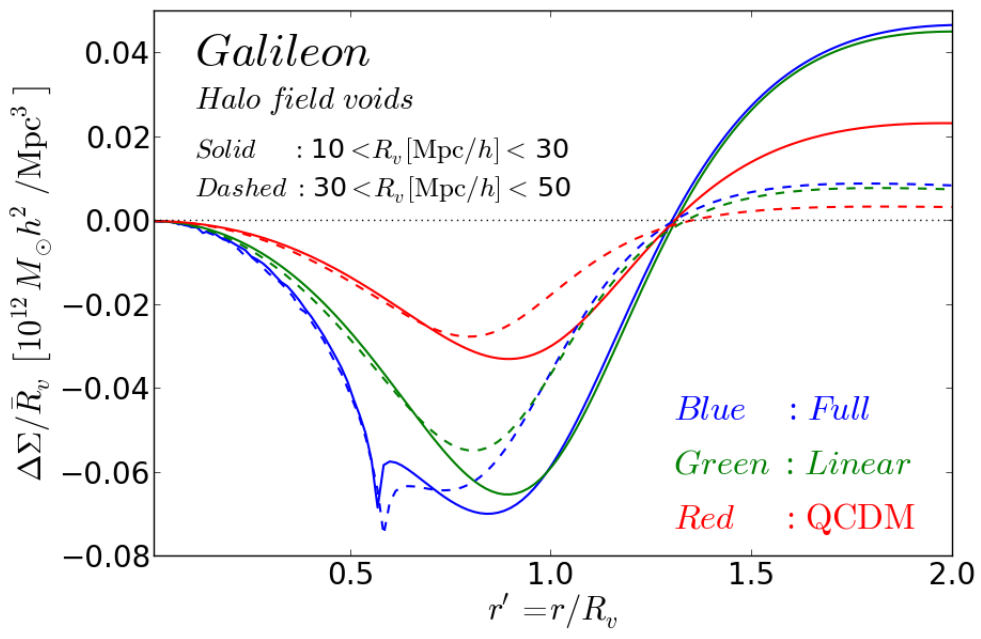}
  }
  \caption{(a) Comparison of the fifth-force (y-axis) and normal GR (x-axis) contributions to the total force felt at particle locations in N-body simulations of the nDGP model. The four panels split the particles by the type of cosmic web structure (haloes, filaments, walls and voids, as identified by the ORIGAMI algorithm \citep{Falck:2012}) where each particle lies. The solid line marks the unscreened linear expectation, which is followed in filaments/walls/voids; in the higher density haloes, the fifth force is suppressed. Figure reproduced from \citet{2014JCAP...07..058F}. (b) Differential surface mass density (which is related to the lensing shear) of spherically symmetric stacks of voids in the Cubic Galileon model. Relative to the GR scenario (red; dubbed ${\rm QCDM}$ because the background is not $\Lambda{\rm CDM}$), the full model result (blue) displays roughly the same enhancement as a linearized unscreened version of it (green). This indicates the screening mechanism is not at play and thus that the fifth force can have prominent observational signatures. Figure reproduced from \citet{2015JCAP...08..028B}.}
  \label{fig:voids}
\end{figure*}

Despite the existence of a few simpler analytic attempts to describe voids in modified gravity \citep{2009arXiv0911.1829M, 2013MNRAS.431..749C, 2015MNRAS.450.3319L, 2017PhRvD..95b4018V}, N-body simulations are still the best available tool to extract void properties such as their profile, abundance, dynamics and screening efficiency (cf.~the left panel of Fig.~\ref{fig:voids}) \citep{2011MNRAS.411.2615L, 2012MNRAS.421.3481L, 2014JCAP...07..058F, 2014PhRvL.112y1302H, 2015MNRAS.451.1036C,  2015JCAP...07..049F, 2016MNRAS.455.3075P, 2018MNRAS.475.3262F, Shao:2019wit}; see also \citet{Cai:2019re} for a recent review.

In practice, a first challenge facing the utility of voids as a cosmological probe lies in how to identify them from the complicated cosmic web. This has opened many possibilities, as a result of which there is no unique or widely-agreed way to define them. For example, voids can be defined according to the type and number density of the tracer field (whether they are identified from a matter field, galaxies, clusters, etc.), and their dimensionality (whether they are identified from a 3D or 2D projected tracer field). On top of this, due to their generally non-regular shapes, there exists a variety of void-finding algorithms to identify them from a given tracer field. These complexities can be portrayed as a source of confusion in void-related works, but one should appreciate also the enrichment of the types of analyses that can be done as a consequence. For example, different void definitions can be more or less sensitive to specific modified gravity signatures, as investigated recently in \citet{2018MNRAS.476.3195C, Paillas2018}.

In the following, we briefly review some of the main recent developments on tests of gravity with cosmic voids:

\begin{itemize}

\item {\it Weak lensing by 3D voids.} The lensing signal from individual voids is too weak to be systematically detected with the currently available precision \citep{1999MNRAS.309..465A}, and so virtually all studies performed to date focus on the signal obtained from stacking many such void lensing profiles, which increases the signal-to-noise \citep{2013ApJ...762L..20K, 2013MNRAS.432.1021H, 2014MNRAS.440.2922M, 2015MNRAS.454.3357C}. Modified gravity theories can impact the observed lensing profiles in two main ways. First, if the growth of structure is boosted, then voids become emptier as the mass outflows towards their surrounding walls/filaments become more efficient \citep{2015MNRAS.451.4215Z, 2015MNRAS.451.1036C}. Second, if the relation between mass and the lensing potential is also modified \citep{2015JCAP...08..028B, 2018PhRvD..98b3511B}, then the lensing signal is affected since photons follow a modified geodesic equation. The right panel of Fig.~\ref{fig:voids} shows how in the Cubic Galileon model the screening is inefficient and a strong signature of the fifth force is imprinted in the void lensing predictions.

\item {\it Weak lensing by 2D voids.} A way to increase the signal-to-noise in lensing observations by underdense regions is to focus the analysis on 2D voids, or equivalently lines-of-sight that are predominantly devoid of structure. This type of analysis has been pioneered by the DES collaboration (who use galaxy troughs \citep{2016MNRAS.455.3367G} or density-split-statistics \citep{2017arXiv171005045G, 2017arXiv171005162F}), in which the lines-of-sight are discriminated by their projected photometric galaxy count. The first few studies of this lensing signal in modified gravity were carried out in \citet{2016MNRAS.459.2762H,2018MNRAS.476.3195C} for $f(R)$ and \citet{2017JCAP...02..031B} for DGP gravity. Recently, \citet{2018MNRAS.476.3195C} proposed two more 2D void definitions: 2D spherical underdensity voids (SVF2; similar to 3D spherical voids but in 2D) and tunnels (circumcicles of triangular 2D Delaunay tessellation cells). These void finders are found to produce stronger signals compared to troughs and 3D voids, and are also better discriminators of GR, $f(R)$ and DGP gravity \citep{2018MNRAS.476.3195C,Paillas2018}. The lensing signal produced by 2D voids found {directly in lensing maps (using peaks as tracers)} allows for even stronger signals and tests of gravity \citep{Davies:2018jpm, 2019arXiv190706657D} because the lensing map is a more direct tracer field of the underlying matter field.

\item  {\it The ISW effect and voids.} The time evolution of the lensing potential in and around voids can be probed via the ISW effect and this is also generically affected by modified gravity. The stacking of CMB maps on top of voids found in foreground galaxy distributions results in {\it cold-spots} that indicate that the potential has been getting shallower with cosmic time (e.g.~\citealt{2008ApJ...683L..99G, 2016A&A...594A..21P}). The modified signal arises from a combination of the same two effects discussed for void lensing above: both modified dynamics (e.g., as in the $f(R)$ model studied in \citet{2014MNRAS.439.2978C}) and modified lensing potentials can contribute to different time-evolutions of the gravitational potential inside voids. For instance, a stronger gravity can work to slow down the decay of the gravitational potential (or even make it grow), thereby leading to a weaker amplitude of the cold spots (or turning them into hot spots). 

\item {\it Growth rate measurements with RSD.} The average radial tracer number density profiles for a large enough number of voids is expected to be spherically symmetric in real space. In redshift space, however, for galaxies used as tracers, peculiar velocities work to distort these profiles in the line-of-sight direction. The degree of anisotropy allows us to put bounds on the growth rate of structure, and consequently, test gravity. For example, \citet{2015JCAP...11..036H, 2016PhRvL.117i1302H, 2017JCAP...07..014H, 2017A&A...607A..54H, 2019MNRAS.483.3472N, 2019PhRvD.100b3504N} developed and applied methods to measure the cross correlation between galaxies and voids in galaxy catalogues to infer  the growth rate of structure (see also \citealt{Cai:2016jek}). In the context of testing modified gravity theories with screening, there is an interest in focusing on void-galaxy cross-correlations (compared to galaxy auto-correlations) that comes from targeting galaxies that lie in the lowest density, less screened regions of the universe. 

\item {\it Void abundances.} While void abundance is in principle a very sensitive probe of modified gravity \citep{2015MNRAS.451.1036C, Li:2011}, its use in cosmological tests is far less popular. A major reason for this is that, in observations, 3D voids are mostly found from a biased tracer field that needs to be also fully understood in order to make accurate predictions of void abundance. Indeed, void abundance is effectively fixed by the number density and clustering of the tracers \citep{2018MNRAS.476.3195C, Paillas2018}.  One interesting work around this difficulty is that proposed in \citet{Davies:2018jpm} to use lensing peaks directly as 2D void tracers. \citet{Davies:2018jpm} showed that, when identified in this way, voids are self-similar in their abundances at various lensing peak signal-noise-ratio values.  More recenlty, \citet{2019arXiv190706657D} verified that 2D voids identified using lensing peaks can indeed be a powerful discriminator of GR and DGP gravity. 

\item {\it In-and-out of void comparisons.} Another powerful test of gravity involves comparing the dynamics and evolution of low-mass objects inside voids (where they would be unscreened in chameleon and symmetron theories) with similar (but screened) objects in higher-density regions \citep{Shi, Li:2011b}. {Unexpected strong differences in properties of objects from these different environments can be a signature of departures from GR.} The study of velocity profiles or rotation curves of screened and unscreened galaxies inside and outside voids is an example of one such tests \citep{Arnold:2016arp}. Another way to search for the unscreened nature of objects in voids is using the marked two-point correlation function, which is similar to the usual correlation function but giving selected objects (e.g., those in voids) a higher weight (see~e.g.,~\citealt{Lombriser:2015axa,Valogiannis:2017yxm,Hernandez-Aguayo:2018yrp,Armijo:2018urs}). 

\end{itemize}

Finally, we mention in passing also Minkowski functionals. Although not direct probes of cosmic voids, these are useful descriptions of the complex morphology of the cosmic web (of which voids occupy most of the volume) that can be used to find interesting signatures from modified gravity effects; see e.g.~\citet{2017PhRvL.118r1301F} who found using $N$-body simulations that Minkowski functionals measured in the three-dimensional matter distribution can discriminate with high significance models such as GR, $f(R)$ and nDGP.


\subsection{Observational screening maps}
\label{sec:screening_maps}

\noindent Testing modified gravity by means of astrophysical objects requires identifying a difference between screened and unscreened subsamples, which in turn requires an observational proxy for the degree of screening. As described in Sec.~\ref{sec:searching}, in thin-shell-screening models the background value of the scalar field sets a threshold in Newtonian potential $|\Phi|$ which marks the onset of screening, as measured for example by a difference between lensing and dynamical masses. This has been verified in modified gravity simulations, where alternative proxies such as the distance to massive nearby neighbours have also been shown to be effective \citep{Shi, Haas, Cabre:2012tq}. In particular, it was shown in \citet{Cabre:2012tq} that the Newtonian potential generated by mass within one Compton wavelength of the scalar field is a good proxy for the degree of screening in thin-shell theories. The degree of other types of screening is also expected to be correlated with simple gravitational variables (e.g. acceleration $a$ for kinetic and curvature $K$ for Vainshtein; \citealt{Khoury:2013tda,Joyce_review}), although in these cases the lack of knowledge of two-body solutions makes exact predictions difficult. The environmental contribution to non-thin-shell screening may be much lower, especially if the test object is much smaller than the screening radius. Nevertheless, to first order the task of determining the degree of screening in a given theory is approximately solved by mapping out the Newtonian potential and its derivatives.

\begin{figure*}
  \subfigure[Newtonian potential]
  {
    \includegraphics[width=0.3\textwidth]{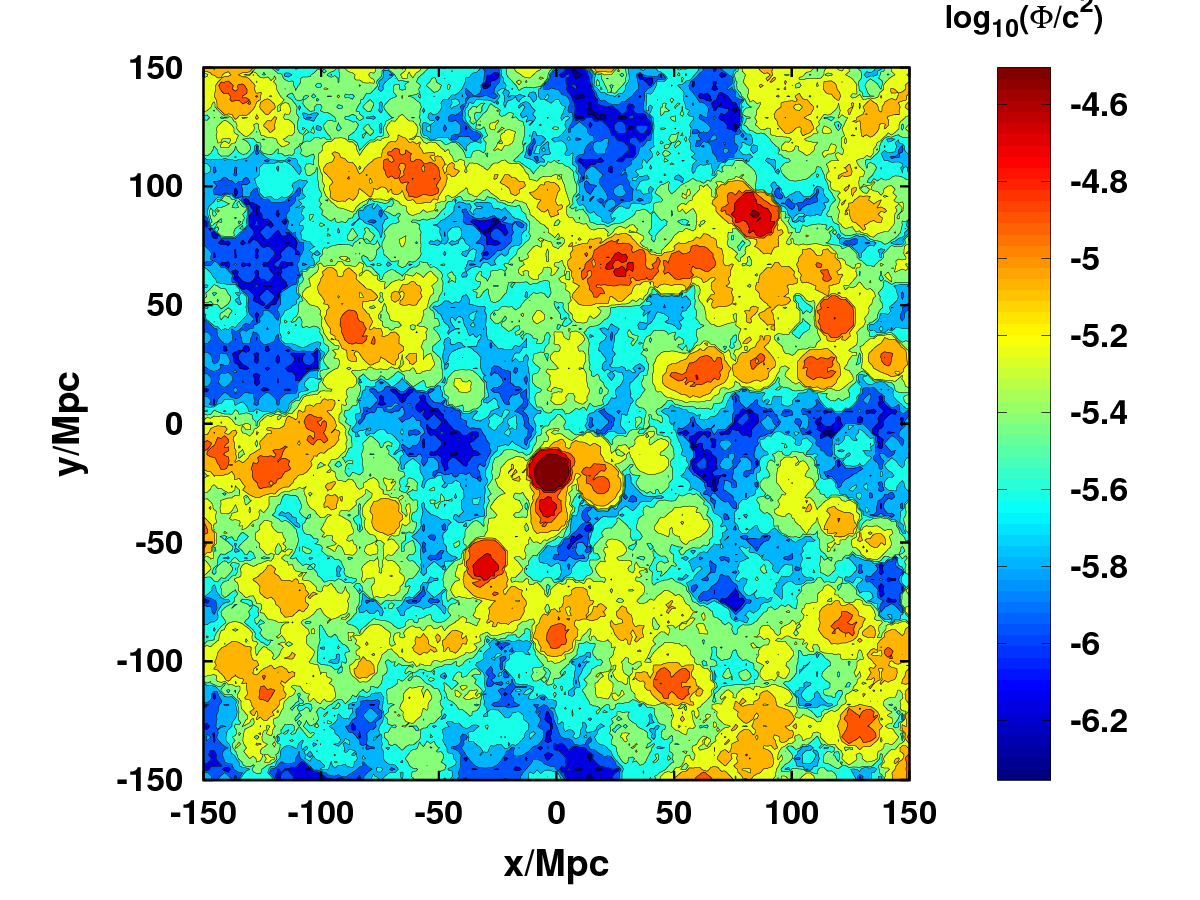}
    \label{fig:map_V}
  }
  \subfigure[Acceleration]
  {
    \includegraphics[width=0.3\textwidth]{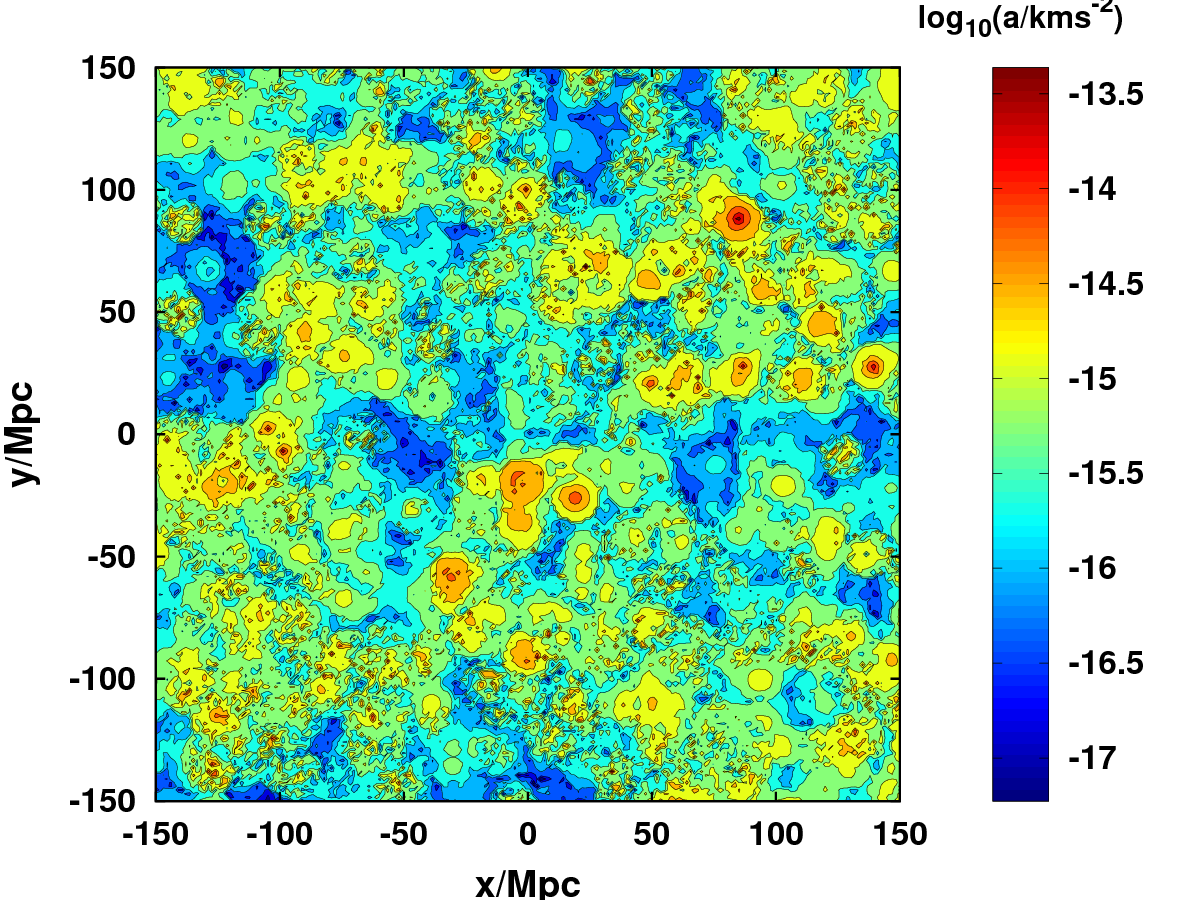}
    \label{fig:map_a}
  }
  \subfigure[Curvature]
  {
    \includegraphics[width=0.3\textwidth]{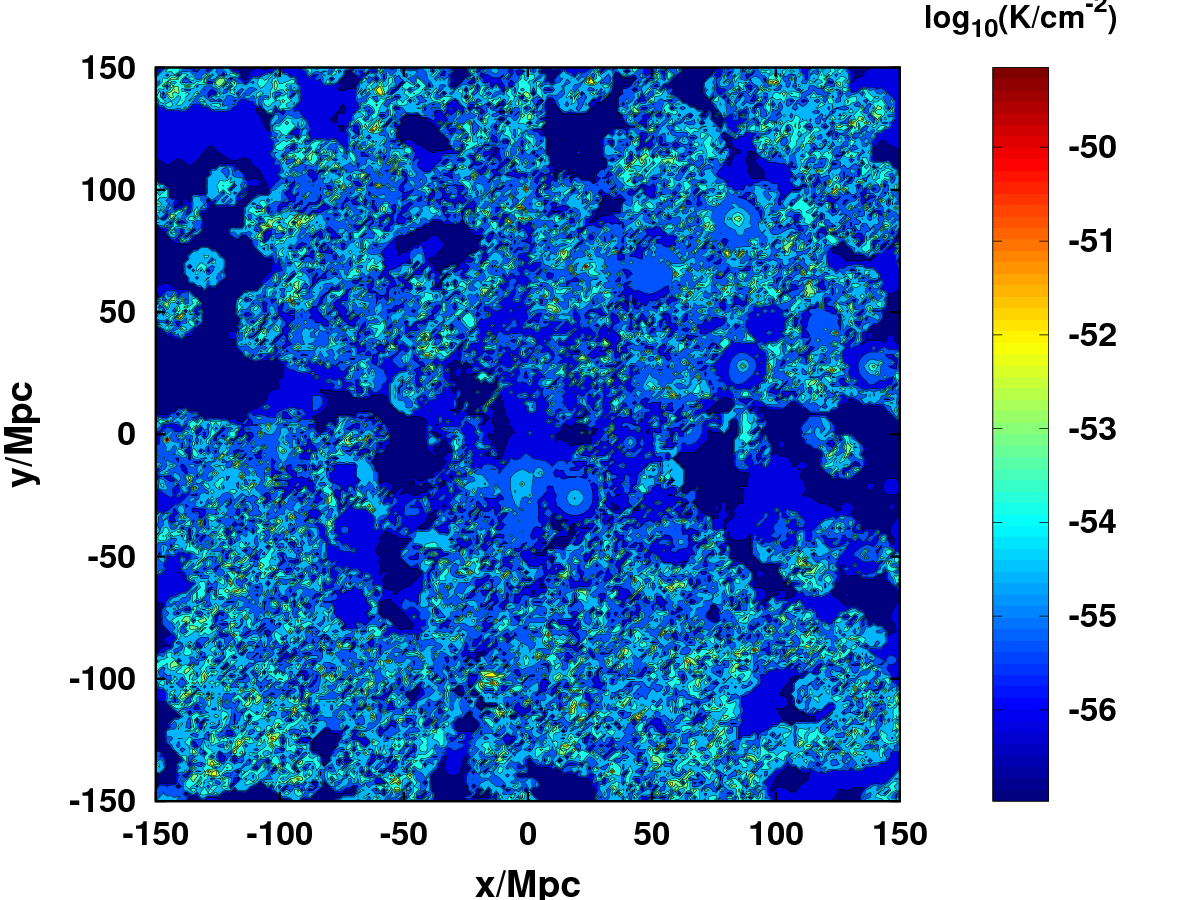}
    \label{fig:map_c}
  }
  \caption{Maps of environmental Newtonian potential $|\Phi|$, acceleration $a$ and curvature $K$ across a $300 \: \text{Mpc} \times 300 \: \text{Mpc}$ slice of the local universe. The Milky Way is located at $x=y=0$. Reproduced from \citet{Desmond}.}
  \label{fig:Maps}
\end{figure*}

For a given object, $\Phi$, $a$ and $K$ receive both an internal contribution, from the object itself, and an external contribution from surrounding mass. Calculating the former requires the mass distribution of the object, but to first order is given simply by $|\Phi|=GM/R$, $a=GM/R^2$ and $K\simeq GM/R^3$, where $M$ is the mass of the object and $R$ its size. The environmental components are more difficult to calculate as they require knowledge of the object's environment. A first estimate may be obtained by summing the contributions from mass associated with observed light within a Compton wavelength of the scalar field from the test point. This may be achieved from a given galaxy catalogue either by estimating group masses from the velocity dispersions of their constituent galaxies \citep{Cabre:2012tq}, or assigning N-body halos to individual galaxies by means of their luminosity in a manner consistent with galaxy--galaxy clustering \citep{Desmond}. However, a significant fraction of the universe's mass is not associated with currently-observable light, either in halos hosting galaxies too faint to see or in diffuse mass not associated with halos at all. The former contribution can be estimated using the halo distribution in N-body simulations, and the latter from a reconstruction of the smooth density field from galaxy number densities and redshifts (given a bias model and fiducial cosmology) using an algorithm such as BORG \citep{Jasche, Jasche_Wandelt1, Jasche_Wandelt2, Lavaux}, or constrained simulations such as ELUCID \citep{Wang2014,Wang2016} or CLUES \citep{Sorce:2015yna}. A full pipeline is constructed in \citet{Desmond}, allowing the degree of environmental screening to be calculated for any object within $\sim 200 \: h^{-1}$ Mpc (Fig.~\ref{fig:Maps}), and the associated code is publicly available on the project website.\footnote{\url{https://www.novelprobes.org/codes}} Alternatively, one can also predict the environmental and self screening of halos by running a modified gravity solver directly on the reconstructed density field to calculate the fifth force \citep{Shao:2019wit} -- this is a model-dependent approach, but the cost is low because for each model and parameter choice one only needs to do the calculation once rather than hundreds of times as in N-body simulations.

The power of any test relying on the distinction between screened and unscreened galaxies is limited by the uncertainty in the degree of environmental screening. This has four main contributions, listed here in approximately decreasing order of importance \citep{Desmond}:

\begin{enumerate}

\item{} Uncertainties in the distribution and masses of halos unassociated with an observable galaxy. This is a factor of a few for $\Phi$ but may be an order of magnitude or more for $a$ and $K$, which depend more sensitively on mass close to the test point. This depends on the magnitude limit of the initial galaxy catalogue as well as the algorithm used to account for unseen mass. (this may not be problematic for non-thin-shell models if environmental screening is suppressed.)

\item{} Uncertainties in the mass distribution of the smooth density field, which accounts for mass outside of halos that are well-resolved in an N-body simulation. The posterior probability distribution for this mass distribution is calculated for example in \citet{Lavaux}, allowing this uncertainty to be straightforwardly propagated into screening maps. This contribution is relatively most significant in low-density regions with few nearby halos, where other contributions to the gravitational field are small.

\item{} Uncertainties in the masses and concentrations of halos hosting galaxies included in the basic galaxy catalogue. In \citet{Desmond} these are calculated using inverse abundance matching, which specifies a range of possible halo properties for a galaxy of given luminosity, even for fixed values of the model parameters which are themselves uncertain.

\item{} Uncertainties in the magnitude and position of the source galaxies. This purely statistical error is subdominant to the foregoing systematic issues.

\end{enumerate}

The lowest background field value that is testable is set by the minimum degree of environmental screening at which a statistically significant sample of objects can be compiled. The most unscreened galaxies are located in the lowest-density regions of the universe, where not only is the number of halos within the Compton wavelength $\lambda_C$ minimised, but also the galaxy distribution indicates little mass situated outside halos in the smooth density field. The analyses of both \citet{Cabre:2012tq} and \citet{Desmond} indicate that objects exist in potentials at least as low as $|\Phi| = 10^{-7}$ for a window radius $\lambda_C \simeq 1$ Mpc, allowing the self-screening parameter $\chi$ to be probed below this level. This gives astrophysical tests the potential for significantly greater constraining power than cosmological tests.

\subsection{Tests of thin-shell screening}
\label{sec:chameleon_tests}

\subsubsection{Stellar evolution}
\label{sec:stars}




Stars are important probes of chameleon/symmetron theories precisely because they have Newtonian potentials that range from $10^{-6}$ (main-sequence) to $10^{-7}$--$10^{-8}$ (post-main-sequence). In this subsection we will discuss the novel effects that can be exhibited in stars under modified gravity. Stars in dense galaxies are subject to environmental screening so the effects described in this section must be tested using unscreened galaxies. As discussed in section \ref{sec:searching}, these are typically dwarf galaxies in voids. These can be found in observational data sets using the maps described in \ref{sec:screening_maps}. We begin by showing how hydrostatic equilibrium is altered, and then present observable effects of this on stellar luminosity, lifetime and pulsations.


A star is a complex system where many different areas of physics play an important role, including nuclear physics, atomic physics, thermodynamics and convection. Despite this, there is only one stellar structure equation where gravitational physics is important, the hydrostatic equilibrium equation, which in GR is given by
\begin{equation}\label{eq:HSE}
\frac{\dd P}{\dd r} = -\frac{GM(r)\rho(r)}{r^2}.
\end{equation}
This equation tells us the pressure profile that a star must assume if the star is to remain in equilibrium, i.e. if the inward gravitational force is to be balanced by the outward pressure, from nuclear burning for example. In the case of chameleons/symmetrons, this is modified to include the fifth force so that one has
\begin{equation}\label{eq:HSE_MG}
\frac{\dd P}{\dd r} = -\frac{GM(r)\rho(r)}{r^2}\left[1+2\alpha^2\left(1-\frac{M(\rs)}{M(r)}\right)\Theta(r-\rs)\right],
\end{equation}
where $\Theta(x)$ is the Heaviside step function; this ensures that the fifth force is only operative in the region exterior to the screening radius (see \eqref{eq:f5chamshell}). Calculating the new properties of stars in chameleon gravity is tantamount to solving this equation simultaneously with the equations describing stellar structure and energy production.


The new term in the hydrostatic equilibrium equation \eqref{eq:HSE_MG} essentially increases the gravitational force in the region outside the screening radius, which has several important consequences. Consider two stars of equal mass $M$, one screened and one unscreened. The unscreened star feels a stronger inward gravitational force and must therefore burn more nuclear fuel per unit time in order to prevent gravitational collapse. This suggests that the unscreened star will deplete its fuel reserves faster than the screened star and will therefore have a shorter lifetime. Furthermore, the increased rate of nuclear burning results in the star being more luminous. To make this more quantitative, consider the extreme case where the star is fully unscreened so that $\rs=0$ and one therefore has $G\rightarrow(1+2\alpha^2)G$. Simple dimensional analysis arguments show that the luminosity of low mass stars scales as $L\propto G^4$ at fixed mass for low-mass main-sequence stars and that $L\propto G$ for high-mass stars \citep{Davis:2011qf}. In the former case, this scaling law arises because low-mass stars are supported by thermodynamic pressure (i.e. the ideal gas law) so that $P\propto \rho T$ and in the latter because high-mass stars are supported by radiation pressure $P\propto T^4$. The ratio of the luminosity of the unscreened to the screened star is then
\begin{align}
\frac{L_{\rm unscreened}}{L_{\rm screened}}=
  \begin{cases}
            (1+2\alpha^2)^4,  &\quad \textrm{low-mass main-sequence}\\
   (1+2\alpha^2), & \quad \textrm{high-mass main-sequence}
  \end{cases}.
\end{align}
Evidently, low-mass stars are more susceptible to the effects of modified gravity than high-mass stars. One explanation for this is that high mass stars need to absorb more of the extra radiation being produced by the enhanced gravitational force in order to support themselves. A more thorough (and technical) analytic treatment of stellar structure in chameleon gravity is given in \citet{Davis:2011qf,Sakstein:2015oqa}.

\begin{figure}[ht]
\centering
{\includegraphics[width=0.5\textwidth]{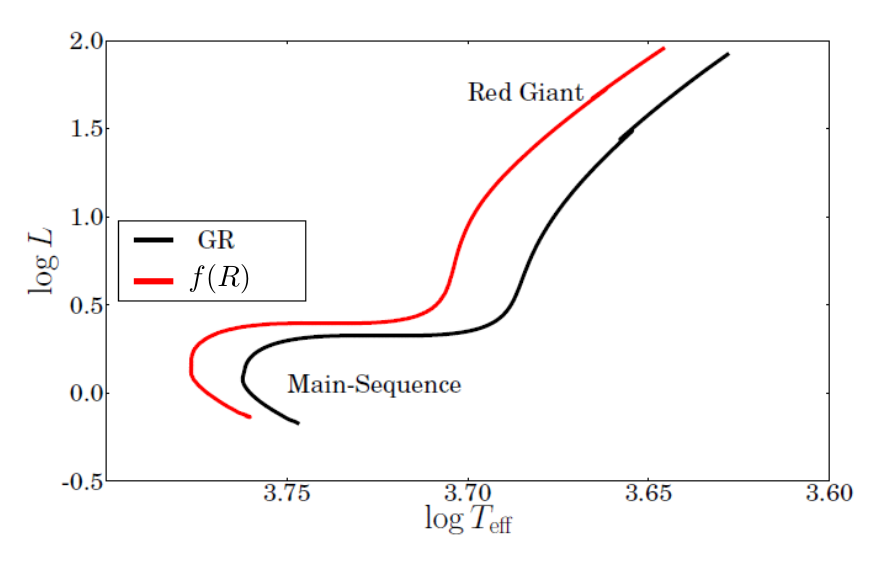}}
\caption{The tracks in the color-magnitude diagram for a solar mass and metallicity star in GR (black) and Hu-Sawicki $f(R)$ gravity (red) with $f_{R0}=10^{-6}$. Here, $L$ is normalized to the solar luminosity and $T_\text{eff}$ is measured in Kelvin. }\label{fig:MESA}
\end{figure}

In practice, the complex nature of stars means that a numerical treatment is necessary in order to produce realistic stellar models that can make realistic observational predictions. For this reason, the publicly available stellar structure code MESA \citep{Paxton:2010ji,Paxton:2013pj} has been modified by \citet{Chang:2010xh,Davis:2011qf} so that the hydrostatic equilibrium equation to be solved for is equation \eqref{eq:HSE_MG} rather than \eqref{eq:HSE}. This is achieved using an iterative procedure: given an initial stellar model, the screening radius is found by solving equation \eqref{eq:rseq}. MESA then uses this to solve the modified hydrostatic equilibrium equation in conjunction with all of the other stellar structure equations in order to find the stellar model resulting from this degree of screening. This is then used to calculate a new screening radius, and the procedure repeated until convergence is reached. \citet{Chang:2010xh} investigated the validity of this approximation by numerically solving for the scalar profile using a Gauss-Seidel algorithm in conjunction with the other stellar structure equations and found excellent agreement between the exact and approximate solutions. For this reason, modern implementations have used the approximation. As an example of the effects of modified gravity, a color-magnitude diagram (or Hertzprung-Russell track) for a solar mass and metallicity ($Z=0.02$) star in Hu-Sawicki $f(R)$ gravity ($2\alpha^2=1/3$) with $f_{R0}=10^{-6}$ is compared to the GR case in Fig. \ref{fig:MESA}. One can see that the simple qualitative predictions made above hold true upon full numerical simulation: the $f(R)$ star is indeed hotter and more luminous than the GR star. If one looks at the age of the stars when they exit the main sequence one also finds that the $f(R)$ star is younger. Note that using $f_{R0}=10^{-6}$ implies that the galaxy hosting the star is unscreened. Chameleon (and similar) searches typically focus on unscreened galaxies (see section \ref{sec:searching}). If one is interested in screened galaxies then one should adjust the value of $f_{R0}$ appropriately to account for environmental screening.

Given that stars are generally brighter in modified gravity, one would expect dwarf galaxies of fixed stellar mass in voids to be brighter than their screened cluster counterparts \citep{Davis:2011qf}. Further, low-mass stars are more affected and the na\"{i}ve expectation is that void galaxies should also be redder in color. To date, no numerical simulations of galactic properties including the effect of modified gravity on stars have been performed. There are many competing effects that could alter this expectation, for example one would expect a larger population of post-main-sequence stars (which are brighter than main-sequence stars and more affected by modified gravity) since low-mass stars would exit the main-sequence sooner. The precise effects of modified gravity on the initial mass function and star formation rate are unknown due to a lack of investigation.

Going beyond hydrostatic equilibrium, the dynamics of small perturbations $\vec{\delta r}$ are governed by the momentum equation
\begin{equation}\label{eq:momentum_eq}
\ddot{\vec{\delta r}}=-\frac{1}{\rho}\frac{\dd P}{\dd r} +\vec{a},
\end{equation}
where $ \vec{a}$ is the force per unit mass. In GR this is $\vec{a}=-\vec{\nabla}\Phi$, but in chameleon and symmetron gravity one instead has $\vec{a}=-\vec{\nabla}\Phi-\alpha\vec{\nabla}\phi$ so that modified gravity also affects the dynamics of stellar oscillations. Stellar oscillations have proven to be valuable tools for testing chameleon gravity \citep{Jain:2012tn,Sakstein:2013pda,Sakstein:2015oqa,Sakstein:2016lyj}. Indeed, the Sun oscillates in over $10^7$ different modes and pulsating stars where the oscillations are driven by some dynamical forcing\footnote{The driving is due to non-gravitational effects and hence are not sensitive to the theory of gravity.} such as Cepheids and RR Lyrae stars can be used as distance indicators \citep{Freedman:2010xv}. Perturbing \eqref{eq:momentum_eq} and the other stellar structure equations, one finds that the frequency of linear radial adiabatic oscillations 
\begin{equation}
\omega^2 \sim \frac{GM}{R^3}.
\end{equation}
Changing the value of $G$ then causes the pulsation period $\Pi$ changes by a fractional amount
\begin{equation}\label{eq:delta_period}
\frac{\Delta \Pi}{\Pi}=-\alpha Q,
\end{equation}
where $Q$ is the scalar charge defined in Eq.~\eqref{eq:Qcham}. This is $\mathcal{O}(1)$ for unscreened stars.

\subsubsection{Distance indicators}
\label{sec:distance_indicators}



Distance indicators are objects which have some known intrinsic property that allows us to calculate their absolute distance. For example, the luminosity distance of an object at non-cosmological distance is given by $d_L^2=\frac{L}{4\pi F}$ for measured flux $F$. Since we cannot measure $L$, only objects for which it is known by some other means (either from theoretical calculations or empirical measurements and calibrations in the local neighborhood) can be used to find the distances to their host galaxies. Distance indicators are invaluable for testing chameleon theories since they are both sensitive to gravity and can be observed in void dwarf galaxies which are least likely to be screened. 

The principle behind distance indicator tests of chameleons is the following: suppose that one attempts to measure the distance to an unscreened void dwarf galaxy using two different distance indicators, one screened (or insensitive to the theory of gravity) and the other unscreened. If the theory of gravity is correct, the two estimates will agree; if not, the estimates will disagree. This is because the formula used to calculate the distance to the unscreened object is incorrect: either it has assumed GR or has been calibrated empirically in the local (screened) neighborhood. As an example, suppose that in GR the luminosity of some object is known to be constant, i.e. it is a standard candle. If unscreened objects are more luminous, then, given a measured flux, the application of the luminosity distance formula to this object will underestimate the distance since one would have used too low a luminosity. 


Three different distance indicators that have been used to constrain chameleon theories are:

\begin{itemize}

\item {\bf Cepheid variable stars}: Cepheid variable stars are post-main-sequence stars with progenitors of mass $3\lsim M/M_\odot \lsim 10$ that have evolved off the main-sequence. Semi-convective processes (convection driven by gradients in the chemical composition) cause them to execute what are referred to as \emph{blue loops} in the color-magnitude diagram where their temperature increases at fixed luminosity. During this looping phase, they enter a narrow vertical strip where they are unstable to pulsations driven by the $\kappa$-mechanism \citep{cox1980theory}. (A helium ionization layer dams up energy because small compressions [density increases] result in the helium becoming doubly ionized rather than increasing the outward pressure gradient.) During this phase, the star pulsates with a period--luminosity relation
\begin{equation}
M_V=a\log \Pi + b\log(B-V) + c, 
\end{equation}
where $a\approx-3$ \citep{Freedman:2010xv} and $\Pi$ is the pulsation period. Using equation \eqref{eq:delta_period}, one finds that applying this to unscreened Cepheids will underestimate the distance by \citep{Jain:2012tn}:
\begin{equation}\label{eq:deltad_cepheid}
\frac{\Delta d}{d} = -0.6 \alpha Q.
\end{equation}
One can make quantitative predictions for individual Cepheids by calculating the average value of 
$2 \alpha Q$ using MESA profiles (see \citealt{Jain:2012tn} for the technical details).
%

\item {\bf Tip of the red giant branch stars}: Stars of mass $1\lsim M/M_\odot \lsim 2$ do not execute blue loops. Instead, they ascend the red giant branch (RGB) where their cores become hotter, denser, and more luminous until the central conditions are such that the triple-$\alpha$ process can begin, at which point helium ignition begins explosively and the star moves very rapidly onto the horizontal branch. This leaves a visible discontinuity in the I-band magnitude $I=4.0\pm0.1$ when the star reaches the tip of the red giant branch (TRGB) \citep{Freedman:2010xv}. The discontinuity is almost independent of mass, with a small spread due to variations in the core masses of RGB stars and metallicity effects. For this reason, the TRGB is a standard candle. During its ascent of the RGB, the star's luminosity is due entirely to a thin hydrogen burning shell around the core and so whether or not the TRGB is sensitive to modified gravity depends on whether the core is unscreened. One finds this to be the case if $\chi\gsim 10^{-6}$. When this happens, the core temperature increases more rapidly than in GR and the helium flash begins earlier, i.e. at a lower luminosity (see Appendix B of \citealt{Jain:2012tn} for technical details). For this reason, unscreened TRGB indicators over-estimate the luminosity distance. When the core is screened ($\chi\lsim 10^{-6}$) their calculated distances give the GR result. The complex structure of RGB stars means that it is best to compute the decrease in the TRGB luminosity using MESA. 


\item {\bf Water Masers}: A third distance indicator that has found some use in testing chameleon theories is based on the distances to water masers \citep{megamasers}. These are clouds of H$_{\rm 2}$O gas in Keplerian orbits in the accretion disks of black holes. Active galactic nuclei cause a population inversion in the clouds resulting in stimulated emission of microwave-frequency radiation. Using this, a simultaneous measurement of the radial velocity, centripetal acceleration, angle on the sky, and inclination of the orbital plane is possible, enabling a geometric distance estimate. As the central regions of galaxies are highly screened, water masers are screened distance indicators.

\end{itemize}


{\bf TRGB vs. Water Masers}: Since TRGB distances are unscreened when $\chi\gsim 10^{-6}$, one can compare their distance estimates with maser estimates to constrain this parameter range. This was done by \citet{Jain:2012tn} for the one galaxy (NGC 4258, a spiral galaxy unscreened when $\chi>10^{-6}$) where there are simultaneous measurements of both TRGB and water maser distances. These both give $7.2$ Mpc within errors, which rules out $\chi \gtrsim 10^{-6}$ with high significance. A later measurement revised the maser distance to $7.6$ Mpc \citep{Humphreys:2013eja} but this is still consistent with the TRGB distance within errors.

\begin{figure}[ht]
\centering
\includegraphics[width=0.5\textwidth]{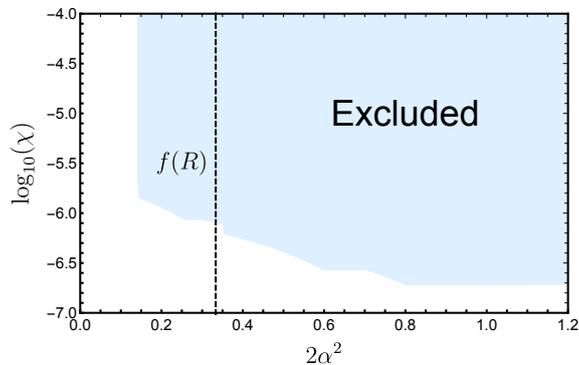}
\caption{The excluded region in the $\chi$--$\alpha$ plane derived by comparing Cepheid and TRGB distance estimates to 22 unscreened galaxies. The black dashed line corresponds to $f(R)$ theories. Recall that $f_{R0}=2\chi/3$. Reproduced from \citet{Jain:2012tn}.}\label{fig:ceph_constraints}
\end{figure}

{\bf Cepheids vs. TRGB}: When $\chi<10^{-6}$ TRGB distances are screened and so one can obtain new constraints by comparing TRGB with Cepheid distances. This is achieved using $\Delta d = d_{\rm Cepheid}-d_{\rm TRGB}$ and $d=d_{\rm TRGB}$ in equation \eqref{eq:deltad_cepheid}. This analysis was performed by \citet{Jain:2012tn} for a sample of 22 unscreened dwarf galaxies in voids (selected using the screening map of \citet{Cabre:2012tq}), producing the constraints shown in Fig. \ref{fig:ceph_constraints}. In particular, $f_{R0}>3\times10^{-7}$ is ruled out at the 68\% confidence level. An updated version of this test using additional data, and generalised to screening mechanisms beyond chameleon, is presented in \citet{Desmond_H0} (figure 5). It is also shown there that the modification to the distance ladder caused by screened fifth forces is able to reduce the Hubble tension.

\vspace{3mm}

\noindent Further discussion of the effect of fifth forces on stars may be found in \citet{rhoDM}.

\subsubsection{Dynamical and structural galaxy properties}
\label{sec:dynamics}



\noindent Besides its effect on individual stars, screening can influence the overall structure and dynamics of galaxies' stellar and gas mass components. This is because the surface Newtonian potential of main sequence stars is greater than the total potential of dwarf galaxies in low-density regions; hence for a range of $\chi$ values the stars will be screened while the gas and dark matter will not. This can give these separate mass components measurably different kinematics. The particular observational signals for thin-shell screening models, along with their expected magnitudes, are the following \citep{Jain_Vanderplas,Desmond_HIOC,Desmond_warp}:

\begin{enumerate}

\item{} If the gas disk feels the fifth force and hence follows the motion of the halo center while the stellar disk does not, the stars lag behind the gas when the system falls in an external field. This would manifest observationally as an offset between the centroids of optical and H\textsc{i} light, which trace the stars and gas respectively. The magnitude of this offset can be calculated by requiring in the equilibrium state that the stars and gas have the same overall acceleration in an external field, so that they remain together. This implies that the acceleration of the stellar disk due to its offset from the halo centre equals the additional acceleration of the gas disk due to the fifth force. Let $\vec{a}$ be the total Newtonian acceleration at the position of the galaxy,\footnote{We assume that the acceleration field varies insignificantly between the star and gas centroids, which will be justified post-facto.} and $\vec{a}_5$ the Newtonian acceleration due to unscreened matter within the Compton wavelength $\lambda_C$ of the scalar field, which sources the fifth force. As unscreened mass couples to this with strength $G(1+2\alpha^2)$ rather than $G$ (see Sec.~\ref{sec:screening}), the total acceleration of the gas and dark matter is $\vec{a}_g = \vec{a} + 2 \alpha^2 \: \vec{a}_5$. 
The acceleration of the stellar disk is

\begin{equation}
\vec{a}_* = \vec{a} + \frac{G M(<r_*)}{r_*^2} \: \hat{r}_*,
\end{equation}

\noindent where $\vec{r}_*$ is the offset between the centre of mass of the stars and gas and $M(<r)$ is the total mass enclosed by a sphere of radius $r$ around the halo centre. Requiring $\vec{a}_g=\vec{a}_*$, we find that the offset $\vec{r}_*$ satisfies


\begin{equation}
\frac{G M(<r_*)}{r_*^2} \: \hat{r}_*= 2 \alpha^2 \: \vec{a}_5,
\end{equation}

\noindent (for a screened galaxy $r_*=0$). This allows $\vec{r}_*$ to be calculated as a function of the scalar field coupling $\alpha$, total Newtonian potential $\Phi$ (which determines whether the galaxy is screened), external fifth-force field $\vec{a}_5$,\footnote{In practice, $\vec{r}_*$ must be measured in the plane of the sky, making only the tangential component of $\vec{a}_5$ relevant.} and the density profile, which may be estimated from the dynamics of the galaxy, empirical relations between baryonic and total mass, or N-body simulations by means of a technique such as halo abundance matching.\footnote{Note that in principle halo density profiles -- as well as estimators for them which utilize the halo mass function, like abundance matching -- are different between $\Lambda$CDM and chameleon or symmetron cosmology. However, the present cosmological constraints on these theories require that any such modifications be small.} For dwarf galaxies, where even the central regions are dominated by dark matter, $M = M_h$ to good approximation.

\item{} The amplitude of the gas rotation curve is enhanced relative to the stellar RC, since the former receives a contribution from the fifth force in the same direction as Newtonian gravity, while the latter does not:

\begin{equation}
\frac{v_g^2}{r} = \frac{G(1+2\alpha^2) M(<r)}{r^2}, \; \; \; \; \; \frac{v_*^2}{r} = \frac{G M(<r)}{r^2},
\end{equation}

so that

\begin{equation}
\frac{v_g}{v_*} = \sqrt{1 + 2 \alpha^2}.
\end{equation}

An increase of $v_g$ over the $\Lambda$CDM expectation (from mass modelling of galaxies and halos) is also observable, but has a strong degeneracy with the dark matter distribution in galaxies.

\item{} The lagging of the stellar disk behind the halo centre induces a potential gradient that causes the disk to warp into a cup shape. This effect will be greatest when the disk normal is parallel to the external field, in which case the shape may be estimated as follows. First, define a cylindrical coordinate system with $z$-axis along the external fifth-force field and origin coincident with the halo centre, and consider a star moving in a circular orbit around the $z$-axis at a distance $z_0$ along it and a height $x_0$ above it (Fig.~\ref{fig:warp}). As before, the enclosed halo mass must provide the additional acceleration in the $z$-direction:

\begin{equation}
2 \alpha^2 \: a_5 = a_{h,z} = a_h \frac{z_0}{r_0}, 
\end{equation}

\noindent where $a_5$ is defined above, $a_h$ the magnitude of the acceleration due to the halo and $a_{h,z}$ its projection along $z$. Substituting $a_h = \frac{G M(<r_0)}{r_0^2}$ yields

\begin{equation}
z_0 = \frac{2 \alpha^2 \: a_5 \: r_0^3}{G M(<r_0)},
\end{equation}

\noindent which for $M_h \gg M_g$ and $x_0 \gg z_0$ (i.e. disk sizes much larger than offsets due to modified gravity, again to be justified post-facto) simplifies to

\begin{equation}
z_0 = \frac{2 \alpha^2 \: a_5 \: x_0^3}{G M_h(<x_0)}.
\end{equation}

\begin{figure}[ht]
  \centering
  \includegraphics[width=0.5\textwidth]{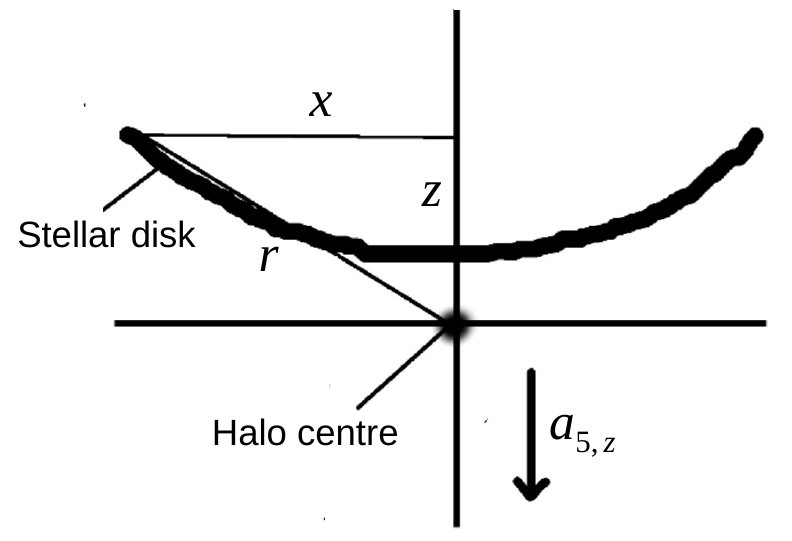}
  \caption{Schematic illustration of disk warping in an unscreened galaxy in thin-shell-screened modified gravity. Reproduced from \citet{Desmond_warp}.}
  \label{fig:warp}
\end{figure}

\noindent The amplitude of this effect therefore depends on the same function of the external fifth-force field, scalar coupling and total density profile as the offset $r_*$ described above, viz $\frac{a_5 \: \alpha^2}{M(<r)}$. For realistic halo density profiles which fall with increasing $r$, $z_0$ is an increasing function of $x_0$, so that the disk acquires a convex shape around the halo centre (as shown in Fig.~\ref{fig:warp}). Heuristically, as the total halo acceleration is lower at larger $x_0$ it must point at a smaller angle to the $z$ axis to compensate for the fixed acceleration difference between screened and unscreened mass. Face-on infall as described here gives maximum warping, and none would be expected in the edge-on case.

\item{} When the infall is near edge-on the stellar and gas disks and rotation curves develop asymmetries: the side of the disk facing the external field becomes more compact than the far side, and the RCs become asymmetric around the galaxy's centre of mass. Deducing the magnitude of these effects requires simulating disk infall under modified gravity \citep{Jain_Vanderplas}.

\end{enumerate}

Similar signals would also be expected from other screening mechanisms such as symmetron and environment-dependent dilaton \citep{Brax_1, 2012JCAP...10..002B}, while Vainshtein screening requires qualitatively different tests (Sec.~\ref{sec:vainshtein_tests}).

While tracing the location of stellar and gas mass through optical and H$\textsc{i}$ photometry is straightforward, care must be taken in identifying appropriate kinematic tracers under modified gravity. In particular, H$\alpha$ emission is unlikely to faithfully trace the stellar component in this case. H$\alpha$ is emitted in the $n=3\rightarrow2$ transition of hydrogen in an ionised sphere around a star (the Str\"{o}mgren sphere). Depending on the star's ionising energy output and mass, the majority of the Str\"{o}mgren sphere may be unscreened, giving it the kinematics of the gas rather than the stars. \citet{Vikram} estimate this to be the case for typical stars and interesting values of $\chi$ ($\sim 10^{-6}-10^{-7}$). In this case, measuring stellar kinematics requires the use of molecular absorption lines originating closer to the surface of the star, such as MgIb or CaII. As these are typically faint, tests 2 and 4 above will benefit from long-exposure observations of these lines with large telescopes.

The present state-of-the-art in searching for these four effects is presented in \citet{Desmond_2, Desmond_HIOC}, \citet{Vikram_RC, Naik:2019moz}, \citet{Desmond_warp} and \citet{Vikram} for effects 1, 2, 3 and 4 respectively; \citet{Vikram_RC} and \citet{Vikram} use the gravitational field reconstruction of \citet{Cabre:2012tq} to determine galaxies' degrees of screening, and the others that of \citet{Desmond} (see Sec.~\ref{sec:screening_maps}). By reconstructing the acceleration as well as potential field, the maps of \citet{Desmond} enable the signals to be forward-modelled as a function of fifth-force coupling $\alpha$, range $\lambda_C$, and screening threshold $\chi$. This was supplemented in \citet{Desmond_HIOC} by structural modelling of galaxies' baryon and dark matter mass profiles in order to estimate $M(<r)$ and hence check for conformity between the observed and predicted signals in terms of their dependence on both the internal properties of galaxies and their gravitational environments. To illustrate the present state of these four tests -- as well as the types of future data needed to advance them -- we list here the samples employed in these studies and the factors limiting their constraining power.


\begin{enumerate}

\item{} \citet{Desmond_2, Desmond_HIOC} used $\sim11,000$ H\textsc{i} detections within 100 Mpc from the ALFALFA survey cross-correlated with optical data to search for a systematic displacement between stellar and gas centroids correlated with gravitational environment. With a highly-conservative assumption for the measurement uncertainty in the H\textsc{i} centroid they set constraints on $2 \alpha^2$ (written there as fifth-force strength relative to gravity, $\Delta G/G$) from $\sim \text{few} \times 10^{-4}$ for $\lambda_C=50$ Mpc to $\sim0.1$ for $\lambda_C=500$ kpc (see Fig.~\ref{fig:HI-OC}). This corresponds to $f_{R0} < \text{few} \times 10^{-8}$. This test is limited by the sample size and H\textsc{i} resolution. It is estimated that future data from a radio survey such as SKA should increase the sensitivity to $\alpha^2$ by around 6 orders of magnitude, rendering this test comparable in strength to solar system fifth-force probes \citep{Sakstein:2017pqi}. With less conservative uncertainties, \citet{Desmond_HIOC} finds evidence at the $\sim6\sigma$ level for a screened fifth force with range $\lambda_C \simeq 2$ Mpc and strength $2 \alpha^2 \simeq 0.02$, although further investigation of possible systematics (e.g. baryonic effects) is necessary to validate this result.

Two further related tests were carried out in \citet{Vikram}, one using 967 galaxies with H$\alpha$ RCs and optical imaging to search for displacement between the optical and H$\alpha$ kinematical centres, and the other using 28 galaxies within 4 Mpc with measured positions for $\sim 10^3-10^5$ Red Giant Branch and main sequence stars to search for a displacement between these populations. (Red giants are expected to remain partially unscreened due to their diffuse outer shells which have lower Newtonian potentials; see Table~\ref{tab:newtonian_potential}.) These tests were limited by the small sample size and lack of knowledge of the external field.

\begin{figure}[ht]
\centering
\includegraphics[width=0.45\textwidth]{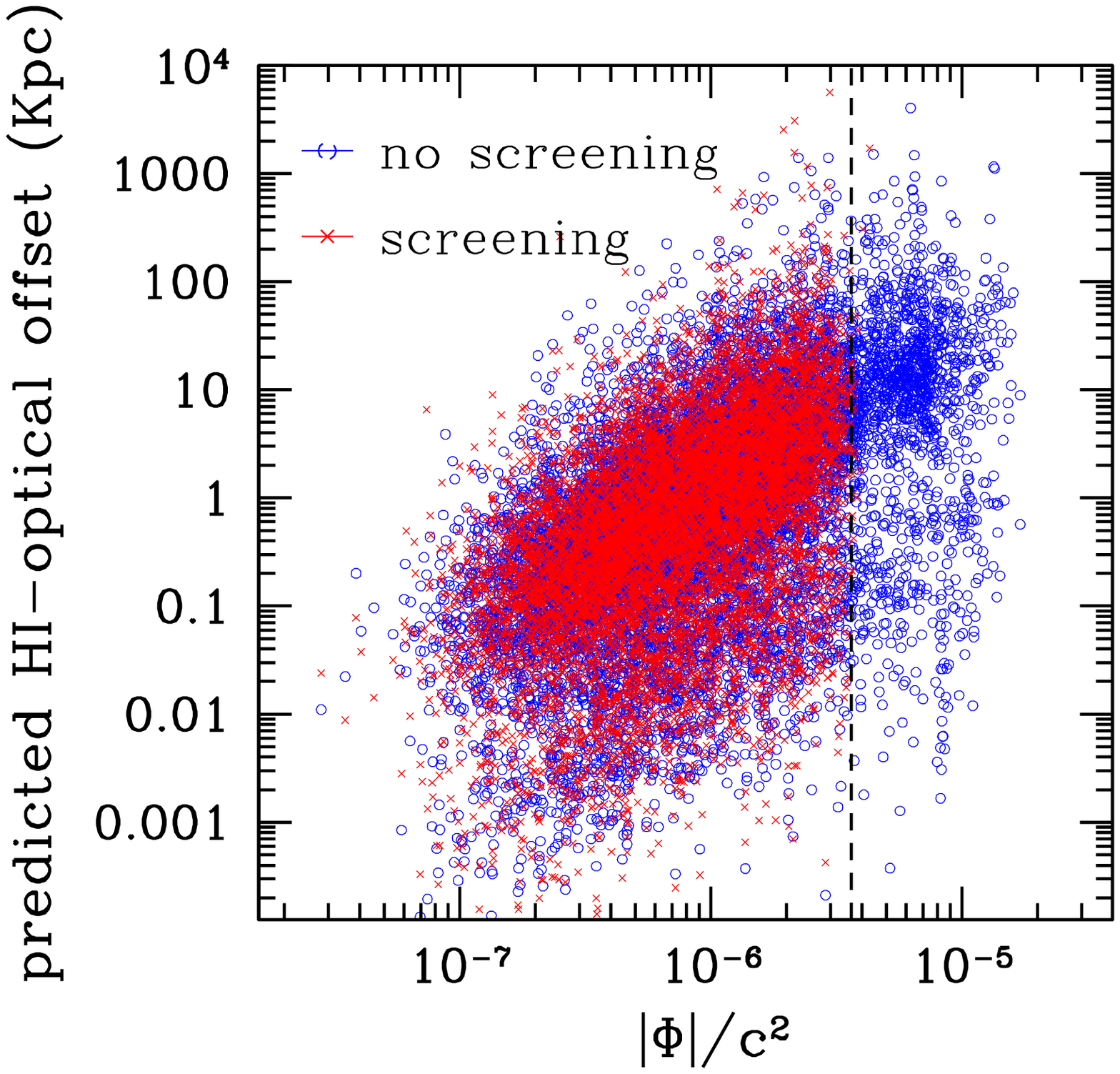}\hfill
\includegraphics[width=0.45\textwidth]{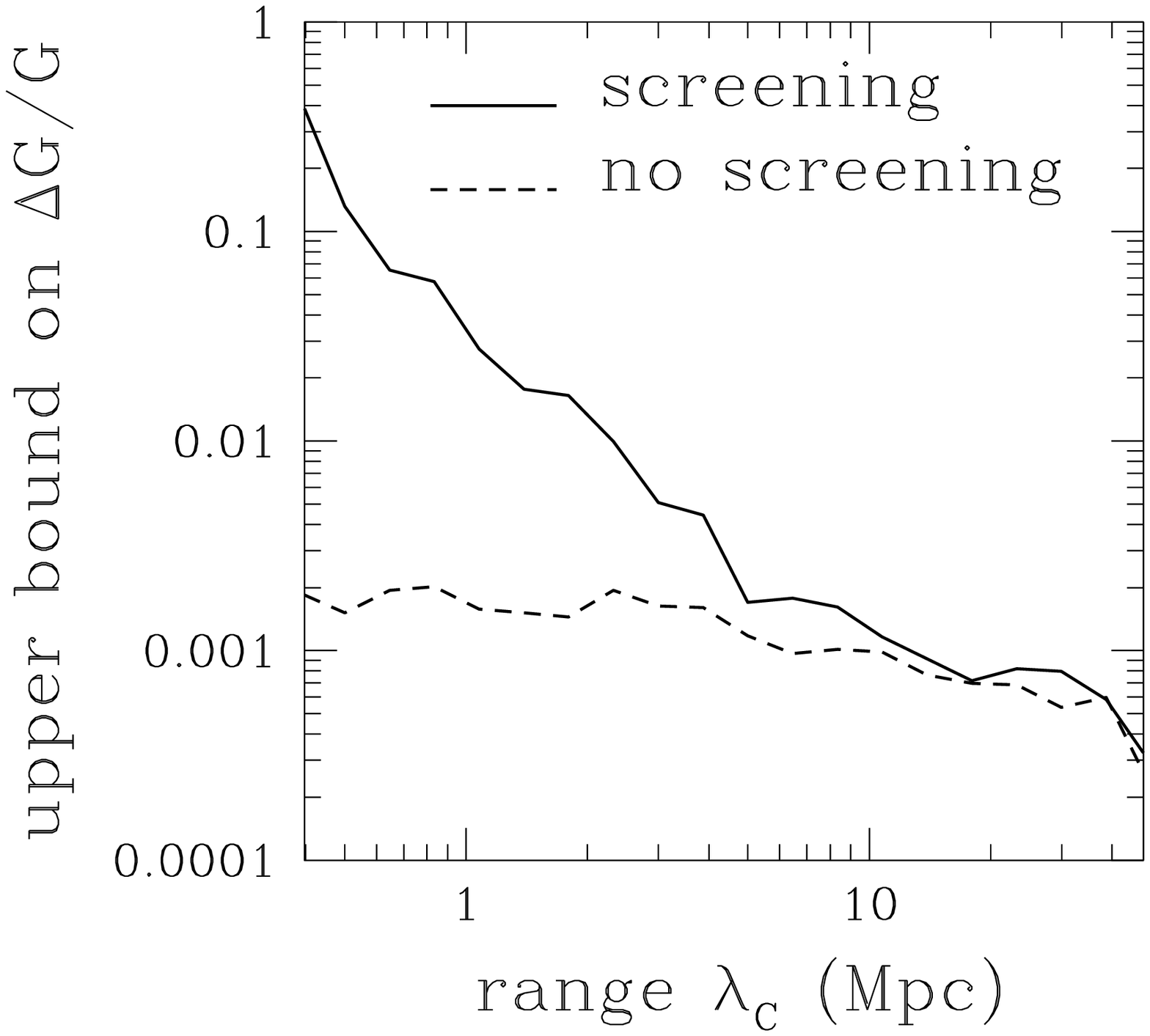}
\caption{\emph{Left}: The displacement between H\textsc{i} and optical centroids for $\sim11,000$ ALFALFA galaxies predicted by a fifth-force model with $\lambda_C = 5$ Mpc, $2 \alpha^2 = 1$, both with and without thin-shell screening.
\emph{Right}: $1\sigma$ constraints in the $2 \alpha^2$ ($\Delta G/G$) vs $\lambda_C$ plane obtained by comparing the above prediction to the measured displacements, using highly conservative measurement uncertainties. Reproduced from \citet{Desmond_2}.}\label{fig:HI-OC}
\end{figure}


\item{} \citet{Desmond_warp} reduce images of $\sim4,200$ galaxies from the \textit{NASA Sloan Atlas} to constrain screened fifth forces by means of warping of stellar disks. This uses a similar inference methodology to \citet{Desmond_HIOC}, but an almost fully orthogonal signal and largely independent data. A similar sensitivity to \citet{Desmond_HIOC} is demonstrated, and further evidence is presented for a screened fifth force model with $\lambda_C \simeq 2$ Mpc, $2 \alpha^2 \simeq 0.02$. These are the most sensitive tests to date of screened fifth forces beyond the Solar System.


\item{} \citet{Vikram_RC} compare the stellar and gas rotation curves of six low surface-brightness galaxies, finding $f_{R0} < 10^{-6}$. This inference is limited by the very small sample size.

\citet{Naik:2019moz} analyse the rotation curves of 85 galaxies from the SPARC sample under an $f(R)$ model, finding evidence for $f_{R0} \simeq 10^{-7}$ if halos are assumed to have an NFW profile, but no evidence for modified gravity if halos are instead assumed to have a cored profile as predicted by some hydrodynamical simulations. This test is limited mainly by uncertainty in the dark matter distributions.

\item{} \citet{Vikram} investigate asymmetries in the H$\alpha$ rotation curves of 200 disk galaxies from the Gassendi H$\alpha$ survey of spirals (GHASP), but were unable to place significant constraints. The limiting factor here is again the paucity of kinematical information with which to calculate the degree of self-screening.

\end{enumerate}



These tests will benefit greatly from the improved radio resolution offered by interferometric surveys such as SKA (and its pathfinders ASKAP and APERTIF), an increase in the number of known dwarfs in low density environments from surveys such as DES and LSST, spatially resolved kinematics of various mass components from IFU surveys such as MaNGA \citep{Bundy:2015}\footnote{\url{https://www.sdss.org/surveys/manga/}}, observations of stellar RCs using stellar absorption lines, and/or increased sample size for any of the above datasets. Uncertainties in the determination of environmental screening proxies (see Sec.~\ref{sec:screening_maps}) also affect the sensitivity of these tests, which may best be reduced by using a base screening catalog from a deeper survey. We discuss future observational prospects further in Sec.~\ref{sec:prospects}.

In the presence of thin-shell screening, the fifth force in the galactic outskirts can make a cuspy matter distribution appear more core-like when reconstructed using Newtonian dynamics because one infers more mass in the outer regions \citep{Lombriser:2014nfa, Naik:2019moz}. 
This could make an underlying NFW profile consistent with observations of the central regions of dwarf galaxies which suggest cores, ameliorating the longstanding ``cusp-core problem'' \citep{deBlok}. For sufficiently strong screening, the inferred density profile may even decrease towards the galaxy center, which would provide a `smoking gun' for modified gravity. The jury is still out, however, on whether NFW is an appropriate profile for halos in the presence of hydrodynamics and stellar feedback in $\Lambda$CDM (e.g. \citealt{Gnedin_2,Maccio,Pontzen:2011ty,DiCintio:2014xia}).

Screening may also give rise to unusual correlations between dynamical galaxy variables; \citet{Burrage} for example use a symmetron model to reproduce the (arguably) unexpected mass discrepancy--acceleration relation of spiral galaxies \citep{McGaugh, RAR}. Testing these effects in detail will require improved theoretical understanding of their origin, magnitude and scope in modified gravity, a systematic investigation of the degeneracies with baryonic physics, and larger and more precise observational datasets.

All of the tests above are subject to potential systematic errors associated with galaxy formation physics, which may induce similar signals even under standard gravity. For example, ram pressure can separate stellar and gas mass and alter their relative kinematics, interactions, mergers, and tidal interactions can warp disks, and baryonic feedback can alter the shape of halo density profiles and rotation curves. Minimizing these effects therefore requires locating galaxies with quiet merger histories that are unaffected by neighbors, in addition to examining the precise dependence of the signal on the relevant galaxy parameters. Modified gravity and baryonic signals may also be distinguished by their dependence on environment: while screening is expected to kick in at a relatively sharp threshold value of Newtonian potential, acceleration or curvature, effects from galaxy formation physics would be expected to have a much more gradual dependence on surrounding density. These degeneracies are similar in origin to those for the power spectrum described in Sec.~\ref{sec:baryons}, and may be investigated further by means of high resolution hydrodynamical simulations.

For dynamical observables within galaxies there is a further degeneracy with dark matter properties, particularly its temperature and possible interactions. Self-interacting dark matter (SIDM), where dark matter particles interact either elastically or inelastically through a new mediator, reduces the density in the central regions of halos through kinetic heating. For appropriate values of the cross-section per unit mass ($\sim1$ cm$^2$/g for a velocity-independent contact interaction) this can turn the cusps predicted at the centers of halos in cold dark matter cosmology into cores \citep{Spergel:1999mh}. Similar effects can be produced by warm dark matter (e.g. \citealt{Bode_WDM}), axions/fuzzy dark matter (e.g. \citealt{Bernal,Hu_FuzzyDM}), and by baryon-dark matter interactions (e.g. \citealt{Khoury_2,Khoury_1}). Other observables of SIDM include spatial offsets between dynamical and galaxy masses in merging clusters, warping and thickening of stellar disks and evaporation of dark matter substructure. Other types of modified gravity theory, which we do not focus on in this review, provide novel explanations of and predictions for galaxy kinematics, e.g. MOND \citep{MOND}.

\subsection{Tests of Vainshtein screening}
\label{sec:vainshtein_tests}





\subsubsection{Vainshtein screening on small scales}

As mentioned above, the efficiency of the Vainshtein mechanism makes small-scale tests very difficult. Theories can self-accelerate cosmologically if $r_c\sim 6000$ Mpc \citep{Fang:2008kc}, which typically gives highly-suppressed fifth forces on smaller scales. Smaller values of $r_c$ represent theories that are less screened but not important cosmologically, meaning one still needs a cosmological constant to self-accelerate. Lunar laser ranging sets a lower bound (for $\alpha=1$) of $r_c\gsim 150$ Mpc \citep{Khoury:2013tda} and most astrophysical tests, including those described below, are sensitive to larger values.

\subsubsection{Strong equivalence principle violations: offset supermassive black holes} \label{sec:SEPviolations}

\begin{figure}[ht]
\centering
\includegraphics[width=0.15\textwidth,angle=-45]{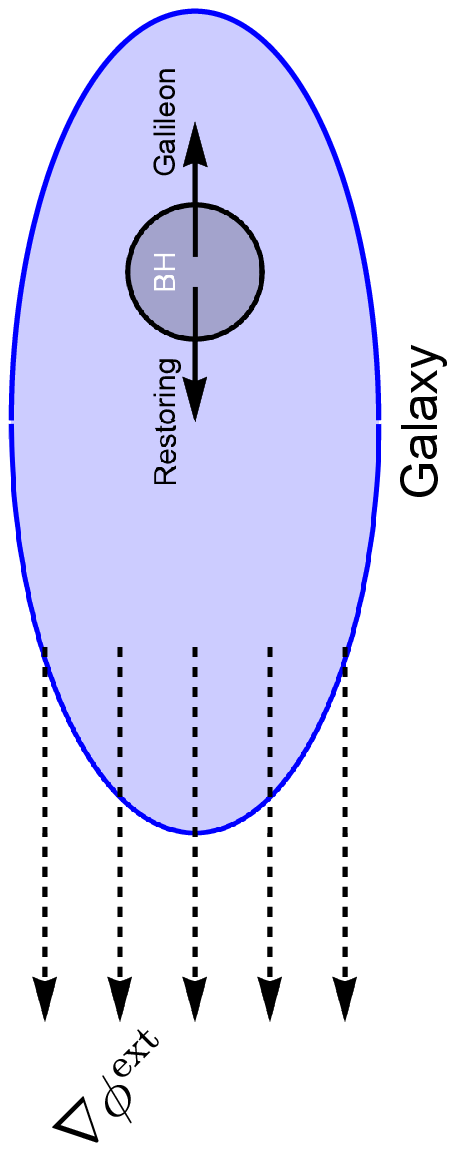}
\caption{A galaxy falling into an external galileon field $\phi^{\rm ext}$. In the rest frame of the galaxy, the SMBH feel two forces: the outward galileon force and the restoring force of the baryons left at the center.}\label{fig:BH_offset}
\end{figure}

One system where the predicted SEP violations could manifest is a galaxy falling in an external galileon field. If the galaxy itself is unscreened, the stars and gas (which have scalar charge-to-mass ratio $Q=\alpha$) will feel the galileon force, while the supermassive black hole (SMBH) at the galaxy's centre, which has no scalar charge ($Q=0$), will not. BHs have no scalar charge in galileon theories because the scalar field couples to the trace of the stress-energy tensor, which excludes gravitational binding energy, while the mass of a BH is purely gravitational. As the galileon fifth force acts in the same direction as gravity, the black hole will lag behind the galactic center (defined as the central density cusp or the potential minimum for cored halos) as the galaxy falls in the external field. At the equilibrium position, the missing galileon force on the SMBH is counterbalanced by the Newtonian force due to its offset from the halo center. (This is analogous to the case of thin-shell screening in Sec.~\ref{sec:dynamics} except that there the offsets are a result of WEP violations.) The situation is sketched qualitatively in Fig. \ref{fig:BH_offset}. 

The behavior of the SMBH depends on the relative strength of the galileon and restoring forces, as shown in Fig. \ref{fig:force_plots}. Typically, halos are either cored or cusped (NFW profile) and the behaviour of the SMBH is different in each case. In the case of cored profiles, shown in the left panel of Fig.~\ref{fig:force_plots}, the restoring force rises outwards from the center due to the constant density core. It then reaches some maximum value and falls as the density begins to fall off. Either the galileon force is larger than the maximum restoring force, in which case the black hole will continue unimpeded and will escape the galaxy eventually, or the galileon force is smaller than the maximum, in which case it reaches a fixed offset at the radius where the two forces balance\footnote{Of course, the general expectation is that the black hole should oscillate about the equilibrium point but, in the situations of interest, the time-scale for these oscillations is smaller than the time-scale over which the galileon force turns on, and so the black hole is expected to adiabatically track the equilibrium point.}. In the case of cusped profiles, which we exemplify using the NFW profile in the right panel of Fig.~\ref{fig:force_plots}, the maximum restoring force is at the center of the halo. In this case, the galileon force is either larger than the restoring force, in which case the SMBH will leave the galaxy unimpeded, or the galileon force is less than the force at the center and the BH will remain there\footnote{One interesting possibility is that the black hole could be initially displaced to some larger radius where the restoring force is smaller than the galileon, in which case the SMBH would begin to exit the galaxy. There are several scenarios for this such as asymmetric AGN jets and gravitational recoil (kick) from binary black hole mergers \citep{Merritt:2004gc}.}. 

In order to utilize this SMBH phenomenon it is necessary to look for situations where the motion of galaxies receives a contribution from a {partially unscreened galileon field. 
In what follows, we will describe two scenarios that have been proposed to test the SEP violation: cosmological field galaxies and satellite galaxies falling towards the center of clusters. 

\begin{figure}[ht]
\centering
{\includegraphics[width=0.45\textwidth]{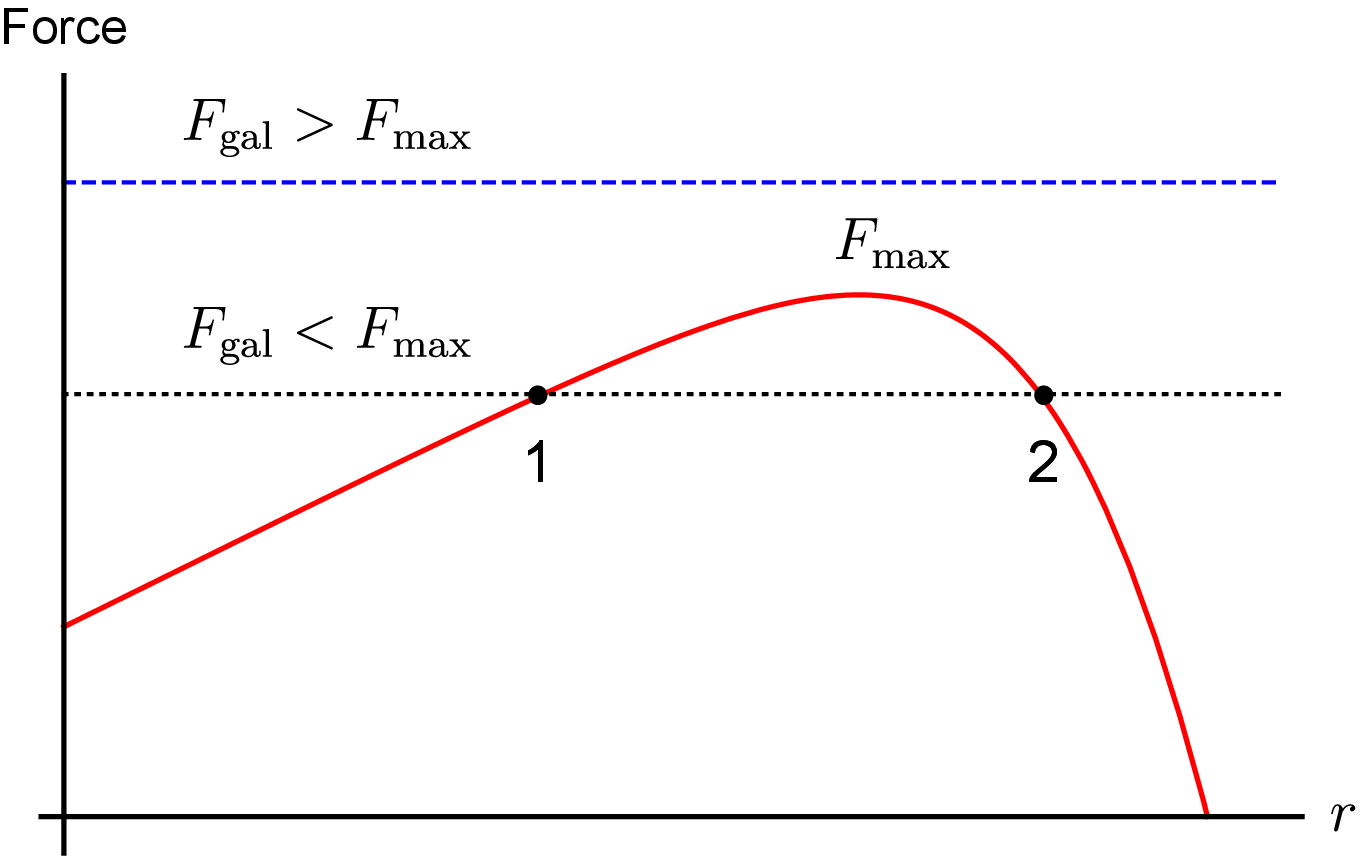}}
{\includegraphics[width=0.45\textwidth]{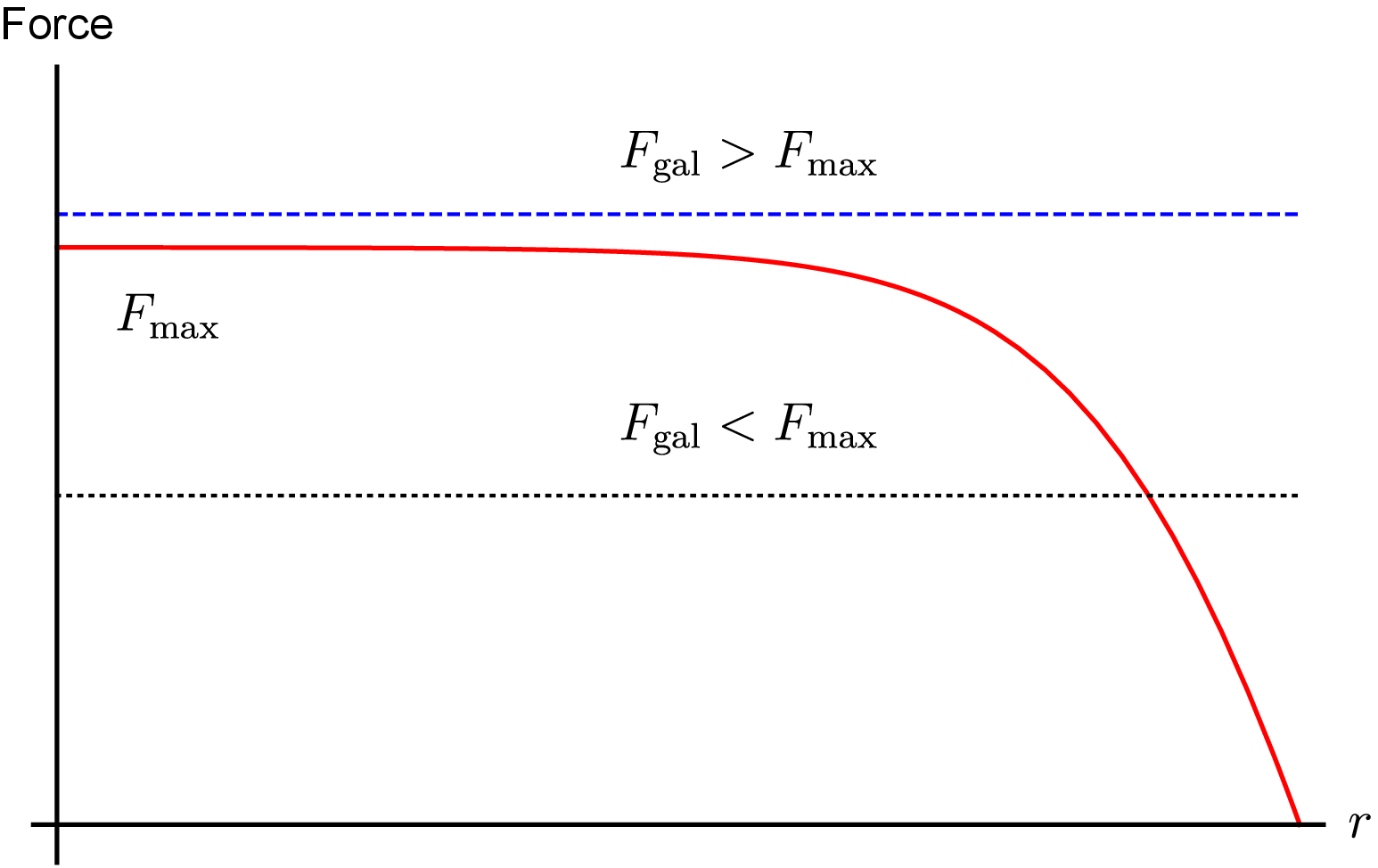}}
\caption{\emph{Left}: The restoring force profile for a cored galaxy (red, solid). The blue, dashed, upper line shows the case where the galileon force exceeds the maximum restoring force and the black, dotted, lower line shows the case where the galileon force is smaller than this. In the latter case, there are two equilibrium positions labelled 1 and 2. The former point represents a stable offset; the latter is unstable. \emph{Right}: The restoring force for a cusped (NFW) profile (red, solid). The blue, dashed, upper line shows the case where the galileon force is larger than the maximum restoring force (the force at the center) and the black, dashed, lower line shows the case where it is smaller than this. }\label{fig:force_plots}
\end{figure}


Numerical simulations have shown that there is an unscreened galileon field on linear cosmological scales (at distances $\gsim 10$ Mpc;  \citealt{Cardoso:2007xc,2009PhRvD..80d3001S,2009PhRvD..80f4023K,2009PhRvD..80j4005C}) and so, as first suggested by \citet{Hui:2012jb}, field galaxies that have peculiar velocities due to the large scale structure in the universe should show offset SMBHs. One can estimate the size of the offset by noting that typical galaxies were accelerated to a peculiar velocity of $300$km/s over a Hubble time so that the Newtonian acceleration is $|\nabla\Phi^{\rm ext}|\sim 20$ (km/s)$^2$/kpc. This value is estimated by assuming a typical peculiar velocity of $300$ km/s, which is close to the RMS value found by averaging the observed matter power spectrum \citep{Hui:2005nm}. One should however bear in mind that this is a statistical variable and precision tests may require screening maps similar to those discussed in Sec.~\ref{sec:screening_maps}. A fully unscreened galileon field has $|\nabla\phi^{\rm ext}|\sim2\alpha|\nabla\Phi^{\rm ext}|$, and, assuming that the density in the center of the halo $\rho_0$ is constant\footnote{This breaks down at some point, but \citet{Hui:2012jb} had low surface brightness Seyfert galaxies in mind, for which this is a good approximation. Furthermore, given the small offset predicted it is sensible to work in the constant density regime. } one finds an offset

\begin{equation}\label{eq:offset_cosmological}
R=0.1 \textrm{ kpc}\left(\frac{2\alpha^2}{1}\right)\left(\frac{|\nabla\Phi^{\rm ext}|}{20\textrm{ (km/s)}^2\textrm{/kpc}}\right)\left(\frac{0.01M_\odot\textrm{/pc}^3}{\rho_0}\right),
\end{equation}
where the fiducial values have been chosen to represent a low surface brightness galaxy. In principle, the position of the black hole could be determined with microarcsecond precision using microwave interferometry \citep{Broderick:2011vt} whereas the optical centroid could be found using the galaxy's isophotes \citep{Asvathaman:2015nna}. In practice, central densities derived from generalized NFW fits produce typical offsets $\lsim0.1$ kpc. Such a small offset is hard to observe with small sample sizes, and would be degenerate with other astrophysical effects such as asymmetric AGN jets and black hole kicks. For these reasons, no constraints on galileon modifications of gravity have been placed using this scenario to date. 


Another situation where the galileon field is partially unscreened is massive galaxy clusters ($M\sim 10^{14}$--$10^{15}M_\odot$). As discussed in Sec.~\ref{sec:vainscreentheory}, an extended mass distribution does not suppress the galileon force as efficiently in its interior \citep{Schmidt:2010jr}. A massive galaxy cluster therefore has a large partially screened galileon field that contributes to the motion of infalling satellite galaxies. As an example, consider a model for the Virgo cluster shown in the left panel of Fig.~\ref{fig:virgo}. Outside the virial radius ($R_{200}$) one can see that the galileon force (for $r_c=500$ Mpc) is a factor of $2\alpha^2$ times the Newtonian force but even inside this it is significant and only a factor of $\sim 5$ smaller than the Newtonian force. Furthermore, it is constant over a large range of radii, which helps to mitigate astrophysical uncertainties on the distances of galaxies from the cluster center. An infalling satellite galaxy will feel a force given by the sum of the red and blue curves, whereas the SMBH will feel only the force shown by the blue curve. We plot the resultant offset for typical cored satellite galaxies falling into massive clusters in the right panel of Fig.~\ref{fig:virgo}. Since the galileon (and Newtonian) force from the cluster is radially-dependent, the offset is a function of how far the satellite galaxy is from the cluster's center. One can see that, owing to the larger galileon force (the cosmological force discussed above is shown in the left panel of Fig.~\ref{fig:virgo}), this offset can be of $\mathcal{O}(\textrm{kpc})$, which is easier to observe using spectroscopic or x-ray observations (if the black hole powers an AGN) and is likely far larger than the black hole can be offset by asymmetric AGN jets or other astrophysical mechanisms. 
It is also significantly larger than the offsets between star and gas centroids produced by thin-shell screening mechanisms (Sec.~\ref{sec:dynamics}) for realistically small values of $\alpha$ and self-screening parameter $\chi$.

\begin{figure}[ht]
\centering
{\includegraphics[width=0.45\textwidth]{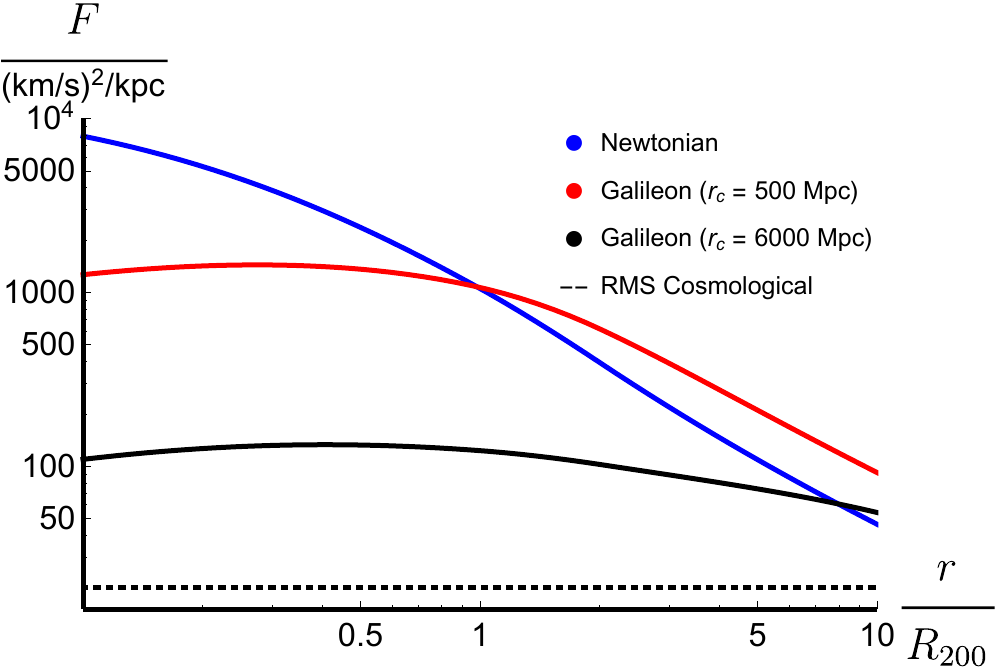}}
{\includegraphics[width=0.45\textwidth]{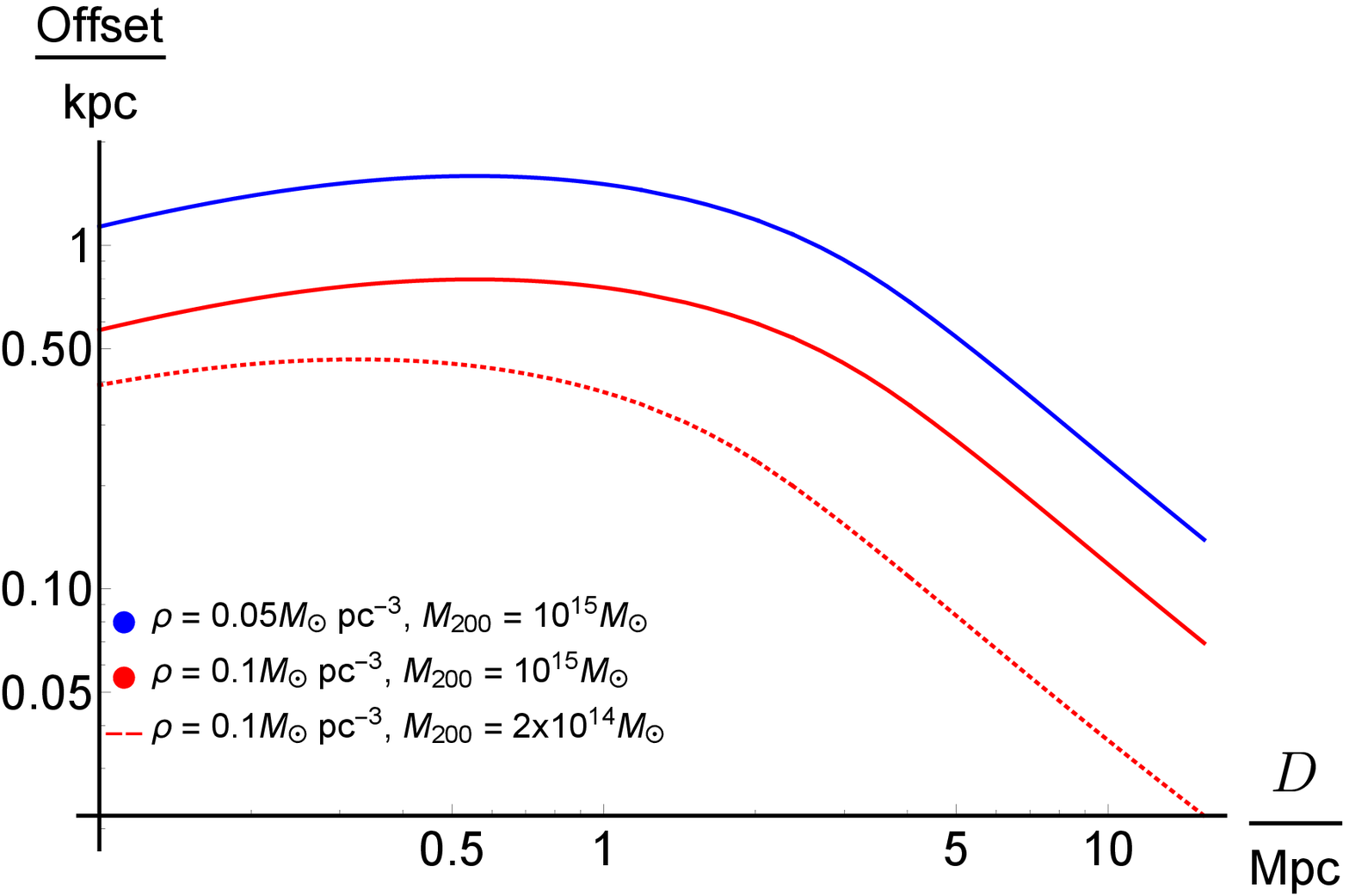}}
\caption{\emph{Left}: The Newtonian and cubic galileon force profile for a Virgo-like galaxy cluster. We have used an NFW profile (with 2-halo term corrections; \citealt{Diemer:2014xya}) with concentration $c=5$ and parameters that give a total mass $M=10^{15}M_\odot$ \citep{Fouque:2001qc,Peirani:2005ti}, consistent with observations. The blue line shows the Newtonian force and the red and black solid lines show the cubic galileon force for $\alpha=1$ and $r_c=500$ and $6000$ Mpc (cosmological galileons) respectively. The black dashed line shows the RMS cosmological galileon force for the fiducial values shown in equation \eqref{eq:offset_cosmological} and is shown for comparative purposes only. 
\emph{Right}: The SMBH offset for typical infalling satellite galaxies for cluster masses and satellite central densities indicated in the caption. The $x$-axis shows the distance between the satellite galaxy and the cluster's center.
Figures reproduced from \citet{Sakstein:2017bws}. }\label{fig:virgo}
\end{figure}

Previously, \citet{Asvathaman:2015nna} had looked at the Virgo cluster to perform general tests of the SEP without a direct focus on galileon theories.\footnote{The SEP violation is not unique to galileon theories. Indeed, all scalar-tensor theories of gravity have this property. What is new is the Vainshtein mechanism, which makes solar system tests of these theories more difficult, and, therefore, SMBH tests more appealing.} Building on their work, \citet{Sakstein:2017bws} have used the model for the Virgo cluster in Fig. \ref{fig:virgo} to place new constraints on the cubic galileon shown in Fig. \ref{fig:SMBH_constraints}. In particular, the SMBH in the galaxy M87 is offset by less than 0.03 arcseconds, implying that the galileon force must be less than $1000$ (km/s)$^2$/kpc; the constraints in Fig. \ref{fig:SMBH_constraints} are found by scanning the parameter space to look for regions that satisfy this bound. One can see that the bounds are stronger than those coming from LLR, but cosmological galileons ($r_c\sim 6000$ Mpc) are a long way from being probed. \citet{Sakstein:2017bws} checked that the bounds are robust to changing some of the model assumptions (halo concentration, central density, functional form of the profile) but of course the exact bound is sensitive to the precise model details. The effects of departures from spherical symmetry were not investigated. One could improve upon this technique by either performing a detailed modelling of the Virgo cluster or by looking at a large number of clusters and performing a statistical analysis to reduce the effects of modelling uncertainties.

\begin{figure}[ht]
\centering
\stackunder{
{\includegraphics[width=0.45\textwidth]{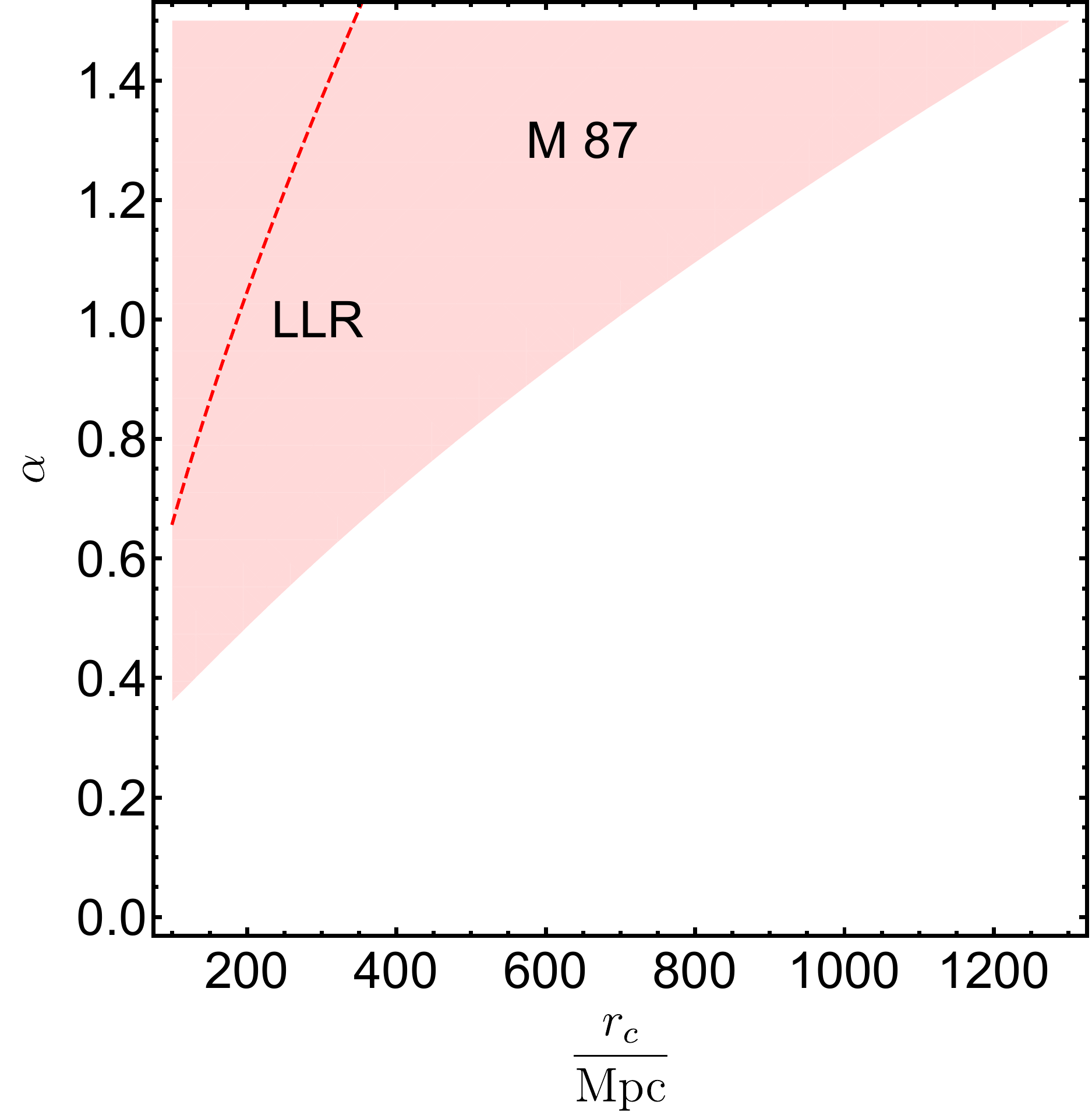}}}{$\phantom{xxxz}$Cubic galileon}
\stackunder{
{\includegraphics[width=0.465\textwidth]{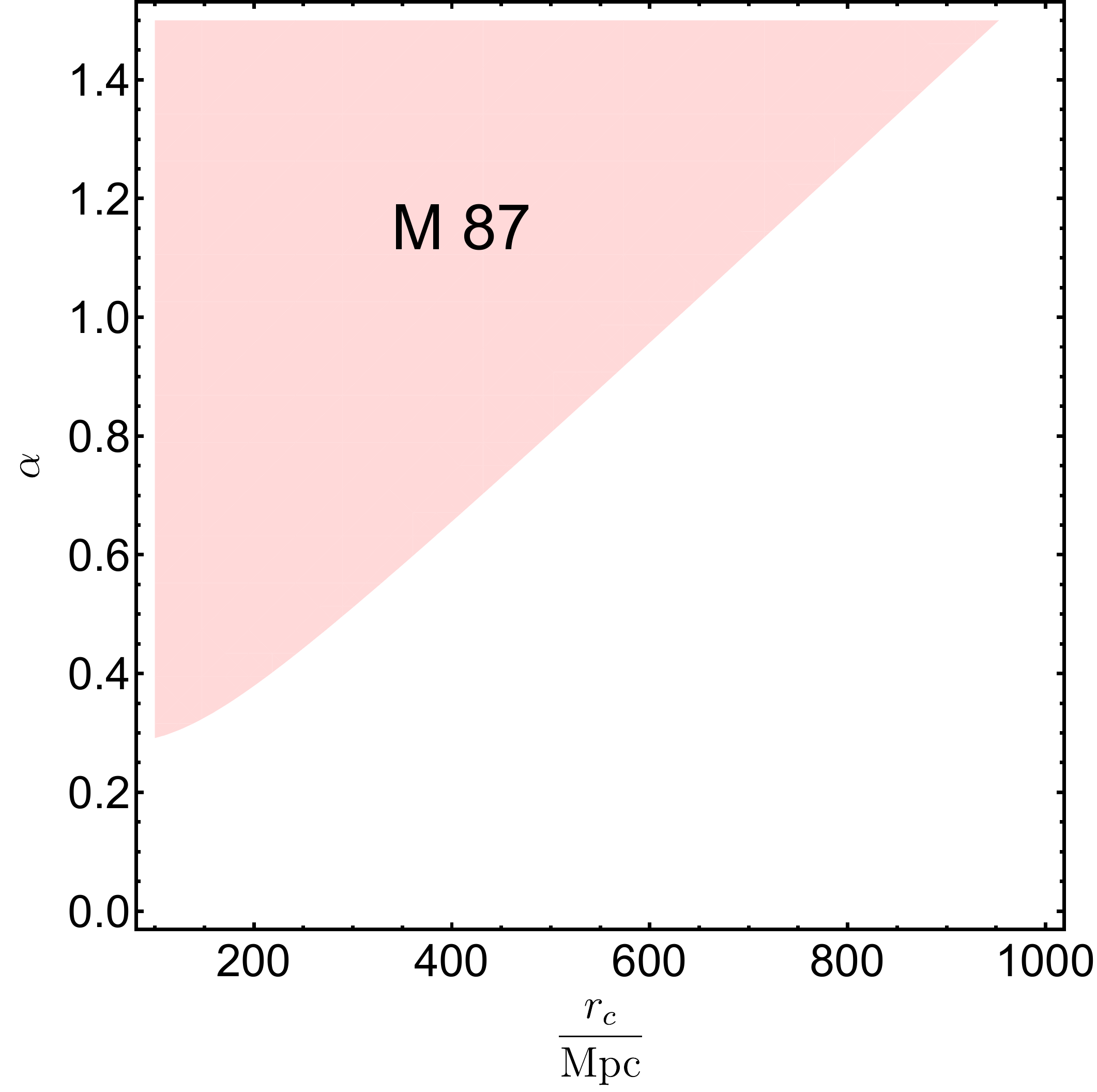}}}{$\phantom{xxx}$Quartic galileon}
\caption{The constraints on the coupling $\alpha$ and crossover scale $r_c$ for cubic (left panel) and quartic (right panel) galileon models. The pink region is excluded because it would predict a larger offset than is observed in M~87, which is falling towards the center of the Virgo cluster. The red dashed line in the left panel indicates the previous constraints from lunar laser ranging \citep{Dvali:2002vf}. LLR constraints on the quartic galileon are not competitive with SMBH constraints, hence they are omitted from the right hand panel.}\label{fig:SMBH_constraints}
\end{figure}

\subsection{Vainshtein breaking}
\label{sec:vainshtein_breaking}

In beyond Horndeski and DHOST theories, the Vainshtein mechanism can be broken in the sense that fifth-forces are efficiently screened outside astrophysical bodies but deviations from GR may appear inside. See section \ref{sec:VBintro} for a description of this. Since Vainshtein breaking is universal and not subject to environmental screening it can be tested using objects in any galaxy including our own Sun.

\subsubsection{Existence of stars}

A necessary condition to form stable stars is that the pressure gradient decreases toward larger radii, $P''(r)<0$. This implies a theoretical condition $\Upsilon_1>-2/3$ (eq.\eqref{eq:BH1}) for DHOST theories \citep{Saito:2015fza,Babichev:2016jom,Sakstein:2016oel}.

\subsubsection{Tests with dwarf stars}

Dwarf stars make particularly strong tests of Vainshtein breaking compared with main-sequence or post--main-sequence stars theories due to their homogeneous structure, small variability and lack of astrophysical degeneracies. Here, we will describe how red, brown, and white dwarf stars can probe theories with Vainshtein breaking.

Red dwarf stars are low mass $0.08M_\odot\le M\le 0.6 M_\odot $ stars that are heavy enough to fuse hydrogen to helium-3 but not all the way to helium-4 on the PP-chains. They are supported by a mixture of degeneracy pressure and ideal gas pressure so their physics is easily calculable using polytropic models \citep{Burrows:1992fg}. The upper bound in mass corresponds to the onset of the PPI chain and the lower to the minimum mass for hydrogen burning (MMHB). The latter is the minimum mass for which the star can burn hydrogen continuously and stably, i.e. the energy from hydrogen burning in the core is balanced by losses at the surface. Stars with masses lighter than the MMHB do not fuse hydrogen and are brown dwarfs. The MMHB is a very strong probe of Vainshtein breaking theories because when $\Upsilon_1>0$ the star's core is cooler and less dense so that one needs larger masses in order to achieve the requisite conditions for hydrogen burning. \citet{Sakstein:2015zoa,Sakstein:2015aac} used a simple analytic model of red dwarfs in Vainshtein breaking theories to show that the MMHB exceeds the mass of the lightest observed red dwarf, GL886C, when $\Upsilon_1>1.6$. Future observations of lighter mass red dwarfs with surveys such as Gaia \citep{Gaia-Collaboration:2016}\footnote{\url{http://sci.esa.int/gaia/}} could improve upon this. Similar to the MMHB, there is also a minimum mass for lithium and deuterium burning, although these have yet to be calculated due to the more complex reaction chains involved. (There is only one relevant reaction for hydrogen burning.)


Brown dwarf stars are inert stars that do not burn nuclear material (they may burn primordial fuel in short bursts but there is no sustained burning). They are primarily supported by the pressure from coulomb scattering so that they have the simple equation of state $P=K\rho^2$. Equations of state such as these predict a `radius plateau' such that the sizes of brown dwarfs are independent of their mass and given by \citet{Sakstein:2015aac}
\begin{equation}
R=0.1 \: R_\odot \: f(\Upsilon_1),
\end{equation}
where $f(0)=1$ corresponding to the GR prediction $R=0.1R_\odot$. The change due to Vainshtein breaking can be $\mathcal{O}(20\%)$ or more. At present, no bounds have been placed due to sparse measurements of the mass-radius relation for brown dwarfs; future data from Gaia and similar surveys could measure this relation empirically and hence allow strong constraints on $\Upsilon_1$, depending on the scatter.


White dwarf stars can probe Vainshtein breaking in two different ways. First, the Chandrasekhar mass decreases when $\Upsilon_1<0$ (this mimics the effects of increasing the strength of gravity) and second, the mass-radius relation is altered. \citet{Jain:2015edg} have investigated both of these effects using a simple fermionic equation of state to describe carbon white dwarfs. They found that the mass of the heaviest presently observed white dwarf ($1.37\pm0.01 M_\odot$) exceeds the Chandrasekhar mass when $\Upsilon_1<-0.22$. Fitting to the mass-radius relation for a sample of 12 white dwarfs they find the bounds $-0.48\le\Upsilon_1\le0.54$ 
at $5\sigma$ ($-0.18\le\Upsilon_1\le0.27$ at $1\sigma$).

\section{Future directions}
\label{sec:prospects}

To conclude this review, we briefly summarize the current state of cosmological gravity theories and enumerate some potential future directions. On the theory side, these are theories that are either under-developed or have not received much attention. On the experimental side, these are directions that may not have been fully-exploited, or for which the theoretical modeling is lacking. 

\subsection{Current status of modified gravity theories}





The present status of the modified gravity models featured prominently in this review can be summarized as follows:

\begin{itemize}
\item {\bf Chameleon and similar theories including $f(R)$}: These are well-constrained and cannot drive the cosmic acceleration without a cosmological constant but may be relevant on small scales. They were not affected by GW170817. Applying the tests described in Sec.~\ref{sec:chameleon_tests} to upcoming data sets should constrain $f_{R0}$ to the $10^{-8}$ level. Beyond $f(R)$ models, smaller strengths of the coupling of the scalar field to matter can be tested. 

\item {\bf Galileons and Vainshtein screened theories:} The cubic galileon is in severe tension with cosmological probes \citep{Renk:2017rzu}, while the quartic and quintic Galileon theories are ruled out as dark energy candidates by the recent bounds imposed by GW170817. Massive gravity theories that screen using the Vainshtein mechanism manifest as a mass in the gravitational wave dispersion relation, and are most tightly constrained by Lunar Laser Ranging \citep{2019arXiv190304467T, 2017RvMP...89b5004D}. On smaller scales, Vainshtein screened theories such as galileons (that do not self-accelerate) may be active. For these theories, small scale probes can be explored further, with supermassive black hole offsets being one example. In the specific case of the nDGP model with the cosmic acceleration driven by a cosmological constant --- which is described by a cubic galileon in the decoupling limit --- a strong bound $r_c>3090$ Mpc ($2\sigma$) was obtained using observations the growth rate of structure compared with modified gravity simulations \citep{2016PhRvD..94h4022B}. It would be interesting to repeat this analysis for more general models, including quartic galileons. Tests of Vainshtein screening are discussed further in Sec.~\ref{sec:vainshtein_tests}.

\item {\bf General scalar tensor theories including beyond Horndeski and DHOST}: The space of viable theories that can drive the cosmic expansion has been dramatically restricted following GW170817 (Sec.~\ref{sec:GWs}). A notable exception is cubic terms but these are strongly constrained by cosmological measurements.  Recently, \citet{Creminelli:2018xsv} have indicated that beyond Horndeski terms are completely ruled out because they would cause gravitational waves to decay to dark energy scalars.

\item{\bf Gravity parameters for large-scale tests:} As detailed in Sec.~\ref{sec:cosmo_constraints}, large-scale measurements of lensing, redshift space distortions, and other  observables have been used to constrain parameters for the deviations of gravity in the linear regime. Figs.~\ref{fig:mu_eta_EG}-\ref{fig:cosmo} summarize some of the current constraints. 

\end{itemize}

\subsection{Future directions for theory}

As summarized above, at present there appears no viable modified gravity model that also provides an alternative to dark energy for cosmic acceleration. Here we summarize potentially interesting directions for future theoretical work, one or more of which may lead to renewed connections between theory and cosmological observations.  
\begin{itemize}

\item {\bf Gravitational wave speed}. The tight bounds on the speed of gravitational waves described in Sec.~\ref{sec:GWs} must be respected by future developments of gravity theories, providing useful restrictions. However, care must be taken if the strong coupling energy scale of a theory is close to that of the gravitational wave observations, in which case the concerns of \citet{deRham:2018red} apply. \\

\item {\bf New screening mechanisms.} To date, only a handful of screening mechanisms are known: the chameleon, symmetron and dilaton mechanisms, k-mouflage, and the Vainshtein mechanism (Sec.~\ref{sec:screening}). Furthermore, the first three of these share some similarity in their implementation at the theoretical level. New methods through which screening could occur would stimulate new tests; however, at present it seems that potentially all the screening mechanisms possible from standard field couplings have been exploited. Perhaps new mechanisms could be linked to a transition scale in spacetime curvature (see Fig.~\ref{fig:landscape1}) or acceleration, for example. Very recently, \citet{rhoDM} has identified a new screening mechanism that screens fifth forces mediated by interactions between dark matter and baryons. They suggest several astrophysical tests that merit further investigation.  \\

\item {\bf Looking beyond standard field theory}. Over the past decade much progress has been achieved in developing extremely general field theories for gravity. For example, the Beyond Horndeski paradigm covers the general second-order theory of a scalar field and a metric, and Generalized Proca theory acts similarly for theories of a vector and a metric (though is not yet proven to be the most general construction possible). If these general `parent theories' are ruled out by observations -- as is beginning to happen for the Horndeski family --- then we will have effectively exhausted the application of regular field theory techniques to the coupling of scalar, vector, and tensor fields. This will likely prompt us to look further afield towards more nonstandard ideas, such as nonlocal Lagrangians, or thermodynamic or emergent viewpoints on gravity \citep{2018JCAP...03..002B, 2015arXiv151206546P, 1995PhRvL..75.1260J}. \\


\item {\bf Multi-field theories}: Current theories typically focus on one new field (the exception is massive gravity where bigravity has been extensively studied) and there is relatively little work on multi-scalar-tensor, multi-vector-tensor, or scalar-vector-tensor theories even though such theories exist and are known to be free of the Ostrogradsky instability (see e.g. \citealt{Padilla:2012dx,Sivanesan:2013tba,Heisenberg:2018vsk}). It would be interesting to study the cosmology of these theories, and determine whether they are subject to the same stringent bounds from GW170817 and graviton decay into dark energy (discussed in Sec.~\ref{sec:GWs}) that are highly-constraining for single filed extensions of GR. 

\item {\bf The cosmological constant problem}: With a vast number of alternatives to $\Lambda$CDM ruled out, now may be a good time to re-examine the cosmological constant problem. After all, as $\Lambda$CDM is still the model that fits the data best the problem is entirely theoretical: the small cosmological constant that we observe is fine-tuned 
\citep{Weinberg:1988cp,Burgess:2013ara,Padilla:2015aaa} because theory predicts a value that is larger by many orders of magnitude. 
A compelling mechanism that explains this would place $\Lambda$CDM on a more solid theoretical footing. In the last decade, significant theoretical progress has been made towards understanding the cosmological constant problem and new models such as supersymmetric large extra dimensions \citep{Aghababaie:2003wz}, vacuum energy sequestering \citep{Kaloper:2013zca,Kaloper:2014dqa,Kaloper:2015jra}, degravitating superfluids \citep{Khoury:2018vdv}, and tempering the cosmological constant \citep{Appleby:2018yci,Emond:2018fvv} have emerged as potential resolutions. An effort aimed towards finding novel small scale tests of these theories akin to the program aimed at testing screened modified gravity theories could help to confirm or refute these models.\\

\item {\bf Novel probes of quintessence/K-essence}: While the landscape of possible modified gravity models has been reduced by the observation of GW170817, models of dark energy where new degrees of freedom do not couple to matter (i.e. their scalar potential drives the cosmological acceleration) have survived unscathed. For example, quintessence models where the acceleration is driven by a scalar field rolling down a shallow potential are phenomenologically viable, as are K-essence theories where the acceleration is driven by kinetic self-interactions\footnote{With the exception of theoretically well-motivated models such as DBI, these models are on shakier theoretical footing since they rely on higher-order terms that are outside the range of validity of the effective field theory so may operate in a regime where the theory is not predictive.} \citep{Copeland2006}. 
Cosmologically, these models can be tested by measuring the dark energy equation of state and other probes of the background expansion as well as the growth of cosmic structure \citep{Copeland2006}. It would be interesting to see if small-scale tests could be devised along the lines of those used to test modified gravity. One such example is the effect of a dynamically changing cosmological constant on black holes \citep{Gregory:2017sor}.\\

\item {\bf Massive gravity}: Massive gravity (and massive bigravity) models are theoretically well-motivated, because they are both natural alternatives to general relativity and 
screen using the Vainshtein mechanism so are phenomenologically viable. Massive gravity itself does not admit flat FRW cosmological solutions \citep{DAmico:2011eto} and those with spatial curvature are unstable \citep{DeFelice:2013bxa}. Massive bigravity does admit flat FRW solutions but they are unstable \citep{Comelli:2012db}. 
It has been speculated that the Vainshtein mechanism can cure this linear instability and some preliminary exploratory work has been performed \citep{Hogas:2019ywm} but further work is needed to asses whether this is indeed the case. Similarly, extensions of massive bigravity have been studied by \citet{Kenna-Allison:2019tbu}, who find that there are regions of parameter space that admits a stable cosmology. This theory certainly merits further investigation. Given that scalar-tensor and vector-tensor models are largely excluded, now may be the time to search for stable massive trigravity models (and possibly even more interacting metrics) that could be candidate benchmarks for upcoming observations and simulations.\\

\item {\bf Baryon-dark matter couplings}: Traditionally, theories that explain the cosmic acceleration fall into two classes: dark energy, where the acceleration is driven by a new fluid with an equation of state $w<-1/3$ that does not result fifth-forces or equivalence principle violations e.g. quintessence, and modified gravity, where the acceleration is driven by modifications of general relativity (the theories discussed in this review). Recently, a third class has been proposed where dark matter is postulated to be a superfluid whose excitations couple to baryons. This coupling could lead to cosmic acceleration at late times \citep{Berezhiani:2016dne,ferreira:2018}. Recently, \citet{rhoDM} investigated this model and showed that it contains a novel screening mechanism where the value of Newton's constant can vary depending on the local dark matter density. The authors also list further potential astrophysical tests that could prove fruitful once the theory is developed (see also \citealt{Desmond_H0,Desmond:2020wep}). 

\item {\bf Massive Galileons}: The galileons studied in this work are massless but, recently, it has been noted that one can add a mass without spoiling the theoretical properties such as non-renormalization that make these appealing effective field theories \citep{Goon:2016ihr}. For some parameters, massive galileons can be derived from a Lorentz-invariant UV-completion \citep{deRham:2017imi}, something which is presently unclear for their massless counterparts. To date, there has been little investigation of the properties of massive galileons (with the exception of \citealt{Sakstein:2018pfd}). Further theoretical investigation of their screening (if it persists) and cosmology could reveal them to be a strong competitor to $\Lambda$CDM.

\item {\bf Degeneracies with baryonic physics}: As mentioned in Sec.~\ref{sec:baryons} (and throughout), many modified gravity predictions are degenerate with the hydronamical effects of baryons in the process of galaxy formation. Overcoming these degeneracies will require improvements in the resolution, accuracy and predictive power of cosmological hydrodynamical simulations, as well as inference frameworks that incorporate more sophisticated models for baryonic effects. Performing such simulations in modified gravity will reduce uncertainty concerning the interplay between baryons and modified gravity, and enable fully self-consistent modified gravity analyses. For current attempts at such simulations see e.g. \citet{Arnold:2019vpg,Arnold:2019zup}.


\end{itemize}

\subsection{Future directions for observations}

\begin{itemize}

\item {\bf Dynamical vs. lensing tests:} A powerful, generic way to test for the existence of deviations from GR is to compare observables that depend on the motion of massive particles versus massless ones. This is because massive particles respond to the Newtonian potential $\Phi$, whilst massless ones respond to the Weyl potential $\Phi + \Psi$. In GR this is equivalent to $2\Phi$; a generic feature of modified gravity theories is that they break this relationship, such that $\Phi+\Psi\neq 2\Phi$. Observationally, the test involves the comparison of dynamical and lensing masses on galaxy and cluster scales (Sec.~\ref{sec:dynmass}), and of larger scale cross-correlations of galaxy clustering and weak lensing (Sec.~\ref{sec:RSD}). In an era in which the landscape of modified gravity theories is constantly evolving, this robust test is valuable independent of particular models.   \\
\item {\bf Voids:} Cosmological voids -- under-dense regions of large-scale structure -- provide new ways to test gravity (Sec.~\ref{sec:universal_tests2}). In particular, the low-density nature of voids means that screening mechanisms should be largely ineffective there, allowing deviations from GR to manifest unmitigated. Both lensing and redshift-space signatures are promising. However, a detailed understanding of void tracers and void finding algorithms is required and still in progress. Planned analyses from surveys with spectroscopy (to enable 3D void finding) or well-calibrated void finders that work with high quality photometric redshifts are expected to yield qualitative advances.\\

\item {\bf Galaxy-scale tests of screening:} Signals in the morphology and kinematics of galaxies have emerged as powerful ways to probe screened modified gravity (Sec.~\ref{sec:dynamics}). These tests typically require one or more of the following: i) galaxies with a wide range of masses and environments and hence degrees of screening (especially low mass galaxies in underdense regions), ii) multi-wavelength observations to locate the various mass components of galaxies at high precision, iii) knowledge of the geometry of the system relative to the external field direction, and iv) kinematic information to determine halo mass profiles and degrees of self-screening in addition to the relative dynamics of stars and gas. Larger galaxy samples may greatly improve constraints, as could smaller surveys specifically targeting unscreened regions. These tests also require knowledge of the environmental screening field, which will be improved by deeper and wider photometric surveys.\\


\item {\bf Future gravitational wave tests:} We have discussed the constraints provided by the LIGO detection of a neutron star merger in Sec~\ref{sec:GWs}. Doubtless there will be more major results to come that use gravitational waves to constrain deviations from GR. Strong-field modifications to gravity can be constrained from the waveform, provided that parameters of the binary can be measured sufficiently well and that that adequate predictions can be made using numerical or semi-analytic relativity calculations. Given more events with electromagnetic counterparts, constraints on the gravitational wave luminosity distance and standard sirens can be brought into play \citep{2019arXiv190601593B}. Alternatively, it may be possible constrain cosmological parameters using `dark sirens' -- mergers without detected electromagnetic counterparts -- by marginalising over likely host galaxies within a the localisation volume of an event \citep{PhysRevD.86.043011, 2018MNRAS.475.3485D, 2019arXiv190806050G}.
\\
\item {\bf Testing more general parameterizations:} The background expansion rate of the universe has been found to be consistent with the predictions of $\Lambda$CDM to a high degree (although there is an intriguing tension between the value of $H_0$ measured locally and that inferred from the CMB and other probes). 
Much of the phenomenology of modified gravity has therefore been geared towards modified perturbation dynamics, while leaving the background expansion rate largely unchanged. As such, modern tests of gravity need to do more than simply constrain the background expansion history.
  In cosmology, parameters such as $\mu$ and $\Sigma$ (Sec.~\ref{subsec:paramvsmodel}) are easier to relate directly to gravity than general measurements of the growth rate $f$. All such parameterizations are however motivated by large-scale tests, and generally involve a redshift and scale dependence which is more challenging to constrain. For smaller scale tests, ideally functions or parameters of the action (or of a representative action in the case of a broader class of theories) may be constrained, as has been done recently for chameleon theories.
\\
\item {\bf Laboratory \& strong-field tests of gravity:} It was realized a few years ago \citep{2015JCAP...03..042B,2016ConPh..57..164B,2016JCAP...12..041B} that sufficiently low-density environments in vacuum chambers should allow tests of the chameleon screening mechanism in the laboratory. Likewise, tests of axion physics in the laboratory have been proposed. Any new ways to constrain screening or other modified gravity effects in the laboratory could be  powerful, considering the repeatable nature and sensitivity achievable with modern apparatus. Greater integration with the results of strong-field tests of gravity may also be achievable.\\



\end{itemize}

Many of the tests described above rely to some degree on an assumption of a background cosmology. The means that taking into account the effect of modifications of gravity on the inference of cosmological parameters becomes crucial. A notable example \citep{Lagos:2019kds} is the interplay between the effects of a running Planck mass and $H_0$. Thus, comprehensive analyses of astrophysical tests of gravity must incorporate a consistent inference of constraints on the parameters describing the background cosmology.

\begin{table*}[h!]
\caption{Summary of tests of modified gravity enabled by current and future surveys.
}
\begin{center}
\begin{threeparttable}
\begin{tabular}{>{\centering}m{2.5cm}>{\centering}m{1.3cm}>{\centering}m{0.5cm}>{\centering}m{1.4cm}>{\centering}m{0.8cm}>{\centering}m{0.6cm}>{\centering}m{0.8cm}>{\centering}m{1.1cm}>{\centering}m{0.7cm}>{\centering}m{0.7cm}>{\centering}m{1.1cm}@{}m{0pt}@{}}
\hline\hline 

Survey & Clusters I\tnote{a} & CMB (incl. ISW) & Clusters II\tnote{b} & Strong lens. & RSDs & Rot. curves & Cepheids (dist. indic.) & Voids & Stellar ev. & Galaxy dyn./ struct. & \\ \hline 

DES \\HSC & \ding{51} & \ding{51} & \ding{51} & \ding{51} & \ding{51} &  &  & \ding{51} &  & \ding{51} & \\ \hdashline

CMB & \ding{51} & \ding{51} & \ding{51} &  &  &  &  & \ding{51} &  &  & \\
\hline

DESI \\PFS &  & \ding{51} & \ding{51} &  & \ding{51} &  &  & \ding{51} &  &  & \\ \hdashline

LSST & \ding{51} & \ding{51} & \ding{51} & \ding{51} & \ding{51} &  & \ding{51} & \ding{51} & \ding{51} & \ding{51} & \\ \hdashline

WFIRST \\Euclid & \ding{51} & \ding{51} & \ding{51} & \ding{51} & \ding{51} &  &  & \ding{51} &  & \ding{51} & \\ \hdashline

Simons Obs. \\CMB S-4 & \ding{51} & \ding{51} & \ding{51} &  &  &  &  & \ding{51} &  &  & \\ \hdashline

HIRAX \\CHIME & ? & \ding{51} &  &  & ? &  &  &? &  & ? & \\ \hdashline

SKA & \ding{51} & \ding{51} & \ding{51} & \ding{51} & \ding{51} & \ding{51} &  & \ding{51} &  & \ding{51} & \\ \hdashline

SPHEREx\tnote{c} &  & \ding{51} & \ding{51} &  & \ding{51} &  &  & \ding{51} &  &  & \\ \hdashline

MaNGA &  &  &  &  &  & \ding{51} &  &  &  & \ding{51} & \\ \hdashline

SDSS-V \citep{Kollmeier:2017}\tnote{d} &  &  &  &  &  & \ding{51} &  &  & ? & ? & \\ \hdashline

4MOST \citep{deJong:2012}\tnote{e} &  & \ding{51} & \ding{51} &  & \ding{51} &  & \ding{51} & \ding{51} & ? & ? & \\ \hdashline

Gaia &  &  &  &  &  & \ding{51} & \ding{51} &  & \ding{51} & \ding{51} & \\ \hdashline

MUSE \citep{MUSE}\tnote{f} &  &  &  &  &  & \ding{51} &  &  & ? & \ding{51} & \\

 \hline \hline
\end{tabular}
\begin{tablenotes}
\item[a] \footnotesize{Cluster density profiles.}
\item[b] \footnotesize{Cluster abundances.}
\item[c] \footnotesize{\url{http://spherex.caltech.edu}}
\item[d] \footnotesize{\url{https://www.sdss.org/future/}}
\item[e] \footnotesize{\url{https://www.4most.eu/cms/}}
\item[f] \footnotesize{\url{https://www.eso.org/sci/facilities/develop/instruments/muse.html}}
\end{tablenotes}
\end{threeparttable}
\end{center}
\label{tab:surveys}
\end{table*}

\subsection{Outlook}

In this review, we have described the motivations for constructing and testing theories of gravity that may be active on weak-field astrophysical and cosmological scales, and have reviewed the theoretical, computational, and observational work that has been undertaken to date. We focus in particular on astrophysical probes. With several cosmological and galaxy surveys already underway and others on the horizon (summarised qualitatively in Table~\ref{tab:surveys}), the next decade of research in this field will be driven by ever richer datasets. The widening gap between the theoretical and observational communities makes enhancing our interaction of paramount importance to utilize fully the potential of this data. The Novel Probes Project is a step towards this goal, and we hope the reader is inspired to participate. Please visit \url{https://www.novelprobes.org} to join our Slack forum and become involved in this exciting opportunity.

\section*{Acknowledgements}

Marisa March helped get this review and the Novel Probes Project off the ground. We are very grateful for her energy, vision, and support in the early stages of this work. We thank Rachel Bean, Eric Linder, and Gong-Bo Zhao for helpful discussions and suggestions on early drafts. We have benefited from discussions with Philippe Brax, Claire Burrage, Justin Khoury, Bill Paxton,  and Mark Trodden.  \newline

TB is supported by the Royal Society, grant no. URF\_ R1\_{180009}. 
AB and FS acknowledge support from the Starting Grant (ERC-2015-STG 678652) ``GrInflaGal'' of the European Research Council.
HD is supported by St John's College, Oxford, and acknowledges financial support from ERC Grant No 693024 and the Beecroft Trust.
PGF is supported by the ERC, STFC and the Beecroft Trust.
BJ is supported in part by the US Department of Energy grant DE-SC0007901 and by NASA ATP grant NNH17ZDA001N.
KK is supported by the European Research Council through 646702 (CosTesGrav) and the UK Science and Technologies Facilities Council (STFC) grants ST/N000668/1 and ST/S000550/1. BL is supported by the European Research Council through Grant No. ERC-StG-PUNCA-716532, and the STFC through Consolidated Grants No. ST/I00162X/1 and No. ST/P000541/1. 
LL acknowledges support by a Swiss National Science Foundation (SNSF) Professorship grant (No.~170547), a SNSF Advanced Postdoc.Mobility Fellowship (No.~161058), and the STFC Consolidated Grant for Astronomy and Astrophysics at the University of Edinburgh.
JS is supported by funds made available to the Center for Particle Cosmology by the University of Pennsylvania. 

\bigskip{}

\bibliography{references.bib}

\end{document}